\documentclass[prd,preprint,longbibliography,
  showpacs,showkeys,lengthcheck,
  nofootinbib,tightenlines,onecolumn,notitlepage,
  preprintnumbers,superscriptaddress]{revtex4-1}

\usepackage{graphicx}
\usepackage[usenames,dvipsnames]{xcolor}
\graphicspath{{figures/}}
\usepackage{circuitikz}

\usepackage{hyperref}
\hypersetup{colorlinks=true,citecolor=blue,linkcolor=blue,urlcolor=blue}

\usepackage{diagbox}
\usepackage{booktabs}

\usepackage{amsmath,amssymb,amsfonts}
\usepackage{mathrsfs}
\usepackage{slashed}
\usepackage{bm}
\usepackage{bbm}

\usepackage{scalerel,stackengine}
\stackMath
\newcommand\reallywidehat[1]{%
\savestack{\tmpbox}{\stretchto{%
  \scaleto{%
    \scalerel*[\widthof{\ensuremath{#1}}]{\kern-.6pt\bigwedge\kern-.6pt}%
    {\rule[-\textheight/2]{1ex}{\textheight}}%WIDTH-LIMITED BIG WEDGE
  }{\textheight}%
}{0.6ex}}%
\stackon[1pt]{#1}{\tmpbox}%
}
\usepackage[draft]{changes} % Use for draft to show changes
\usepackage[utf8]{inputenc}
\usepackage{newtxtext}
\usepackage[mathcal]{euscript}

\renewcommand*{\d}{\mathop{}\!\mathrm{d}}
\newcommand*{\e}{\mathop{}\!\mathrm{e}}
\renewcommand*{\i}{\mathop{}\!\mathrm{i}}

\newcommand*{\E}{\mathcal{E}}
\renewcommand*{\P}{\mathcal{P}}
\newcommand*{\Efl}{\bm{\mathcal{F}}_E}

\begin{document}

\title{Synchronization effects on rest frame energy and momentum densities in the proton}

\author{Adam Freese}
\email{afreese@jlab.org}
\address{Theory Center, Jefferson Lab, Newport News, Virginia 23606, USA}
\address{Department of Physics, University of Washington, Seattle, WA 98195, USA}

\author{Gerald A. Miller}
\email{miller@uw.edu}
\address{Department of Physics, University of Washington, Seattle, WA 98195, USA}

\begin{abstract}
  We obtain two-dimensional relativistic densities and currents of
  energy and momentum in a proton at rest.
  These densities are obtained at surfaces of fixed light front time,
  which physically corresponds to using an alternative synchronization convention.
  Mathematically, this is done using tilted light front coordinates,
  which consist of light front time and ordinary spatial coordinates.
  In this coordinate system, all sixteen components of the
  energy-momentum tensor
  obtain clear physical interpretations,
  and the nine Galilean components reproduce results from
  standard light front coordinates.
  We find angular modulations in several densities that are absent
  in the corresponding instant form results,
  which are explained as optical effects arising from
  using fixed light front time when motion is present within the target.
  Additionally, transversely-polarized spin-half targets
  exhibit an energy dipole moment---which evaluates to
  $-1/4$ for all targets if the Belinfante EMT is used,
  but which is target dependent and vanishes for pointlike fermions
  if the asymmetric EMT is instead used.
\end{abstract}

\preprint{JLAB-THY-23-3889}
\preprint{NT@UW-23-10}

\maketitle

%%%%%%%%%%%%%%%%%%%%%%%%%%%%%%%%%%%%%%%%%%%%%%%%%%%%%%%%%%%%%%%%%%%%%%%%%%%%%%%%

\section{Introduction}
\label{sec:intro}

Significant attention has been placed on
the energy momentum tensor (EMT)
and the associated gravitational form factors~\cite{Kobzarev:1962wt}
over the past few years.
Major questions in the field of hadron physics,
such as the
proton mass puzzle~\cite{Ji:1994av,Ji:1995sv,Lorce:2017xzd,Hatta:2018sqd,Metz:2020vxd,Ji:2021mtz,Lorce:2021xku}
and proton spin puzzle~\cite{Ashman:1987hv,Ji:1996ek,Leader:2013jra,Wakamatsu:2014zza,Ji:2020ena}
are directly related to the EMT.
Additionally, there has been much discussion over how (and whether)
the EMT encodes distributions of static forces
within hadrons~\cite{Polyakov:2002yz,Polyakov:2018zvc,Lorce:2018egm,Freese:2021czn,Lorce:2021xku,Ji:2021mfb}.
This attention is especially pertinent with the anticipated construction
of the Electron Ion Collider~\cite{Boer:2011fh,Accardi:2012qut,AbdulKhalek:2021gbh},
since the measurement of generalized parton distributions~\cite{Ji:1996nm,Radyushkin:1997ki,Belitsky:2005qn}
is the most promising means of empirically accessing
the gravitational form factors.

Currently the literature is filled with a variety of perspectives on how
to obtain spatial distributions of local currents in composite systems,
including those encoded by the EMT
(see for instance Refs.~\cite{Fleming:1974af,Burkardt:2000za,Polyakov:2018zvc,Miller:2018ybm,Lorce:2018egm,Epelbaum:2022fjc,Li:2022ldb,Chen:2022smg,Freese:2022fat,Panteleeva:2022uii,Freese:2023jcp,Chen:2023dxp,Panteleeva:2023evj}).
The light front formalism stands out among these as
providing relativistically exact 2D densities~\cite{Burkardt:2002hr,Miller:2010nz}
that are obtained from elementary field-theoretic definitions~\cite{Burkardt:2000za,Freese:2021czn}
in a wave-packet-independent way~\cite{Freese:2022fat}.
Misgivings have been expressed about the light front densities
with the understanding that they constitute a description of the system
moving at infinite momentum~\cite{Lorce:2020onh}.
However,
in a recent work~\cite{Freese:2023jcp},
we showed that light front densities constitute
\emph{rest frame densities}
within hadrons at a fixed light front time
by utilizing a coordinate system called
tilted light front coordinates (or tilted coordinates):
\begin{subequations}
  \label{eqn:tilted}
  \begin{align}
    \tau = x^0 &\equiv t_{\text{IF}} + z_{\text{IF}} \\
    x    = x^1 &\equiv x_{\text{IF}} \\
    y    = x^2 &\equiv y_{\text{IF}} \\
    z    = x^3 &\equiv z_{\text{IF}}
    \,,
  \end{align}
\end{subequations}
first proposed by Blunden, Burkardt and Miller~\cite{Blunden:1999wb}.
By using light front time $\tau$ but ordinary Cartesian spatial coordinates
$(x,y,z)$,
the Galilean subgroup of the Poincar\'{e} group can be exploited
while utilizing everyday intuition about space,
including that a target is at rest when
$(v_x, v_y, v_z) = (0,0,0)$.

Operationally, the use of tilted coordinates corresponds to synchronizing
spatially distant clocks under the assumption that the speed of light
is infinite in the $-z$ direction,
and consequently the light front densities constitute a literal picture
of what an observer looking in the $+z$ direction sees
when their local time is $\tau$.
In our prior work~\cite{Freese:2023jcp} we refer to this
synchronization rule as \emph{light front synchronization}.
Light front synchronization stands in contrast to
the standard Einstein synchronization convention~\cite{Einstein:1905ve},
under which spatially distant clocks are synchronized by assuming
that the one-way speed of light is isotropic and equal to $c$ in all directions.
Using Einstein synchronization
results in the standard Minkowski (or instant form) coordinate system,
in which the observer is understood to see a past state of
the system they are observing.
(See Refs.~\cite{reichenbach2012philosophy,gruenbaum2012philosophical,Zhang:1995test,Anderson:1998mu,Veritasium:2020oct} for detailed discussions of synchronization conventions.)

Previously, we obtained the rest frame electromagnetic currents
of the proton and neutron in tilted light front coordinates~\cite{Freese:2023jcp}.
The purpose of the present work is to obtain the energy and momentum currents
encoded by the EMT within the same formalism.
A variety of EMT densities already exist in the literature in different formalisms,
but the tilted coordinate framework offers a number of advantages
that make the presentation of new EMT densities worthwhile.
Much like the standard light front densities,
the densities obtained in tilted coordinates are relativistically exact,
while the more commonly-used Breit frame densities are
leading-order contributions that dominate for
spatially diffuse wave packets~\cite{Li:2022ldb,Freese:2022fat},
and are as such subject to relativistic corrections~\cite{Polyakov:2018zvc}.
Moreover, when localizing wave packets in instant form coordinates,
the resulting densities differ from the Breit frame
densities~\cite{Panteleeva:2022khw,Panteleeva:2022uii,Panteleeva:2023evj},
since the dominating term in an infinite series differs for
localized wave packets~\cite{Li:2022ldb}.
The standard light front and tilted light front densities, by contrast,
are fully independent of the target's wave packet~\cite{Freese:2022fat,Freese:2023jcp}.

There are also several advantages to using tilted light front coordinates
over standard light front coordinates when obtaining densities.
One of these is
the ability to clearly show that the results are rest frame densities.
Additionally,
for local currents such as the electromagnetic current $j^\mu(x)$
and the energy-momentum tensor $T^{\mu\nu}(x)$,
every component of the current obtains a clear physical interpretation
in tilted coordinates.
By contrast, the components $j^-(x)$, $T^{i-}(x)$ and $T^{--}(x)$
do not have clear interpretations in standard light front coordinates,
and accordingly are typically ignored.
In this work, we will present results for all sixteen components of
the proton's EMT density.

One last benefit of tilted coordinates over standard light front
coordinates is that the tilted energy $E$ is exactly equal to
the standard instant form energy,
and that the tilted energy density
\begin{multline}
  \label{eqn:energy:half}
  \E(\bm{b}_\perp,\hat{\bm{s}})
  =
  m
  \int \frac{\d^2\bm{\varDelta}_\perp}{(2\pi)^2}
  \Bigg(
  A(-\bm{\varDelta}_\perp^2)
  +
  \frac{ \bm{\varDelta}_\perp^2 }{4m^2}
  D(-\bm{\varDelta}_\perp^2)
  \\
  +
  \frac{
    (\hat{\bm{s}} \times \i \bm{\varDelta}_\perp) \cdot \hat{e}_z
  }{2m}
  \left\{
    B(-\bm{\varDelta}_\perp^2)
    -
    J(-\bm{\varDelta}_\perp^2)
    -
    \frac{ \bm{\varDelta}_\perp^2 }{4m^2}
    D(-\bm{\varDelta}_\perp^2)
    \right\}
  \Bigg)
  \e^{-\i\bm{\varDelta}_\perp\cdot\bm{b}_\perp}
\end{multline}
is thus an exact 2D relativistic distribution of the
usual energy $E$, rather than of $P^-$.
The tilted energy density
is thus more pertinent to debates about the proton mass decomposition,
which typically frame the mass decomposition as an energy
decomposition~\cite{Ji:1994av,Ji:1995sv,Lorce:2017xzd,Metz:2020vxd,Ji:2021mtz,Lorce:2021xku}.
(Eq.~(\ref{eqn:energy:half}) will be proved below in Sec.~\ref{sec:half}
after the necessary formalism has been developed.
Table~\ref{tab:EMT:compare} can be consulted to quickly find
explicit results to the EMT densities,
as well as their analogues in standard light front coordinates
and the Breit frame formalism.)

Tilted coordinates have several unfamiliar mathematical properties,
and this work is not intended as an introduction to them.
We have compiled a collection of helpful basic properties
and identities in Appendix~\ref{sec:basic}
for easy access,
but a full exposition of the coordinate system is given in
Ref.~\cite{Freese:2023jcp}.
The remainder of this work uses tilted coordinates,
and contains occasional reminders of their idiosyncratic properties.

\begin{table}[t]
  \setlength{\tabcolsep}{0.5em}
  \renewcommand{\arraystretch}{1.3}
  \begin{tabular}{@{} rcccc @{}}
    \toprule
    % ~~~~~~~~~~~~~~~~~~~~~~~~~~~~~~~~~~
    % Column labels
    ~ &
    Component &
    Breit frame &
    Standard light front &
    Tilted coordinates \\
    \hline
    % ~~~~~~~~~~~~~~~~~~~~~~~~~~~~~~~~~~
    % Energy density
    Energy density &
    $T^0_{\phantom{0}0}(\bm{x})$ &
    Eq.~(17a) of \cite{Polyakov:2018zvc} &
    --- &
    Eq.~(\ref{eqn:energy:half}) \\
    \hline
    % ~~~~~~~~~~~~~~~~~~~~~~~~~~~~~~~~~~
    % Momentum density
    Momentum density &
    $-T^0_{\phantom{0}i}(\bm{x})$ &
    Eq.~(17c) of \cite{Polyakov:2018zvc} &
    \begin{tabular}{@{}c@{}}
    Eqs.~(11) \& (20) of Ref.~\cite{Freese:2021mzg} [long.] \\
    same as Eq.~(\ref{eqn:momentum:half:perp}) [trans.]
    \end{tabular}
    &
    Eq.~(\ref{eqn:momentum:half}) \\
    \hline
    % ~~~~~~~~~~~~~~~~~~~~~~~~~~~~~~~~~~
    % Energy fluxes
    Energy flux density &
    $T^i_{\phantom{i}0}(\bm{x})$ &
    same as Eq.~(17c) of \cite{Polyakov:2018zvc} &
    same as Eq.~(\ref{eqn:flux:half}) [trans.\ only] &
    Eq.~(\ref{eqn:flux:half}) \\
    \hline
    % ~~~~~~~~~~~~~~~~~~~~~~~~~~~~~~~~~~
    % Stress tensor
    Stress tensor &
    $-T^i_{\phantom{i}j}(\bm{x})$ &
    Eq.~(17b) of \cite{Polyakov:2018zvc} &
    Eq.~(21) of \cite{Freese:2021mzg} [trans.\ only] &
    Eq.~(\ref{eqn:stress:half}) \\
    \bottomrule
  \end{tabular}
  \caption{
    Explicit results for EMT densities of spin-half targets in
    the Breit frame formalism,
    the standard light frame formalism,
    and in tilted coordinates can be found in the references
    and equations provided in this table.
    The references have been chosen for easy consultation
    and for providing formulas for arbitrary polarization,
    rather than for original discovery.
    In several cases, standard light front results do not exist,
    or only exist for transverse components.
    Reference~\cite{Lorce:2018egm} provides a light front $P^-$ density
    in its Eq.~(107),
    but is excluded from the table because $P^- \neq E$ and because
    the result is only for unpolarized targets.
    In several other cases, standard light front densities
    are obtainable, but we could not find results for them in the literature,
    so we have pointed to equivalent formulas in the present work.
  }
  \label{tab:EMT:compare}
\end{table}

This work is organized as follows.
In Sec.~\ref{sec:currents},
we explain how components of the energy-momentum tensor are interpreted
as furnishing densities and flux densities of energy and momentum,
and provide a dictionary for converting components of the EMT
into energy and momentum currents.
In Sec.~\ref{sec:convolution},
we explore how expectation values of the EMT for physical states
can be decomposed into an internal rest-frame distributions
and state-dependent smearing functions,
the latter of which absorbs dependencies on the target's overall motion.
Next, in Sec.~\ref{sec:zero}, we obtain the rest frame energy and momentum
currents for a spin-zero target as a warm-up exercise.
Sec.~\ref{sec:half} then provides expressions for the rest frame
EMT densities of a spin-half target
as well as numerical examples for a proton.
Finally, we conclude in Sec.~\ref{sec:end}.

Throughout this work---and in contrast to our previous
work on the subject~\cite{Freese:2023jcp}---we do not include any special
markings (such as a tilde) to indicate that tilted coordinates are being used.
Unless explicitly indicated otherwise
(such as by a subscript or superscript IF for ``instant form''),
all non-invariant quantities should be assumed to signify a quantity in tilted coordinates.

%%%%%%%%%%%%%%%%%%%%%%%%%%%%%%%%%%%%%%%%%%%%%%%%%%%%%%%%%%%%%%%%%%%%%%%%%%%%%%%%

\section{Energy and momentum currents in tilted coordinates}
\label{sec:currents}

The energy-momentum tensor (EMT) is a local operator characterizing
the distribution and flow of energy and momentum of a system.
In quantum chromodynamics (QCD), the operator is formally given by~\cite{Kugo:1979gm,Leader:2013jra}:
\begin{align}
  \label{eqn:emt:qcd}
  \hat{T}_{\text{QCD}}^{\mu\nu}
  =
  \sum_q \frac{i}{4} \bar q \gamma^{\{\mu} \overleftrightarrow{D}^{\nu\}} q
  + F_a^{\mu\rho} F_{\rho}^{a\,\nu}
  - A_a^{\{\mu}(\partial^{\nu\}} B_a
  -i (D^{\{\mu} c) (\partial^{\nu\}} \bar{c})
  -
  g^{\mu\nu}
  \mathscr{L}_{\mathrm{QCD}}
  \,,
\end{align}
where
$\mathscr{L}_{\mathrm{QCD}}$ is the QCD Lagrangian:
\begin{align}
  \label{eqn:lagrangian:qcd}
  \mathscr{L}_{\mathrm{QCD}}
  =
  \sum_q
  \bar{q}
  \left(
  \frac{i}{2}
  \overleftrightarrow{\slashed{\partial}}
  +
  g\slashed{A}_a T^a
  -
  m_q
  \right)
  q
  -
  \frac{1}{4} F^a_{\mu\nu} F_a^{\mu\nu}
  - (\partial_\mu B_a) A^\mu_a
  +
  \frac{\alpha_0}{2} B_a^2
  -
  i (\partial_\mu \bar{c}^a) (D^\mu_{ab} c^b)
  \,.
\end{align}
Here $A^\mu_a$ is the gluon four-potential,
$B_a$ are Lagrange multiplier fields and $c_a$ and $\bar{c}_a$
are the Faddeev-Popov ghosts.
The Lagrange multiplier and ghost fields are unphysical and annihilate
physical states,
but are necessary to quantize and renormalize the theory~\cite{Kugo:1979gm}.
The different representations of the gauge-covariant derivative are:
\begin{subequations}
  \begin{align}
    \overrightarrow{D}_\mu q
    &=
    \overrightarrow{\partial_\mu} q
    -
    ig A^a_\mu T_a q
    \\
    \bar{q} \overleftarrow{D}_\mu
    &=
    \bar{q} \overleftarrow{\partial_\mu}
    +
    ig \bar{q} A^a_\mu T_a
    \\
    D_\mu^{ab} c^b
    &=
    \Big(
    \delta_{ab} \partial_\mu
    + g f_{acb} A_\mu^c
    \Big)
    c^b
    \,,
  \end{align}
\end{subequations}
and the gluon field strength tensor is:
\begin{align}
  F_{\mu\nu}^a
  =
  \partial_\mu
  A_\nu^a
  -
  \partial_\nu
  A_\mu^a
  +
  g f_{abc} A_\mu^b A_\nu^c
  \,.
\end{align}
Here, $T_a$ are the generators of the color $\mathfrak{su}(3,\mathbb{C})$
algebra and $f_{abc}$ are the totally antisymmetric structure constants
defined by:
\begin{align}
  [T_a, T_b] = i f_{abc} T_c
  \,.
\end{align}
The EMT can be derived through several methods.
Noether's first theorem and invariance of the QCD action under global spacetime translations
infamously results in an EMT that is not gauge
invariant~\cite{Kugo:1979gm,Belitsky:2005qn,Leader:2013jra},
but this is rectified through the Belinfante improvement procedure~\cite{Belinfante:1939emt},
which adds a trivially conserved quantity to the EMT in order to restore gauge invariance.
The trivially conserved quantity is usually chosen to reproduce Eq.~(\ref{eqn:emt:qcd}).
However, Leader and Lorc\'{e}~\cite{Leader:2013jra} show that an alternative EMT can be obtained,
with an additional antisymmetric piece $\hat{T}^{\mu\nu}_{\text{A}}(x)$:
\begin{align}
  \label{eqn:emt:anti}
  \hat{T}^{\mu\nu}_{\text{asym}}(x)
  &=
  \hat{T}^{\mu\nu}_{\text{QCD}}(x)
  +
  \sum_q \left\{
    \frac{1}{2}\bar{q}(x) \gamma^{[\mu} \i\overleftrightarrow{D}^{\nu]} q(x)
    \right\}
  \equiv
  \hat{T}^{\mu\nu}_{\text{QCD}}(x)
  +
  \hat{T}^{\mu\nu}_{\text{A}}(x)
  \,.
\end{align}
The antisymmetric piece is interpreted as describing intrinsic fermion spin;
see Ref.~\cite{Leader:2013jra} for further details.

The EMT can alternatively be derived using Noether's second theorem
while assuming invariance of the QCD action under local spacetime translations\cite{Freese:2021jqs}.
If fermion fields transform according to their Lie derivative under these local translations,
the resulting EMT is exactly that in Eq.~(\ref{eqn:emt:qcd}).
The EMT in Eq.~(\ref{eqn:emt:qcd}) can also be obtained by taking the functional derivative
of the QCD action with respect to the metric tensor~\cite{Belitsky:2005qn}
or with respect to the vierbein~\cite{Kugo:1979gm}.
These methods avoid the need for an improvement procedure to ensure gauge invariance,
and lack an ambiguity about the resulting EMT.

\begin{figure}
  \centering
  \includegraphics[scale=1]{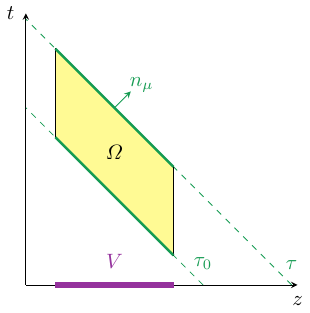}
  \caption{
    A finite spacetime region $\varOmega$ bounded by two hypersurfaces
    of equal light front time $\tau_0$ and $\tau$,
    drawn in terms of instant form coordinates.
    Each slice of fixed light front time contains the same spatial region $V$.
    The future-directed normal $n_\mu$ to the equal-light-front-time hypersurfaces
    is also indicated in this diagram.
  }
  \label{fig:spacetime}
\end{figure}

Regardless of whether the antisymmetric piece
$\hat{T}^{\mu\nu}_{\text{A}}(x)$ is included in the EMT,
integrals of the EMT over equal-time surfaces reproduce the generators of spacetime translations,
as a consequence of being the conserved Noether current associated with
spacetime translation symmetry.
If $V$ is a fixed-time hypersurface and $n_\mu$ is a unit forward-directed
normal to this surface:
\begin{align}
  \label{eqn:P}
  \hat{P}^\nu(\tau)
  =
  \int_{V} \d^3 \bm{x} \,
  n_\mu
  \hat{T}^{\mu\nu}(x,\tau)
  \,,
\end{align}
where $\tau$ is the time variable under consideration.
If instant form time (the time resulting from Einstein synchronization)
is used to define equal-time surfaces,
then $n_\mu$ is a timelike vector pointing in the forward-$t_{\text{IF}}$ direction.
If light front time $t_{\text{IF}} + z$ is instead used to define
equal-time surfaces,
$n_\mu$ is a lightlike vector pointing along the light cone.
The latter scenario is depicted with a finite hypersurface in Fig.~\ref{fig:spacetime}.
If $V$ is extended to all of space,
then $\hat{P}^\nu$ is conserved, and thus time-independent,
by virtue of Noether's theorems.

The four-vector operator $\hat{P}_\nu$ plays the role of
a spacetime translation generator,
specifically in its covariant form (with a lower index):
\begin{align}
  \i [ \hat{P}_\nu, \hat{O}(x) ]
  =
  \partial_\nu \hat{O}(x)
  \,.
\end{align}
The contravariant (upper-index) components of the four-momenta
are related to the covariant components through
$\hat{P}^\nu = g^{\nu\rho} \hat{P}_\rho$.
In instant form coordinates, this gives a trivial relationship
for components of the vector momentum:
\begin{align}
  \hat{\bm{P}}_{\text{IF}}
  =
  (\hat{P}^1_{\text{IF}}, \hat{P}^2_{\text{IF}}, \hat{P}^3_{\text{IF}})
  =
  (-\hat{P}_1^{(\text{IF})}, -\hat{P}_2^{(\text{IF})}, -\hat{P}_3^{(\text{IF})})
  \,,
\end{align}
but in tilted coordinates the relationship is more complicated---see
Eq.~(\ref{eqn:appa:metric}) for the metric tensor in tilted coordinates
and Eq.~(\ref{eqn:appa:covcon}) for the covariant-contravariant relations.
In order to play their proper role as space translation generators,
components of vector momentum are identified through
\emph{covariant} components of the four-momentum:
$\hat{\bm{P}} = (-\hat{P}_1, -\hat{P}_2, -\hat{P}_3)$.
Likewise, the Hamiltonian (as the time translation generator)
is given by $\hat{P}_0$.
Accordingly, the energy and momentum densities are associated
with the mixed upper-lower form of the EMT,
$\hat{T}^{\mu}_{\phantom{\mu}\nu}(x)$,
which can be interpreted as a $\hat{P}_\nu$ current.
As such,
$\hat{T}^{\mu}_{\phantom{\mu}0}(x)$
gives an energy-four current---a combination of an energy density
and energy flux density---while
$-\hat{T}^{\mu}_{\phantom{\mu}i}(x)$
encodes three vector momentum currents.

As is standard in continuum mechanics~\cite{Fetter:1980str,Batchelor:2000int,Irgens:2008con,Friedman:2013xza,flugge2013tensor,liu2013continuum},
flux densities of momentum can be interpreted as stresses.
We review the rationale behind this.
By virtue of Noether's theorems,
the EMT obeys a continuity equation:
\begin{align}
  \partial_\mu
  T^{\mu}_{\phantom{\mu}\nu}(x)
  =
  0
\end{align}
which is the differential form of energy-momentum conservation.
If we integrate this differential form over the spacetime region
$\varOmega$ depicted in Fig.~\ref{fig:spacetime}
and use the divergence theorem,
we obtain the integral form of the conservation law:
\begin{align}
  \frac{\d}{\d\tau}
  \left[
    T^{0}_{\phantom{0}\nu}(\bm{x},\tau)
    \right]
  =
  -
  \oint_{\partial V} \d A_i \,
  T^{i}_{\phantom{i}\nu}(\bm{x},\tau)
  \,,
\end{align}
where $\partial V$ is the boundary of the spatial region $V$
and $\d \bm{A}$ is an area element with outward-pointing normal.
This equation describes the amount of $P_\nu$
in a spatial region $V$ changing due to the flux of $P_\nu$
through the boundary of this region.
For this reason,
$T^{0}_{\phantom{0}\nu}(\bm{x},\tau)$
is the $P_\nu$ density
and
$T^{i}_{\phantom{i}\nu}(\bm{x},\tau)$
is the $P_\nu$ flux density.
For $\nu \in \{1,2,3\}$ this equation describes a net change of momentum
in the region:
\begin{align}
  \bm{F}_V(\tau)
  \equiv
  -
  \frac{\d}{\d\tau}
  \left[
    \bm{P}_V(\tau)
    \right]
  =
  -
  \hat{e}_j
  \oint_{\partial V} \d A_i \,
  T^{i}_{\phantom{i}j}(\bm{x},\tau)
  \,.
\end{align}
Since momentum is leaving (or entering) the region $V$,
this will be felt by the region's surroundings as a force
$\bm{F}_V(\tau)$, exerted \emph{by} the region,
which would be measured for instance by a hypothetical pressure gauge
placed at the boundary $\partial V$.
Accordingly,
$ - \hat{e}_j T^{i}_{\phantom{i}j}(\bm{x},\tau) $
is a force per unit area on a surface with a unit normal $\hat{e}_i$,
and thus has a straightforward interpretation as a pressure.
More generally,
$ - T^{i}_{\phantom{i}j}(\bm{x},\tau) $
is referred to as the stress tensor,
and encodes the pressures that would be measured on a surface
in any orientation.

For a system in equilibrium,
one will have zero net force exerted by any region $V$,
and thus equal fluxes of momentum into and out of any region.
In integral form, the equilibrium condition is:
\begin{align}
  \hat{e}_j
  \oint_{\partial V} \d A_i \,
  T^{i}_{\phantom{i}j}(\bm{x},\tau)
  =
  0
  \,,
\end{align}
but the divergence theorem can be used to require this in differential form:
\begin{align}
  \partial_i
  T^{i}_{\phantom{i}j}(\bm{x},\tau)
  =
  0
  \,.
\end{align}
This is possible even when
$T^{i}_{\phantom{i}j}(\bm{x},\tau) \neq 0$.
If the stress tensor is non-zero in an equilibrium system,
this means that static pressures will be felt, and
in general the pressures will be anisotropic.
The components of
$S^{ij}(\bm{x},\tau)\equiv-T^{i}_{\phantom{i}j}(\bm{x},\tau)$
are referred to as stresses
and
$S^{ij}(\bm{x},\tau)$ itself as the stress tensor,
and these have an interpretation as furnishing mechanical pressures in a variety
of continuum systems~\cite{Fetter:1980str,Irgens:2008con,flugge2013tensor,liu2013continuum},
including
fluids~\cite{Batchelor:2000int},
solids~\cite{bower2009applied},
liquid crystals~\cite{10.1122/1.548883,PhysRevLett.26.1016,PhysRevE.103.012705},
and
neutron stars~\cite{doi:10.1098/rsta.1992.0074,Friedman:2013xza}.
(See also Ref.~\cite{PhysRevA.6.2401} for a unified treatment of liquids,
crystals, and liquid crystals.)
Since the fundamental ontological objects of quantum field theory are fields
rather than particles,
it is sensible to interpret QCD as a theory of a continuous medium as well,
and to interpret components of the operator
$-\hat{T}^{i}_{\phantom{i}j}(x)$
as stresses in this medium.

Although the mixed upper-lower form of the EMT has the most straightforward
interpretation in terms of energy-momentum four-currents,
it is convenient to work with tensors having all upper indices.
The rules for raising and lowering indices in Eq.~(\ref{eqn:appa:covcon})
can be used to rewrite the energy-momentum four-currents entirely in terms of
$T^{\mu\nu}(x)$.
In light of this,
the following dictionary can be quickly consulted to
ascribe physical meanings to components of the EMT in tilted coordinates:
\begin{subequations}
  \label{eqn:dictionary}
  \begin{itemize}
    \item
      \textbf{Energy density}:
      \begin{align}
        \E(x)
        =
        T^0_{\phantom{0}0}(x)
        =
        T^{00}(x)
        -
        T^{03}(x)
      \end{align}
    \item \textbf{Energy flux density}:
      \begin{align}
        \Efl(x)
        =
        T^i_{\phantom{0}0}(x)
        \hat{e}_i
        =
        \big(
        T^{i0}(x)
        -
        T^{i3}(x)
        \big)
        \hat{e}_i
      \end{align}
    \item \textbf{Momentum density}:
      \begin{align}
        \bm{\P}(x)
        =
        -
        T^0_{\phantom{0}i}(x)
        \hat{e}_i
        =
        T^{01}(x)
        \hat{e}_x
        +
        T^{02}(x)
        \hat{e}_y
        +
        T^{00}(x)
        \hat{e}_z
      \end{align}
    \item \textbf{Stress tensor} (i.e., momentum flux densities):
      \begin{align}
        S^{ij}(x)
        =
        -
        T^i_{\phantom{0}j}(x)
        =
        T^{i1}(x)
        \delta^{j1}
        +
        T^{i2}(x)
        \delta^{j2}
        +
        T^{i0}(x)
        \delta^{j3}
      \end{align}
  \end{itemize}
\end{subequations}

%%%%%%%%%%%%%%%%%%%%%%%%%%%%%%%%%%%%%%%%%%%%%%%%%%%%%%%%%%%%%%%%%%%%%%%%%%%%%%%%

\section{Convolution formalism for physical currents}
\label{sec:convolution}

In our previous work~\cite{Freese:2022fat,Freese:2023jcp},
we suggested that physical relativistic densities be identified as
expectation values of local currents for physical states,
which include information about how the system is prepared
(in particular, its wave packet).
If a physical state is described by a density matrix $\hat{\rho}$
(which is equal to $|\varPsi\rangle\langle\varPsi|$ for a pure state),
this expectation value can written (in the Heisenberg picture):
\begin{align}
  \label{eqn:j:phys}
  \langle J^\mu(x) \rangle
  =
  \mathrm{Tr}\big[
    \hat{\rho}
    \hat{J}^\mu(x)
    \big]
  \xrightarrow[\text{pure state}]{}
  \langle\varPsi|\hat{J}^{\mu}(x)|\varPsi\rangle
  \,.
\end{align}
The central idea of Ref.~\cite{Freese:2023jcp} is that if the density is considered
at fixed \emph{light front time} $\tau$ rather than fixed Minkowski time,
and if the longitudinal coordinate is integrated out,
then the physical density can be factorized
into a piece that depends only on its intrinsic structure
and a universal smearing function,
the latter of which absorbs all wave packet dependence.
In particular\footnote{
  Ref.~\cite{Freese:2023jcp} used three-dimensional smearing functions
  depending on $\bm{R}$, but since the $R^3$ coordinate does not appear
  in the intrinsic density, no information is lost by using
  $\mathscr{P}^\mu_{\phantom{\mu}\nu}(\bm{R}_\perp,\tau)
  \equiv
  \int \d R^3 \,
  (\mathscr{P}_{\text{3D}})^\mu_{\phantom{\mu}\nu}(\bm{R},\tau)$
  instead.
}:
\begin{align}
  \label{eqn:conv:j}
  \langle J^\mu(\bm{x}_\perp,\tau) \rangle_{\text{2D}}
  \equiv
  \int \d x^3 \,
  \langle J^\mu(x) \rangle
  =
  \sum_{\lambda,\lambda'}
  \int \d^2\bm{R}_\perp \,
  \mathscr{P}^\mu_{\phantom{\mu}\nu}(\bm{R}_\perp,\tau;\lambda,\lambda')
  j^\nu(\bm{x}_\perp-\bm{R}_\perp;\lambda,\lambda')
  \,,
\end{align}
where
$j^\nu$
is the intrinsic four-current density,
and
$\mathscr{P}^\mu_{\phantom{\mu}\nu}$
is the smearing function.
The intrinsic density retains no information about the system's wave packet,
encoding only information about its internal structure
(which will differ between different hadron species),
while the smearing function is independent of hadron species
and absorbs all wave packet dependence.
The very possibility of this factorization requires the use of
light front synchronization,
as proved in Appendix B of Ref.~\cite{Freese:2023jcp}.

This relation should generalize to the energy-momentum tensor:
\begin{align}
  \label{eqn:conv:emt}
  \langle T^{\mu\nu}(\bm{x}_\perp,\tau) \rangle_{\text{2D}}
  =
  \int \d x^3 \,
  \mathrm{Tr}\big[
    \hat{\rho}
    \hat{T}^{\mu\nu}(x)
    \big]
  =
  \sum_{\lambda,\lambda'}
  \int \d^2\bm{R}_\perp \,
  \mathscr{Q}^{\mu\nu}_{\phantom{\mu\nu}\alpha\beta}(\bm{R}_\perp,\tau;\lambda,\lambda')
  t^{\alpha\beta}(\bm{x}_\perp-\bm{R}_\perp;\lambda,\lambda')
  \,,
\end{align}
where here
$t^{\alpha\beta}$ is the intrinsic EMT and
$\mathscr{Q}^{\mu\nu}_{\phantom{\mu\nu}\alpha\beta}$ is the smearing function.

In Ref.~\cite{Freese:2023jcp}, we gave explicit formulas
for individual components of the intrinsic current and associated
smearing functions.
The formulas given therein can be written more compactly as:
\begin{align}
  \label{eqn:smear:current}
  \mathscr{P}^\mu_{\phantom{\mu}\nu}(\bm{R}_\perp,\tau;\lambda,\lambda')
  =
  \int \frac{\d^3\bm{P}}{2P_z(2\pi)^3}
  \int \frac{\d^2\bm{\varDelta}_\perp}{(2\pi)^2}
  \langle \bm{p},\lambda | \hat{\rho} | \bm{p}',\lambda' \rangle
  \frac{m}{P_z}
  \bar{\varLambda}^\mu_{\phantom{\mu}\nu}
  \e^{\i\varDelta_0\tau}
  \e^{-\i\bm{\varDelta}_\perp\cdot\bm{R}_\perp}
  \Bigg|_{\varDelta_z=0}
\end{align}
for the smearing function,
where $\varDelta_0 = p'_0 - p_0 = (\bm{P}\cdot\bm{\varDelta}_\perp)/P_z$,
and:
\begin{align}
  \label{eqn:current}
  j^\mu(\bm{b}_\perp;\lambda,\lambda')
  =
  \int \frac{\d^2\bm{\varDelta}_\perp}{(2\pi)^2}
  (\bar\varLambda^{-1})^{\mu}_{\phantom{\mu}\nu}
  \frac{
    \langle \bm{p}',\lambda' | \hat{J}^\nu(0) | \bm{p},\lambda \rangle
  }{2m}
  \e^{-\i\bm{\varDelta}_\perp\cdot\bm{b}_\perp}
  \Bigg|_{\varDelta_z=0}
\end{align}
for the intrinsic electromagnetic current.
Here, $\bm{b}_\perp = \bm{x}_\perp - \bm{R}_\perp$ is the impact parameter,
and additional momentum variables are given by
$P = \frac{1}{2}\big( p + p' \big)$ and $\varDelta = p' - p$.
The matrix $\bar{\varLambda}$ appearing in these formulas is given by:
\begin{align}
  \label{eqn:boost}
  \bar{\varLambda}^{\mu}_{\phantom{\mu}\nu}
  =
  \left[
    \begin{array}{cccc}
      P_z/m & 0 & 0 & 0 \\
      P_x/m & 1 & 0 & 0 \\
      P_y/m & 0 & 1 & 0 \\
      (P_z-P_0)/m & -P_x/P_z & -P_y/P_z & m/P_z
    \end{array}
    \right]
  %=
  %\frac{1}{2}\Big(
  %  \varLambda^{\mu}_{\phantom{\mu}\nu}(\bm{p})
  %  +
  %  \varLambda^{\mu}_{\phantom{\mu}\nu}(\bm{p}')
  %  \Big)
  \,.
\end{align}
It should be noted that despite its similar appearance
to Eq.~(\ref{eqn:appa:boost}),
$\bar{\varLambda}^\mu_{\phantom{\mu}\nu}$ is not a Lorentz boost,
as demontrated by Eq.~(\ref{eqn:notboost}).

For the energy-momentum tensor,
the analogues of the equations above are:
\begin{align}
  \label{eqn:smear:emt}
  \mathscr{Q}^{\mu\nu}_{\phantom{\mu\nu}\alpha\beta}(\bm{R}_\perp,\tau;\lambda,\lambda')
  =
  \int \frac{\d^3\bm{P}}{2P_z(2\pi)^3}
  \int \frac{\d^2\bm{\varDelta}_\perp}{(2\pi)^2}
  \langle \bm{p},\lambda | \hat{\rho} | \bm{p}',\lambda' \rangle
  \frac{m}{P_z}
  \Big(
  \bar{\varLambda}^\mu_{\phantom{\mu}\alpha}
  \bar{\varLambda}^\nu_{\phantom{\nu}\beta}
  -
  \frac{\bm{\varDelta}_\perp^2}{4P_z^2}
  \delta^\mu_3
  \delta^\nu_3
  \delta^3_\alpha
  \delta^3_\beta
  \Big)
  \e^{\i\varDelta_0\tau}
  \e^{-\i\bm{\varDelta}_\perp\cdot\bm{R}_\perp}
  \Bigg|_{\varDelta_z=0}
\end{align}
for the smearing function and:
\begin{subequations}
  \label{eqn:emt}
  \begin{align}
    t^{\mu\nu}(\bm{b}_\perp;\lambda,\lambda')
    &=
    \int \frac{\d^2\bm{\varDelta}_\perp}{(2\pi)^2}
    \mathscr{R}^{\mu\nu}_{\phantom{\mu\nu}\alpha\beta}
    \frac{
      \langle \bm{p}', \lambda' | \hat{T}^{\alpha\beta}(0) | \bm{p}, \lambda \rangle
    }{2m}
    \e^{-\i\bm{\varDelta}_\perp\cdot\bm{b}_\perp}
    \Bigg|_{\varDelta_z=0}
    \\
    \mathscr{R}^{\mu\nu}_{\phantom{\mu\nu}\alpha\beta}
    &=
    (\bar\varLambda^{-1})^{\mu}_{\phantom{\mu}\alpha}
    (\bar\varLambda^{-1})^{\nu}_{\phantom{\nu}\beta}
    \times
    \left\{
      \begin{array}{ccl}
        1 &:& \mu\neq3~\text{or}~\nu\neq3 \\
        \left( 1 - \frac{\bm{\varDelta}_\perp^2}{4m^2} \right)^{-1}
        &:& \mu=\nu=3
      \end{array}
      \right.
  \end{align}
\end{subequations}
for the intrinsic EMT.
%As with the current, the intrinsic EMT can be derived
%from Eq.~(\ref{eqn:smear:emt}) by inverting Eq.~(\ref{eqn:conv:emt}).

Eq.~(\ref{eqn:emt}) is the primary formula we shall employ
throughout this work to obtain intrinsic EMT densities.
A proof of this formula,
along with proofs of Eqs.~(\ref{eqn:smear:current}),
(\ref{eqn:current}), and (\ref{eqn:smear:emt})
can be found in Appendix~\ref{sec:proof}.

%%%%%%%%%%%%%%%%%%%%%%%%%%%%%%%%%%%%%%%%%%%%%%%%%%%%%%%%%%%%%%%%%%%%%%%%%%%%%%%%

\section{Energy-momentum tensor of spin-zero targets}
\label{sec:zero}

Although we are primarily interested in the proton in this work,
we consider spin-zero target first as a warm-up exercise,
since the resulting densities are simpler.

The standard form factor breakdown for the EMT of a spin-zero target is~\cite{Polyakov:2018zvc}:
\begin{align}
  \label{eqn:gff:zero}
  \langle \bm{p}' | \hat{T}^{\mu\nu}(0) | \bm{p} \rangle
  =
  2 P^\mu P^\nu
  A(\varDelta^2)
  +
  \frac{1}{2}
  \Big( \varDelta^\mu \varDelta^\nu - \varDelta^2 g^{\mu\nu} \Big)
  D(\varDelta^2)
  \,.
\end{align}
From Eq.~(\ref{eqn:emt}), it follows that the intrinsic EMT
of a spin-zero target is:
\begin{align}
  t^{\mu\nu}(\bm{b}_\perp)
  =
  m
  \int \frac{\d^2\bm{\varDelta}_\perp}{(2\pi)^2}
  \left\{
    \bar{n}^\mu
    \bar{n}^\nu
    A(-\bm{\varDelta}_\perp^2)
    +
    \left(
    \frac{
      \varDelta_\perp^\mu \varDelta_\perp^\nu + \bm{\varDelta}_\perp^2 g^{\mu\nu}
    }{4m^2}
    \right)
    D(-\bm{\varDelta}_\perp^2)
    %-
    %\left( \frac{\bm{\varDelta}_\perp^2}{4m^2} \right)^2
    %n^\mu n^\nu
    %D(-\bm{\varDelta}_\perp^2)
    \right\}
  \,,
\end{align}
where $\bar{n}^\mu = (1;0,0,0)$ is defined to
project out the $0$th component of
a four-vector when written in covariant form
(with a lower Lorentz index), e.g., $\bar{n}^\mu p_\mu = p_0$.
The interpretations of individual components of the EMT
were described in Sec.~\ref{sec:currents};
we shall presently analyze results for the components
in terms of those interpretations.

%%%%%%%%%%%%%%%%%%%%%%%%%%%%%%%%%%%%%%%%

\subsection{Energy density}

Consulting the dictionary of Eq.~(\ref{eqn:dictionary}),
the rest frame energy density for a spin-zero target is:
\begin{align}
  \label{eqn:energy:zero}
  \E(\bm{b}_\perp)
  =
  t^{00}(\bm{b}_\perp)
  -
  t^{03}(\bm{b}_\perp)
  &=
  m
  \int \frac{\d^2\bm{\varDelta}_\perp}{(2\pi)^2}
  \left\{
    A(-\bm{\varDelta}_\perp^2)
    +
    \frac{ \bm{\varDelta}_\perp^2 }{4m^2}
    D(-\bm{\varDelta}_\perp^2)
    \right\}
  \e^{-\i\bm{\varDelta}_\perp\cdot\bm{b}_\perp}
  \,.
\end{align}
This is a new result.
Although a prior results exist for energy densities of
spin-zero~\cite{Freese:2022fat}
and spin-half~\cite{Lorce:2018egm}
targets in standard light front coordinates,
the tilted coordinate energy is different from the light front energy,
so naturally Eq.~(\ref{eqn:energy:zero})
differs from the light front energy density in Ref.~\cite{Freese:2022fat}.
In fact, the energy in tilted light front coordinates is equal
to the more familiar instant form energy~\cite{Freese:2023jcp},
for which Eq.~(\ref{eqn:energy:zero}) provides
a two-dimensional relativistic density in the target's rest frame.

Integrating
Eq.~(\ref{eqn:energy:zero})
gives $m$ as the total energy, as expected for a system at rest.
Additionally,
Eq.~(\ref{eqn:energy:zero})
can be used to define a rest-frame energy radius:
\begin{align}
  \label{eqn:radius:energy:zero}
  \langle \bm{b}_\perp^2 \rangle_{\text{energy}}
  \equiv
  \frac{1}{m}
  \int \d^2\bm{b}_\perp \,
  \bm{b}_\perp^2
  \E(\bm{b}_\perp)
  =
  4 \frac{\d A(\varDelta^2)}{\d \varDelta^2}\bigg|_{\varDelta^2=0}
  -
  \frac{1}{m^2}
  D(0)
  \,.
\end{align}
%This radius is notably not zero for pointlike particles.
%Since $D_{\text{free}}(\varDelta^2) = -1$~\cite{Polyakov:2018zvc},
%the free spin-zero boson an energy radius of $1/m$.
%
%Although $D_{\text{free}}(\varDelta^2) = -1$ follows naturally for
%Klein-Gordon fields by applying Noether's theorems~\cite{Itzykson:1980rh}
%or by differentiating the action with respect to the metric~\cite{Polyakov:2018zvc},
%we might wonder how the energy radius would differ if
%$D_{\text{free}}(\varDelta^2)$
%were a different constant.
%In fact, several authors~\cite{Polyakov:2018zvc,Lorce:2020bsg,Li:2022ldb}
%argue that the EMT is only defined up to a trivially conserved quantity,
%and that the form factor $D(\varDelta^2)$ can be altered by adding such a quantity.
%However, Eq.~(\ref{eqn:radius:energy:zero})
%will only give a non-negative squared radius if:
%\begin{align}
%  D_{\text{free}}(0)
%  \leq
%  0
%  \,.
%\end{align}
%The negativity of $D(0)$ has previously been postulated by Polyakov~\cite{Perevalova:2016dln,Polyakov:2018zvc}
%as a mechanical stability criterion.
%Within the context of an energy radius, however,
%this constraint can be relaxed for composite particles,
%since $A(\varDelta^2)$ is no longer constant.
%More generally, the criterion of positive squared radius imposes:
%\begin{align}
%  D(0)
%  \leq
%  4 m^2
%  \frac{\d A(\varDelta^2)}{\d \varDelta^2}\bigg|_{\varDelta^2=0}
%  \,,
%\end{align}
%which may allow $D(0) > 0$ for composite particles,
%but still imposes a maximum value.

%%%%%%%%%%%%%%%%%%%%%%%%%%%%%%%%%%%%%%%%

\subsection{Momentum density}

From the dictionary in Eq.~(\ref{eqn:dictionary}),
the momentum density is given by:
\begin{align}
  \label{eqn:momentum:zero}
  \bm{\P}(\bm{b}_\perp)
  =
  t^{01}(\bm{b}_\perp)
  \hat{e}_x
  +
  t^{02}(\bm{b}_\perp)
  \hat{e}_y
  +
  t^{00}(\bm{b}_\perp)
  \hat{e}_z
  &=
  m
  \hat{e}_z
  \int \frac{\d^2\bm{\varDelta}_\perp}{(2\pi)^2}
  A(-\bm{\varDelta}_\perp^2)
  \e^{-\i\bm{\varDelta}_\perp\cdot\bm{b}_\perp}
  \,,
\end{align}
which is non-zero only for the longitudinal momentum.
Significantly, the $z$ momentum in tilted coordinates
is equal to the plus momentum in light front coordinates:
$P_z = P^+_{\text{LF}}$.
It is thus not surprising that the momentum density we find
is equal to prior results for the light front momentum
density~\cite{Freese:2021czn}
upon setting $ P^+_{\text{LF}} \rightarrow m$.

It is worth stressing
(see Ref.~\cite{Freese:2023jcp} and Appendix~\ref{sec:basic})
that---in tilted coordinates---$P_z$ is equal to $m$ rather than $0$ at rest.
Thus, a non-zero $P_z$ density does not indicate motion within the system.
Rather than a density for a quantity of motion,
the $P_z$ density can be interpreted as an inertia density,
since classically, contravariant components of the tilted momentum
and the velocity are related by $p^i = p_z v^i$.
On the other hand, the energy fluxes $t^i_{\phantom{i}0}$
have a clearer interpretation as encoding motion within the system.
We shall look at these next.

%%%%%%%%%%%%%%%%%%%%%%%%%%%%%%%%%%%%%%%%

\subsection{Energy flux density}

From the dictionary in Eq.~(\ref{eqn:dictionary}),
the energy flux density is given by:
\begin{align}
  \label{eqn:flux:zero}
  \Efl(\bm{b}_\perp)
  =
  \big(
  t^{i0}(\bm{b}_\perp)
  -
  t^{i3}(\bm{b}_\perp)
  \big)
  \hat{e}_i
  =
  0
  \,,
\end{align}
which is identically zero.

%%%%%%%%%%%%%%%%%%%%%%%%%%%%%%%%%%%%%%%%

\subsection{Stress tensor}
\label{sec:stress:zero}

Using the dictionary in Eq.~(\ref{eqn:dictionary}),
the stress tensor for a spin-zero target is:
\begin{align}
  \label{eqn:stress:zero}
  S^{ij}(\bm{b}_\perp)
  =
  t^{i1}%(\bm{b}_\perp)
  \delta^{j1}
  +
  t^{i2}%(\bm{b}_\perp)
  \delta^{j2}
  +
  t^{i0}%(\bm{b}_\perp)
  \delta^{j3}
  =
  \frac{ 1 }{4m}
  \int \frac{\d^2\bm{\varDelta}_\perp}{(2\pi)^2}
  \Big(
  \varDelta_\perp^i
  \varDelta_\perp^j
  -
  \delta^{ij}
  \bm{\varDelta}_\perp^2
  \Big)
  D(-\bm{\varDelta}_\perp^2)
  \e^{-\i\bm{\varDelta}_\perp\cdot\bm{b}_\perp}
  \,.
\end{align}
The transverse components $i,j\in\{1,2\}$ of the stress tensor
by themselves reproduce prior results for the transverse
light front stress tensor~\cite{Freese:2021czn}
if one sets $P^+_{\text{LF}} \rightarrow m$.

More remarkably, however,
there is apparently a new longitudinal pressure:
\begin{align}
  p_z(\bm{b}_\perp)
  \equiv
  S^{33}(\bm{b}_\perp)
  =
  -
  \int \frac{\d^2\bm{\varDelta}_\perp}{(2\pi)^2}
  \frac{ \bm{\varDelta}_\perp^2 }{4m}
  D(-\bm{\varDelta}_\perp^2)
  \e^{-\i\bm{\varDelta}_\perp\cdot\bm{b}_\perp}
  \,,
\end{align}
where, as is standard in continuum mechanics~\cite{Fetter:1980str,Batchelor:2000int,Irgens:2008con,bower2009applied,flugge2013tensor,liu2013continuum},
normal stresses are called pressures.
Shear stresses involving the $z$ direction, which would correspond
to fluxes of $P_z$ in the transverse plane or longitudinal fluxes of $\bm{P}_\perp$,
vanish for the spin-zero target.
This is likely a consequence of the $z$ coordinate dependence being integrated out,
as integrating out $x$ or $y$ likewise leads to the elimination of shear stresses
in integrated-out direction.

%%%%%%%%%%%%%%%%%%%%%%%%%%%%%%%%%%%%%%%%%%%%%%%%%%%%%%%%%%%%%%%%%%%%%%%%%%%%%%%%

\section{Energy-momentum tensor of spin-half targets}
\label{sec:half}

Since our primary objective is to obtain energy-momentum densities
and currents for the proton,
we proceed to consider spin-half targets.
We primarily focus on the symmetric Belinfante EMT,
but will briefly consider how the formalism changes
when the asymmetric EMT is instead used in Sec.~\ref{sec:asym}.

The standard form factor breakdown for the symmetric EMT
for spin-half targets is~\cite{Polyakov:2018zvc}:
\begin{align}
  \label{eqn:gff:half}
  \langle \bm{p}',\lambda' |
  \hat{T}^{\mu\nu}(0)
  | \bm{p}, \lambda \rangle
  =
  \bar{u}(\bm{p}',\lambda')
  \bigg\{
    \frac{\gamma^{\{\mu}P^{\nu\}}}{2}
    A(\varDelta^2)
    +
    \frac{\i P^{\{\mu}\sigma^{\nu\}\rho} \varDelta_\rho}{4m}
    B(\varDelta^2)
    +
    \frac{\varDelta^\mu \varDelta^\nu -  g^{\mu\nu}\varDelta^2}{4m}
    D(\varDelta^2)
    \bigg\}
  u(\bm{p},\lambda)
  \,.
\end{align}
Using formulas from Appendix A of Ref.~\cite{Freese:2023jcp},
we can explicitly evaluate matrix elements of the EMT when $\varDelta_z = 0$ to be:
\begin{align}
  \langle \bm{p}',\lambda' |
  \hat{T}^{\mu\nu}(0)
  | \bm{p}, \lambda \rangle
  &=
  2 {P}^\mu P^\nu
  \left\{
    (\sigma_0)_{\lambda'\lambda}
    A(-\bm{\varDelta}_\perp^2)
    -
    \frac{
      \i
      \epsilon^{\alpha\beta\gamma\delta}
      {n}_\alpha {\bar{n}}_\beta
      {\varDelta}_\gamma (\sigma_\delta)_{\lambda'\lambda}
    }{2m}
    B(-\bm{\varDelta}_\perp^2)
    \right\}
  \notag \\
  &+
  \left\{
    -
    \frac{
      \i
      P^{\{\mu}
      \epsilon^{\nu\}\rho\sigma\tau}
      {n}_\rho
      {P}_\sigma {\varDelta}_\tau
    }{({P}\cdot{n})}
    (\sigma_3)_{\lambda'\lambda}
    +
    \frac{mP^{\{\mu}{n}^{\nu\}}}{({P}\cdot{n})}
    \i
    \epsilon^{\alpha\beta\gamma\delta}
    {n}_\alpha {\bar{n}}_\beta
    {\varDelta}_\gamma (\sigma_\delta)_{\lambda'\lambda}
    \right\}
  J(-\bm{\varDelta}_\perp^2)
  \notag \\
  &
  +
  \frac{\varDelta^\mu \varDelta^\nu -  g^{\mu\nu}\varDelta^2}{2}
  \left(
  (\sigma_0)_{\lambda'\lambda}
  +
  \frac{
    \i
    \epsilon^{\alpha\beta\gamma\delta}
    {n}_\alpha {\bar{n}}_\beta
    {\varDelta}_\gamma (\sigma_\delta)_{\lambda'\lambda}
  }{2m}
  \right)
  D(-\bm{\varDelta}_\perp^2)
  \,,
\end{align}
where
\begin{align}
  J(\varDelta^2) &= \frac{1}{2}\big( A(\varDelta^2) + B(\varDelta^2)\big)
\end{align}
is the angular momentum form factor~\cite{Polyakov:2018zvc}.
Using Eq.~(\ref{eqn:emt}) for the intrinsic EMT of a general system then gives:
\begin{multline}
  \label{eqn:monstrosity}
  t^{\mu\nu}(\bm{b}_\perp;\lambda,\lambda')
  =
  m
  \int \frac{\d^2\bm{\varDelta}_\perp}{(2\pi)^2}
  \Bigg\{
    \bar{n}^\mu \bar{n}^\nu
    \left(
      (\sigma_0)_{\lambda'\lambda}
      A(-\bm{\varDelta}_\perp^2)
      +
      \frac{
        (\bm{\sigma}_{\lambda'\lambda} \times \i \bm{\varDelta}_\perp) \cdot \hat{e}_z
      }{2m}
      B(-\bm{\varDelta}_\perp^2)
      \right)
    \\
    +
    \frac{
      \bar{n}^{\{\mu}
      \big( \bm{\sigma}_{\lambda'\lambda} \times \i\bm{\varDelta}_\perp \big)^{\nu\}}
    }{2m}
    J(-\bm{\varDelta}_\perp^2)
    +
    \left(
    \frac{\varDelta_\perp^\mu \varDelta_\perp^\nu +  g^{\mu\nu} \bm{\varDelta}_\perp^2}{4m^2}
    %-
    \right)
    \left(
    (\sigma_0)_{\lambda'\lambda}
    -
    \frac{
      (\bm{\sigma}_{\lambda'\lambda} \times \i \bm{\varDelta}_\perp) \cdot \hat{e}_z
    }{2m}
    \right)
    D(-\bm{\varDelta}_\perp^2)
    \Bigg\}
  \,.
\end{multline}
As with the spin-zero target, we will use the dictionary in
Eq.~(\ref{eqn:dictionary}) to obtain energy and momentum
densities and currents from the intrinsic EMT.
We will find the nine Galilean components of the EMT---that is,
the momentum densities and transverse stress tensor---reproduce
results from standard light front coordinates,
but the energy density, $P_z$ fluxes,
and longitudinal energy flux are newly-found.

%%%%%%%%%%%%%%%%%%%%%%%%%%%%%%%%%%%%%%%%

\subsection{Momentum densities}
\label{sec:momentum:half}

We consider the momentum densities first.
It is instructive to consider the $P_z$ density and $\bm{P}_\perp$
densities in separate equations,
as their behavior is quite different.
From Eqs.~(\ref{eqn:emt}) and (\ref{eqn:monstrosity}),
we find these densities to be:
\begin{subequations}
  \label{eqn:momentum:half}
  \begin{align}
    \label{eqn:momentum:half:z}
    \P_z(\bm{b}_\perp,\hat{\bm{s}})
    &=
    -t^0_{\phantom{0}3}%(\bm{b}_\perp\hat{\bm{s}})
    =
    m
    \int \frac{\d^2\bm{\varDelta}_\perp}{(2\pi)^2}
    \left(
    A(-\bm{\varDelta}_\perp^2)
    +
    \frac{
      (\hat{\bm{s}} \times \i \bm{\varDelta}_\perp) \cdot \hat{e}_z
    }{2m}
    B(-\bm{\varDelta}_\perp^2)
    \right)
    \e^{-\i\bm{\varDelta}_\perp\cdot\bm{b}_\perp}
    \\
    \label{eqn:momentum:half:perp}
    \bm{\P}_\perp(\bm{b}_\perp,\hat{\bm{s}})
    &=
    -t^0_{\phantom{0}1}%(\bm{b}_\perp,\hat{\sm{s}})
    \hat{e}_x
    -t^0_{\phantom{0}2}%(\bm{b}_\perp,\hat{\sm{s}})
    \hat{e}_y
    =
    m
    (\hat{\bm{s}}\cdot\hat{e}_z)
    \int \frac{\d^2\bm{\varDelta}_\perp}{(2\pi)^2}
    \frac{
      \hat{e}_z \times \i \bm{\varDelta}_\perp
    }{2m}
    J(-\bm{\varDelta}_\perp^2)
    \e^{-\i\bm{\varDelta}_\perp\cdot\bm{b}_\perp}
    %%\qquad :
    %%i \in \{1,2\}
    \,.
  \end{align}
\end{subequations}
The $P_z$ density reproduces prior results for the
$P^+$ density in standard light front coordinates~\cite{Freese:2021qtb}
if one sets $P^+\rightarrow m$.
For a free point fermion, $A(\varDelta^2)=1$ and $B(\varDelta^2)=0$,
so the $P_z$ density just becomes $m\delta^{(2)}(\bm{b}_\perp)$.
This is expected because $P_z$ is the central charge of the Galilean subgroup
and is preserved by transverse boosts,
so the barycentric coordinate $\bm{R}_\perp$ is a center-of-$P_z$ coordinate
and the intrinsic densities are relative to the center-of-$P_z$.
(This has been explained in terms of standard light front coordinates by Lorc\'{e}~\cite{Lorce:2018zpf}.)

For non-pointlike targets with $B(\varDelta^2) \neq 0$,
transversely-polarized states will exhibit azimuthal modulations in the $P_z$ density.
The behavior is analogous to the modulations in its charge density~\cite{Freese:2023jcp},
and is likewise induced by the synchronization scheme---specifically by
modulations in the apparent clock rate of matter revolving around the target's center.
It is worth pointing out that $F_2(\varDelta^2)$---which controls the charge density
modulations---and $B(\varDelta^2)$ are Mellin moments of the same generalized parton distribution (GPD)
$E(x,\xi,\varDelta^2)$, suggesting that this GPD has an interpretation in terms of encoding
partonic motion.

Despite the presence of these modulations, there is not a $P_z$ dipole moment.
This would of course contradict the barycenter being the center-of-$P_z$.
If one calculates the $P_z$ dipole moment from Eq.~(\ref{eqn:momentum:half}):
\begin{align}
  \langle \bm{b}_\perp \rangle_{p_z}
  =
  \int \d^2 \bm{b}_\perp \,
  \bm{b}_\perp
  \P_z(\bm{b}_\perp, \hat{\bm{s}})
  =
  \frac{\hat{e}_z \times \hat{\bm{s}}}{2}
  B(0)
  \,.
\end{align}
However, $B(0) = 0$, a fact known as the vanishing of the anomalous gravitomagnetic
moment~\cite{Teryaev:1999su}.
It follows from the simultaneous sum rules $A(0) = 1$
(the momentum sum rule)
and $J(0) = \frac{1}{2}$
(the angular momentum sum rule).

A radius can be associated with the $P_z$ density:
\begin{align}
  \label{eqn:radius:momentum:half}
  \langle \bm{b}_\perp^2 \rangle_{p_z}
  =
  \frac{1}{m}
  \int \d^2\bm{b}_\perp \,
  \bm{b}_\perp^2
  \P_z(\bm{b}_\perp,\hat{\bm{s}})
  =
  4 \frac{\d A(\varDelta^2)}{\d \varDelta^2}\bigg|_{\varDelta^2=0}
  \,.
\end{align}
This radius has appeared in the literature before.
It has been called a $P^+$ radius in standard light front coordinates~\cite{Freese:2021czn},
and occasionally called a mass radius.
(As pointed out by Lorc\'{e} \textsl{et al.}~\cite{Lorce:2021xku},
mass plays several roles in relativity,
and the $P^+$ radius could be considered a kind of mass radius.
This is distinct, however, from the energy radius,
which we give in Eq.~(\ref{eqn:radius:energy:half}).)

The $\bm{P}_\perp$ density in Eq.~(\ref{eqn:momentum:half})
would be the same in standard light front coordinates
being a Galilean component of the EMT,
but to the best of our knowledge this result has
not been previously reported.
This density is related to and tracks the
$z$ component of the
\emph{total} angular momentum density, the latter being
\begin{align}
  \label{eqn:Jz:half}
  \mathcal{J}_z(\bm{b}_\perp,\hat{\bm{s}})
  =
  \big( \bm{b}_\perp \times \bm{\P}_\perp(\bm{b}_\perp,\hat{\bm{s}}) \big)\cdot\hat{e}_z
  \,.
\end{align}
This may appear counter-intuitive on first sight,
as it superficially resembles the formula
$\bm{r} \times \bm{p}$ for the \emph{orbital} angular momentum of a body.
However, the symmetric Belinfante EMT appears as the source of gravitation in general relativity,
and the equivalence principle implies that all angular momentum should gravitate the same way.
Thus, neither the Belinfante EMT nor its associated densities should be able to distinguish between
spin and orbital angular momentum.

Moreover, despite the superficial resemblance, the right-hand side of Eq.~(\ref{eqn:Jz:half})
does not give an OAM density---at least not in terms of how OAM is usually defined.
The amount of momentum
$\bm{\P}_\perp(\bm{b}_\perp,\hat{\bm{s}}) \d^2\bm{b}_\perp$
contained in a small spatial region $\d^2\bm{b}_\perp$ %located around $\bm{b}_\perp$
is not necessarily the momentum carried by a constituent of the target.
This is especially clear if the target under consideration is a pointlike particle.
The particle itself is an excitation of a field,
which is the more fundamental object in quantum field theory.
The momentum element
$\bm{\P}_\perp(\bm{b}_\perp,\hat{\bm{s}}) \d^2\bm{b}_\perp$
is the amount of momentum carried \emph{by the field} in this small spatial region.
However the only particle present is the target itself, which is at rest,
so the OAM is zero.
Thus Eq.~(\ref{eqn:Jz:half}) does not encode an OAM density.

%%%%%%%%%%%%%%%%%%%%%%%%%%%%%%%%%%%%%%%%

\subsection{Energy fluxes}
\label{sec:flux:half}

From Eqs.~(\ref{eqn:emt}) and (\ref{eqn:monstrosity}),
we find the energy fluxes to be:
\begin{align}
  \label{eqn:flux:half}
  \Efl(\bm{b}_\perp, \hat{\bm{s}})
  =
  t^i_{\phantom{i}0}%(\bm{b}_\perp, \hat{\bm{s}})
  \hat{e}_i
  =
  m
  \int \frac{\d^2\bm{\varDelta}_\perp}{(2\pi)^2}
  \frac{
    \hat{\bm{s}} \times \i \bm{\varDelta}_\perp
  }{2m}
  J(-\bm{\varDelta}_\perp^2)
  \e^{-\i\bm{\varDelta}_\perp\cdot\bm{b}_\perp}
  \,.
\end{align}
There is an energy flux due to the presence of angular momentum
in the system, which is intuitively sensible.
The transverse energy flux in particular is
equal to the transverse momentum density.
This mimics the well-known fact that the symmetric EMT in instant form coordinates
has identical momentum densities and energy fluxes.

%%%%%%%%%%%%%%%%%%%%%%%%%%%%%%%%%%%%%%%%

\subsection{Stress tensor and momentum fluxes}
\label{sec:stress:half}

We next consider the intrinsic stress tensor of a spin-half target.
From Eqs.~(\ref{eqn:emt}) and (\ref{eqn:monstrosity}),
\begin{multline}
  \label{eqn:stress:half}
  S^{ij}(\bm{b}_\perp,\hat{\bm{s}})
  \equiv
  -
  t^i_{\phantom{i}j}(\bm{b}_\perp)
  =
  m
  \delta^{3j}
  \int \frac{\d^2\bm{\varDelta}_\perp}{(2\pi)^2}
  \frac{
    (\hat{\bm{s}} \times \i \bm{\varDelta}_\perp)^i
  }{2m}
  J(-\bm{\varDelta}_\perp^2)
  \e^{-\i\bm{\varDelta}_\perp\cdot\bm{b}_\perp}
  \\
  +
  \frac{ 1 }{4m}
  \int \frac{\d^2\bm{\varDelta}_\perp}{(2\pi)^2}
  \Big(
  \varDelta_\perp^i
  \varDelta_\perp^j
  -
  \delta^{ij}
  \bm{\varDelta}_\perp^2
  \Big)
  \left[
    1
    -
    \frac{
      (\hat{\bm{s}} \times \i \bm{\varDelta}_\perp) \cdot \hat{e}_z
    }{2m}
    \right]
  D(-\bm{\varDelta}_\perp^2)
  \e^{-\i\bm{\varDelta}_\perp\cdot\bm{b}_\perp}
  \,.
\end{multline}
We have broken the result into two pieces,
the first of which depends on the angular momentum form factor $J(\varDelta^2)$,
and the other of which depends on the form factor $D(\varDelta^2)$.

The first angular momentum
piece of the stress tensor introduces asymmetric shear stresses.
Recalling that the stress tensor consists of momentum fluxes,
this piece of the stress tensor encodes fluxes of $P_z$ in all three spatial directions,
but not of $\bm{P}_\perp$.
Now, a major difference between $P_z$ and $\bm{P}_\perp$ in tilted coordinates
is that the former is non-zero even if the velocity is zero;
a $P_z$ flux cannot be interpreted as a flux of some quantity of motion.
Since classically $\bm{p}_\perp = p_z \bm{v}_\perp$
and $p_z - E = p_z v_z$
(these relations can be found in Appendix~\ref{sec:basic}),
it should perhaps be the $P_z$ flux minus the energy flux that is compared
to the $\bm{P}_\perp$ flux:
\begin{align}
  \bm{\mathcal{F}}_{p_z}(\bm{b}_\perp,\hat{\bm{s}})
  -
  \Efl(\bm{b}_\perp,\hat{\bm{s}})
  =
  -
  m
  \hat{e}_z
  \int \frac{\d^2\bm{\varDelta}_\perp}{(2\pi)^2}
  \frac{ \bm{\varDelta}_\perp^2 }{4m}
  \left[
    1
    -
    \frac{
      (\hat{\bm{s}} \times \i \bm{\varDelta}_\perp) \cdot \hat{e}_z
    }{2m}
    \right]
  D(-\bm{\varDelta}_\perp^2)
  \e^{-\i\bm{\varDelta}_\perp\cdot\bm{b}_\perp}
  \,.
\end{align}
For comparison, the transverse components of the stress tensor are:
\begin{align}
  \label{eqn:stress:half:trans}
  S_\perp^{ij}(\bm{b}_\perp,\hat{\bm{s}})
  =
  \frac{ 1 }{4m}
  \int \frac{\d^2\bm{\varDelta}_\perp}{(2\pi)^2}
  \Big(
  \varDelta_\perp^i
  \varDelta_\perp^j
  -
  \delta_\perp^{ij}
  \bm{\varDelta}_\perp^2
  \Big)
  \left[
    1
    -
    \frac{
      (\hat{\bm{s}} \times \i \bm{\varDelta}_\perp) \cdot \hat{e}_z
    }{2m}
    \right]
  D(-\bm{\varDelta}_\perp^2)
  \e^{-\i\bm{\varDelta}_\perp\cdot\bm{b}_\perp}
  \,.
\end{align}
Both of these include only the form factor $D(\varDelta^2)$,
which vanishes in the case of a pointlike particle~\cite{Hudson:2017oul}
and can be interpreted as encoding internal dynamics in composite fermions~\cite{Hudson:2017oul}.
For pointlike fermions, then, the energy and $P_z$ fluxes are not zero,
but are instead equal,
and differences between them are an indication of dynamics
and internal motion.

As a last remark, we note that the transverse $P_z$ flux is equal
to the $\bm{P}_\perp$ density, i.e.,
${\bm{\mathcal{F}}_{p_z}^{(\perp)}(\bm{b}_\perp,\hat{\bm{s}})
= \bm{\mathcal{P}}_\perp(\bm{b}_\perp,\hat{\bm{s}})}$.
This seems to comport with the classical tilted coordinate relation
$\bm{p}_\perp = p_z \bm{v}_\perp$
if it is applied to the momentum carried by the target in any
small region of space.
Since this was a classical relation derived for observable bodies
that obey the mass-shell relation $p^2 = m^2$,
it is not a formal necessity that small elements of momenta supported
by an infinitesimal region of space obey this relation,
but it is interesting to note.

%%%%%%%%%%%%%%%%%%%%%%%%%%%%%%%%%%%%%%%%

\subsection{Energy density}
\label{sec:energy:half}

The last component of the intrinsic EMT to consider is the energy density.
From Eqs.~(\ref{eqn:emt}) and (\ref{eqn:monstrosity}),
using
$
  \E(\bm{b}_\perp,\hat{\bm{s}})
  =
  t^0_{\phantom{0}0}(\bm{b}_\perp,\hat{\bm{s}})
$
gives exactly Eq.~(\ref{eqn:energy:half}) given in the introduction.
%the rest frame energy density for a spin-half target is:
%\begin{multline}
%  \label{eqn:energy:half}
%  \E(\bm{b}_\perp,\hat{\bm{s}})
%  =
%  t^0_{\phantom{0}0}%(\bm{b}_\perp,\hat{\bm{s}})
%  =
%  m
%  \int \frac{\d^2\bm{\varDelta}_\perp}{(2\pi)^2}
%  \Bigg(
%  A(-\bm{\varDelta}_\perp^2)
%  +
%  \frac{ \bm{\varDelta}_\perp^2 }{4m^2}
%  D(-\bm{\varDelta}_\perp^2)
%  \\
%  +
%  \frac{
%    (\hat{\bm{s}} \times \i \bm{\varDelta}_\perp) \cdot \hat{e}_z
%  }{2m}
%  \left\{
%    B(-\bm{\varDelta}_\perp^2)
%    -
%    %\Big[
%    J(-\bm{\varDelta}_\perp^2)
%    %-
%    %S(-\bm{\varDelta}_\perp^2)
%    %\Big]
%    -
%    \frac{ \bm{\varDelta}_\perp^2 }{4m^2}
%    D(-\bm{\varDelta}_\perp^2)
%    \right\}
%  \Bigg)
%  \e^{-\i\bm{\varDelta}_\perp\cdot\bm{b}_\perp}
%  \,.
%\end{multline}
The spin-independent piece of the spin-half energy density is identical to
the spin-zero energy density of Eq.~(\ref{eqn:energy:zero}).
The spin-dependent piece does not contribute to the energy radius,
which is therefore identical to the spin-zero case:
\begin{align}
  \label{eqn:radius:energy:half}
  \langle \bm{b}_\perp^2 \rangle_{\text{energy}}
  \equiv
  \frac{1}{m}
  \int \d^2\bm{b}_\perp \,
  \bm{b}_\perp^2
  \E(\bm{b}_\perp,\hat{\bm{s}})
  =
  4 \frac{\d A(\varDelta^2)}{\d \varDelta^2}\bigg|_{\varDelta^2=0}
  -
  \frac{1}{m^2}
  D(0)
  \,.
\end{align}
The spin-dependent part of the EMT introduces angular modulations through
several of the form factors,
which must be attributed to distinct physical effects.
The angular modulations introduced through the form factor $B(\varDelta^2)$
can be attributed to clock rate modulations,
as explained for the $P_z$ density in Sec.~\ref{sec:momentum:half}.
In particular, modulations due to $B(\varDelta^2)$
will increase density on the side of the hadron moving away from the observer.

The angular momentum form factor
$J(\varDelta^2)$ contributes
to angular modulations with the opposite sign from $B(\varDelta^2)$,
thus causing the density to increase on the side moving
towards the observer.
This is effectively an artifact of the density being defined
with respect to the center-of-$P_z$.
Since $B(0) = 0$, the amount of $P_z$ on each side of the spin axis
in a transversely-polarized fermion is the same.
If the classical relation $p_z - E = p_z v_z$ is assumed to hold
for each half of the fermion,
the half moving towards the observer has $v_z < 0$ and thus should have greater energy,
and the half moving away should have less energy.
The modulations in the energy density arising from
$J(\varDelta^2)$
accomplish just this.
In fact, by comparing Eqs.~(\ref{eqn:momentum:half}), (\ref{eqn:stress:half})
and (\ref{eqn:energy:half}) we find:
\begin{align}
  \label{eqn:pez}
  \P_z(\bm{b}_\perp, \hat{\bm{s}})
  -
  \E(\bm{b}_\perp, \hat{\bm{s}})
  =
  \hat{e}_z\cdot \bm{\mathcal{F}}_{p_z}(\bm{b}_\perp, \hat{\bm{s}})
  \,,
\end{align}
meaning that $p_z - E = p_z v_z$ apparently holds for infinitesimal elements
of the hadron everywhere on the transverse plane.

In contrast to the $P_z$ density,
the energy density entails a synchronization-induced energy dipole moment:
\begin{align}
  \label{eqn:energy:dipole}
  \bm{d}_E
  \equiv
  \langle \bm{b}_\perp \rangle_{E}
  =
  \int \d^2\bm{b}_\perp \,
  \bm{b}_\perp
  \E(\bm{b}_\perp,\hat{\bm{s}})
  =
  -
  \frac{\hat{e}_z\times\hat{\bm{s}}}{2}
  J(0)
  =
  -\frac{1}{4} \hat{e}_z\times\hat{\bm{s}}
  \,.
\end{align}
We reiterate and stress that this apparent dipole moment
arises from the proton's ``center'' being given
by the center-of-$P_z$ rather than the center-of-energy.
However, the remaining modulations from the form factors
$B(t)$ and $D(t)$---which do not contribute to the dipole
moment---may be due to clock rate modulations,
similar to the modulations in the $P_z$ density.

%%%%%%%%%%%%%%%%%%%%%%%%%%%%%%%%%%%%%%%%

\subsection{Numerical estimates for the proton}
\label{sec:proton}

We now consider what the energy and momentum densities
and currents might look like for the proton.
Currently, high-precision empirical results for the
proton's gravitational form factors do not exist.
We shall thus utilize reasonable model estimates instead.

It is known from holographic QCD models~\cite{Mamo:2019mka,Mamo:2021krl,Fujita:2022jus}
that the pole strucutres of the gravitational form factors
are dominated by an infinite tower of $J^{PC}=0^{++}$ and $2^{++}$
glueball resonances---referred to as ``glueball dominance''
by Fujita \textsl{et al.}~\cite{Fujita:2022jus}, in analogy to the well-known
vector meson dominance~\cite{Sakurai:1960ju,Bauer:1977iq}
of electromagnetic form factors.
Fujita \textsl{et al.} also note that
simple multipole forms can reasonably approximate such an infinite tower,
as has been shown explicitly for holographic models
of electromagnetic form factors~\cite{Hashimoto:2008zw}.
Mamo and Zahed~\cite{Mamo:2021krl} in particular find that the
gravitational form factors in their holographic model
can be reasonably estimated by the simple functional forms\footnote{
  We note that Mamo and Zahed~\cite{Mamo:2021krl} use $A(0)=0.53$ in their work
  rather than $A(0)=1$ as we use here because they are describing
  only the gluonic contributions to the gravitational form factors.
  The estimates we present are for the total energy-momentum tensor,
  including both quark and gluon contributions, so we simply use
  their functional form for the $\bm{\varDelta}_\perp$ dependence
  while imposing the momentum sum rule $A(0)=1$.
}:
\begin{subequations}
  \label{eqn:gff:holography}
  \begin{align}
    A(-\bm{\varDelta}_\perp^2)
    &\approx
    \frac{
      1
    }{
      (1 + \bm{\varDelta}_\perp^2/\tilde{m}_T^2)^2
    }
    \\
    B(-\bm{\varDelta}_\perp^2)
    &\approx
    0
    \\
    D(-\bm{\varDelta}_\perp^2)
    &\approx
    D(0)
    \frac{
      1
      + \frac{\bm{\varDelta}_\perp^2}{4\tilde{m}_T^2}
      + \frac{\bm{\varDelta}_\perp^2}{4\tilde{m}_S^2}
    }{
      (1 + \bm{\varDelta}_\perp^2/\tilde{m}_T^2)^2
      (1 + \bm{\varDelta}_\perp^2/\tilde{m}_S^2)^2
    }
    \,,
  \end{align}
\end{subequations}
where $D(0) = -4$, $\tilde{m}_T = 1.124$~GeV, and $\tilde{m}_S = 1$~GeV.
We remark that these form factors are compatible with
lattice data~\cite{Shanahan:2018pib}
and with recent empirical measurements of $J/\psi$
photoproductionn the $J/\psi$-007 experiment~\cite{Duran:2022xag}.
Additionally, the large-$t$ falloff of these form factors
follows the expected behavior from perturbative
QCD~\cite{Masjuan:2012sk}.

Because $B(t)=0$ in the exmaple we consider,
the $P_z$ density is independent of polarization,
and is equal to:
\begin{align}
  \P_z(\bm{b}_\perp,\hat{\bm{s}})
  =
  \frac{ \tilde{m}_T^3 m b_\perp}{4\pi}
  K_1(\tilde{m}_T b_\perp)
  \,,
\end{align}
where $K_\nu(x)$ is a modified Bessel function of the second kind~\cite{NIST:DLMF}.
It's worth noting that this density is finite at the origin:
\begin{align}
  \lim_{b_\perp\rightarrow0}
  \P_z(\bm{b}_\perp,\hat{\bm{s}})
  =
  \frac{\tilde{m}_T^2 m}{4\pi}
  \,,
\end{align}
in contrast to the 2D Fourier transform of a monopole form~\cite{Miller:2009qu}.

\begin{figure}
  \centering
  \includegraphics[width=0.33\textwidth]{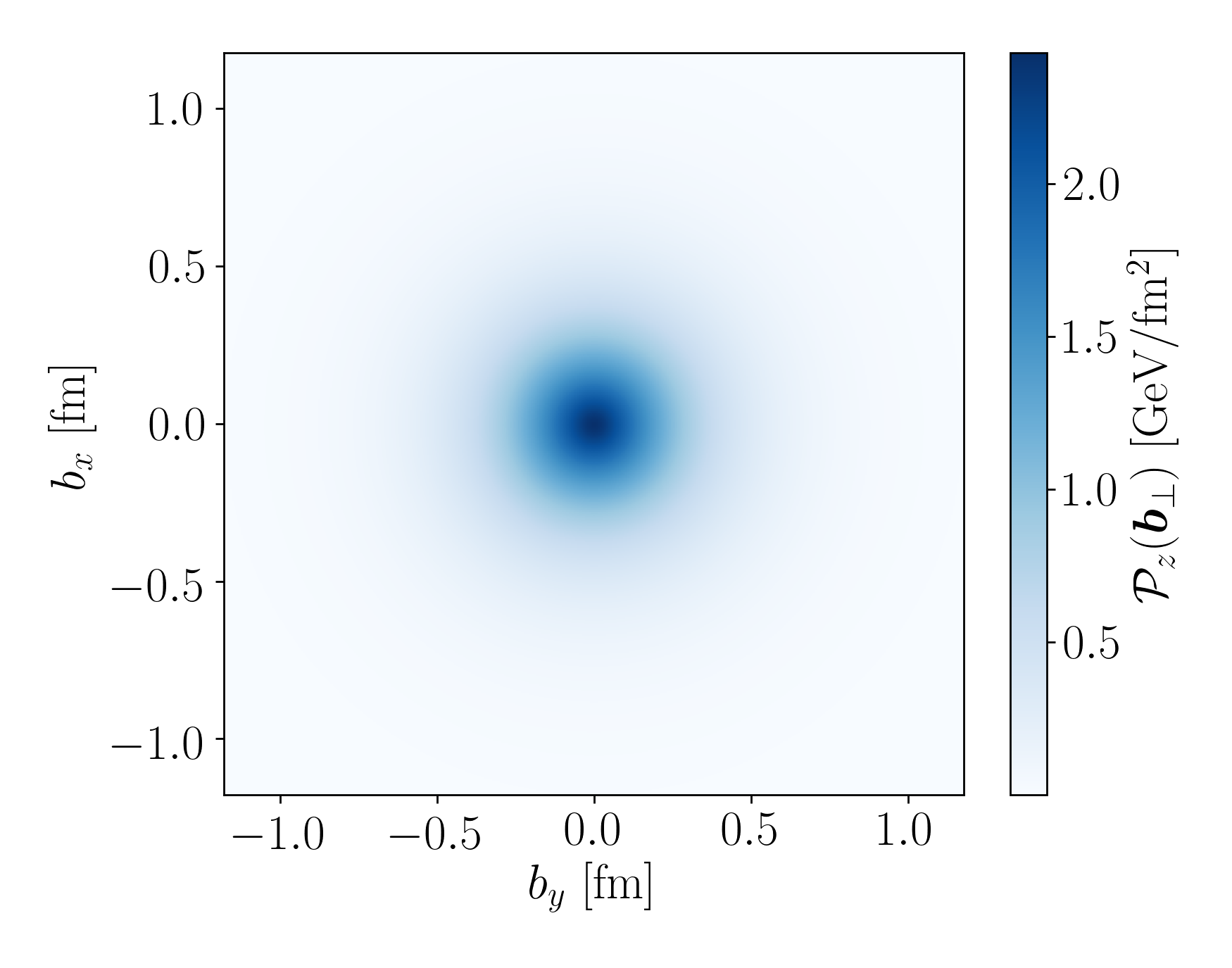}
  \includegraphics[width=0.33\textwidth]{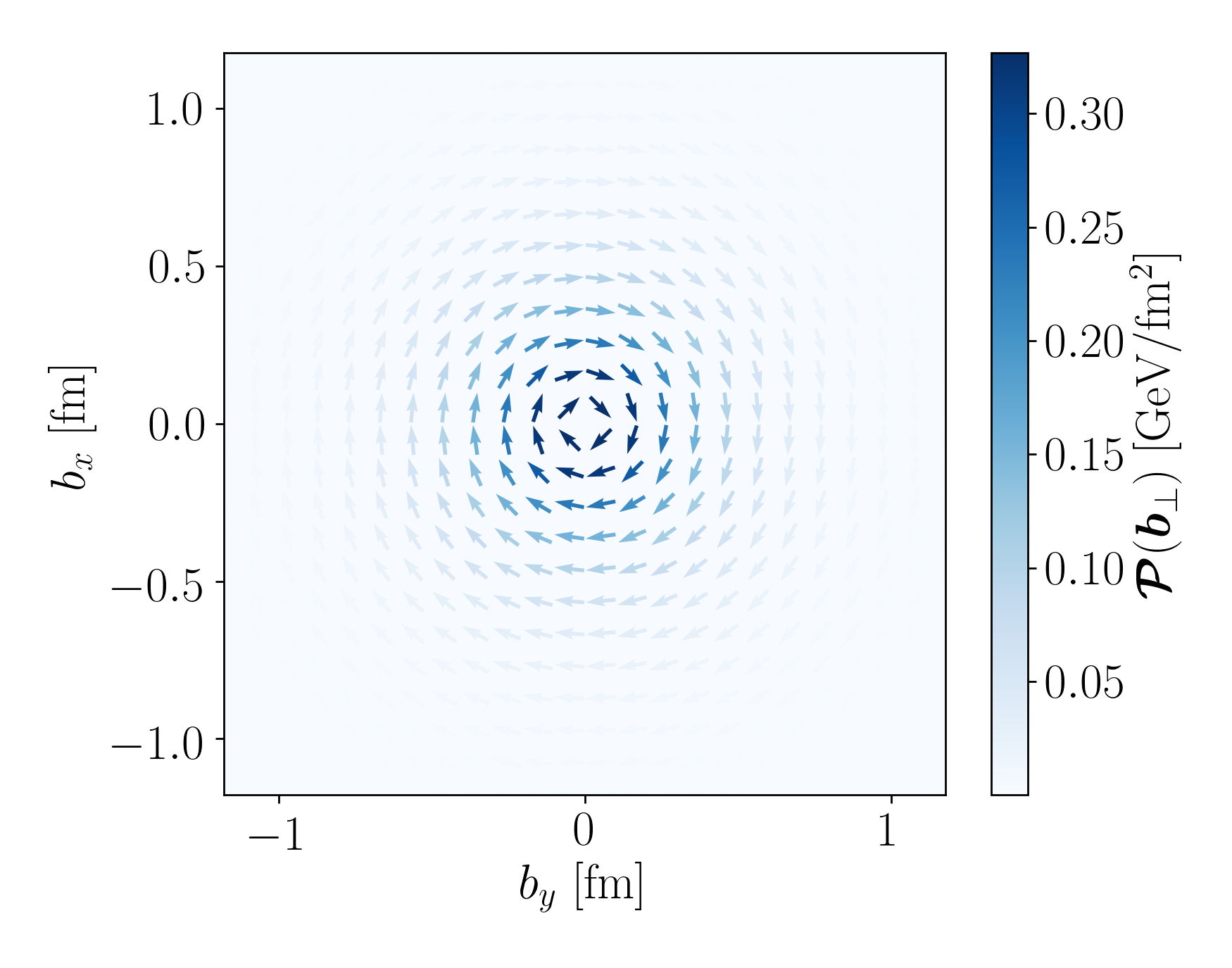}
  \includegraphics[width=0.33\textwidth]{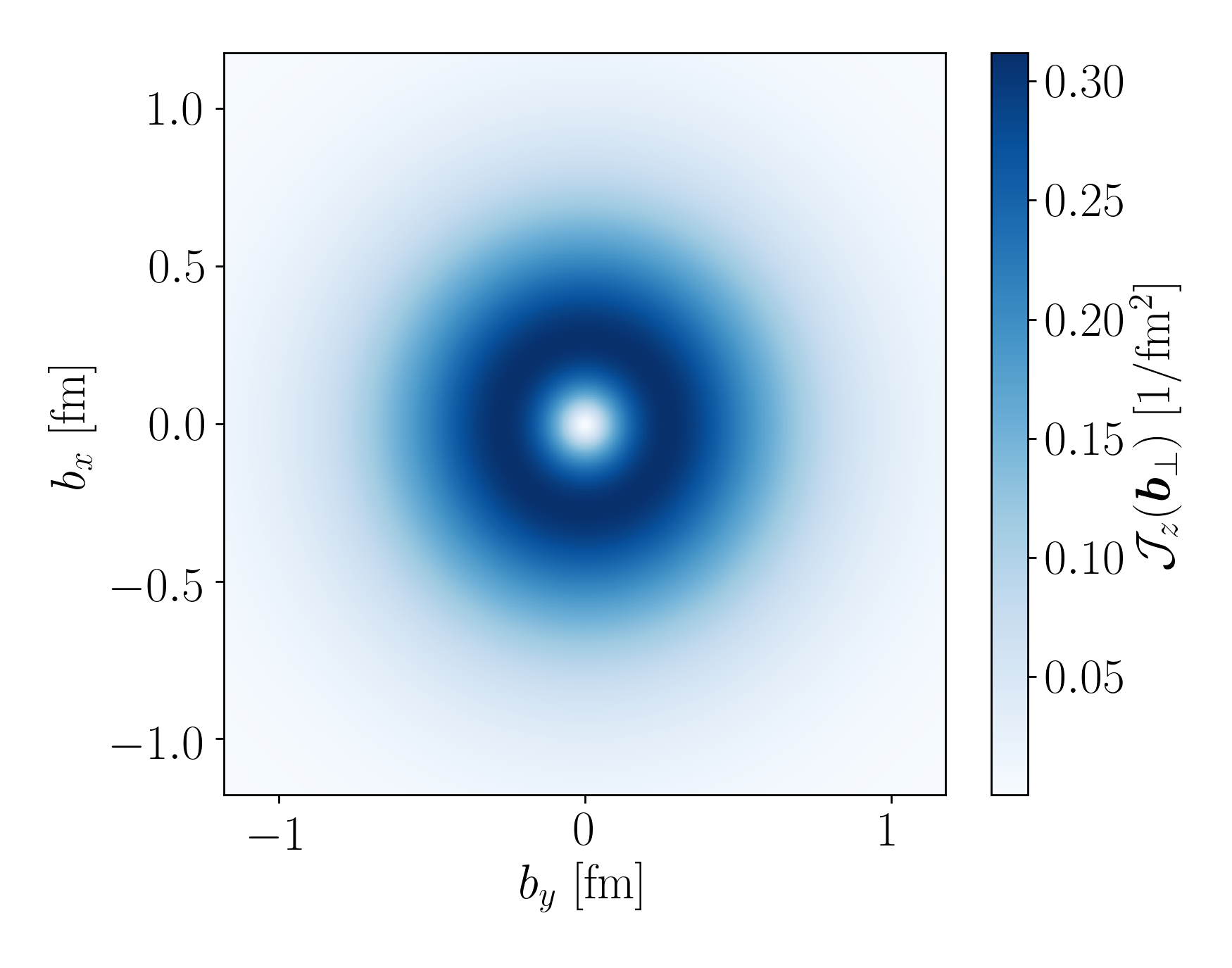}
  \caption{
    Momentum and angular momentum densities in a proton.
    (Left panel) is the $P_z$ density,
    which is independent of polarization under the assumption $B(t)=0$.
    (Middle panel) is the $\bm{P}_\perp$ density in a proton
    that is spin-up along the $z$-axis.
    (Right panel) is the $J_z$ (angular momentum) density in a proton
    that is spin-up along the $z$-axis.
    In these plots, the $x$-axis is oriented vertically
    and the $y$-axis horizontally so that the $z$ axis points into the page,
    allowing plots to mimic what an observer would see
    at fixed light front time.
  }
  \label{fig:proton:momentum}
\end{figure}

The transverse momentum density is given by:
\begin{align}
  \bm{\P}_\perp(\bm{b}_\perp,\hat{\bm{s}})
  =
  \frac{
    (\hat{\bm{s}}\cdot\hat{e}_z)
    (\hat{e}_y\cos\phi - \hat{e}_x\sin\phi)
  }{16\pi}
  \tilde{m}_T^4
  b_\perp
  K_0(\tilde{m}_T b_\perp)
  \,,
\end{align}
from which it follows that the density
of the $z$ component of angular momentum is:
\begin{align}
  \mathcal{J}_z(\bm{b}_\perp,\hat{\bm{s}})
  =
  \frac{
    (\hat{\bm{s}}\cdot\hat{e}_z)
    b_\perp^2
    \tilde{m}_T^4
  }{16\pi}
  K_0(\tilde{m}_T b_\perp)
  \,.
\end{align}
Numerical results for the momentum and angular momentum densities
for a longitudinally-polarized proton
are presented in Fig.~\ref{fig:proton:momentum}.
The $P_z$ density is unchanged for transversely-polarized protons
owing to the assumption $B(t)=0$,
while the $\bm{P}_\perp$ and $J_z$ densities vanish identically
for transversely-polarized protons.
The angular momentum density has an apparent hole in it because
$b_\perp^2 K_0(\tilde{m}_T b_\perp)$ vanishes at the origin.

The energy and momentum flux densities are our next consideration.
The transverse components of the stress tensor have previously
been considered in Refs.~\cite{Lorce:2018egm,Freese:2021czn,Panteleeva:2021iip,Freese:2021mzg},
with Ref.~\cite{Freese:2021mzg} in particular
exploring the distortions in eigenpressure directions
that occur in transversely-polarized states.
Since tilted coordinates newly allow access to energy and $P_z$
flux densities, we will focus on these specifically.

As explained in Sec.~\ref{sec:stress:half},
the transverse $P_z$ flux is equal to the $\bm{P}_\perp$ density;
and as pointed out in Sec.~\ref{sec:flux:half}, it is also
equal to the transverse energy flux due to symmetry of the EMT.
We thus point the reader to the middle panel of Fig.~\ref{fig:proton:momentum}
for an estimate of these quantities for a longitudinally-polarized proton.

\begin{figure}
  \centering
  \includegraphics[width=0.33\textwidth]{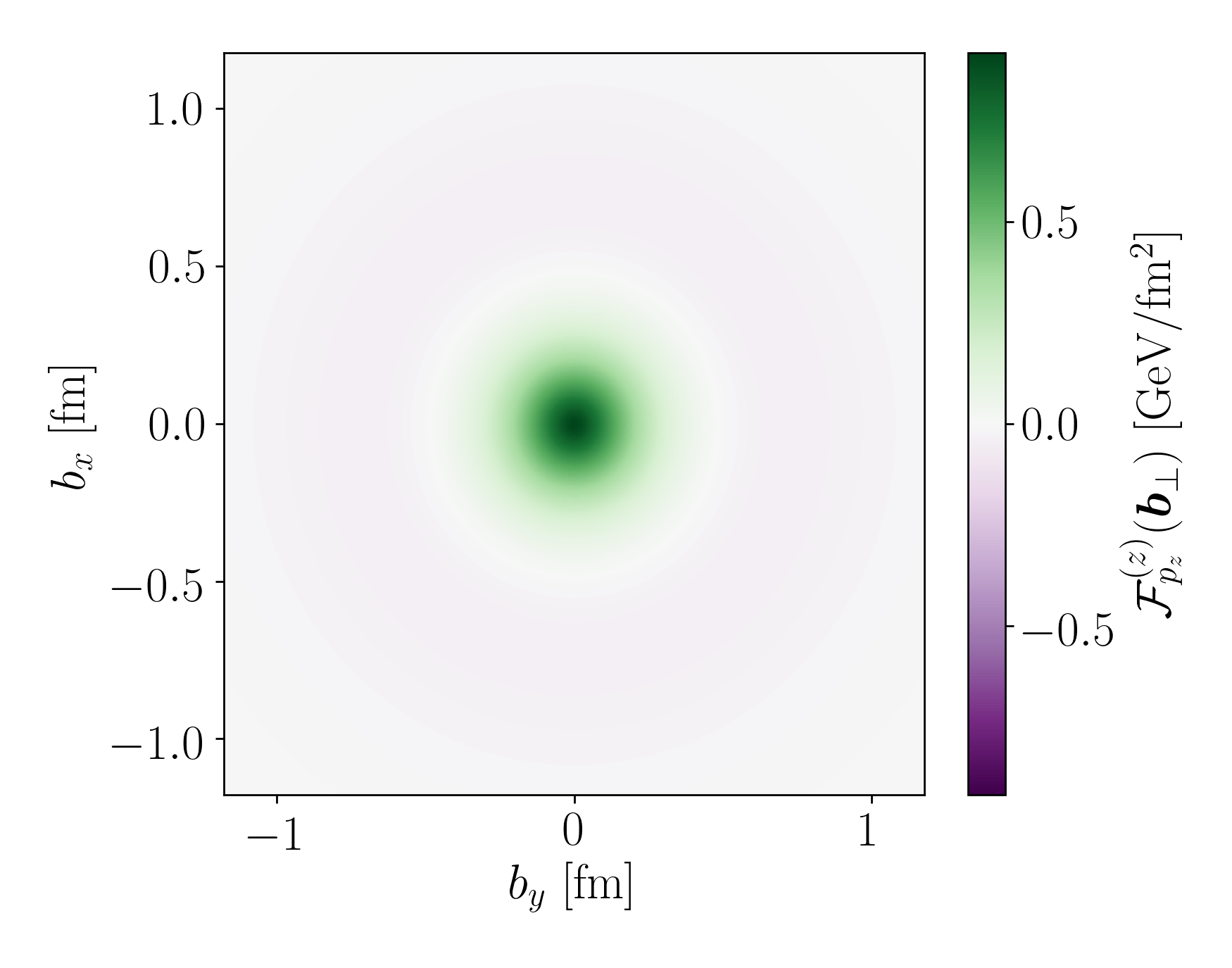}
  \includegraphics[width=0.33\textwidth]{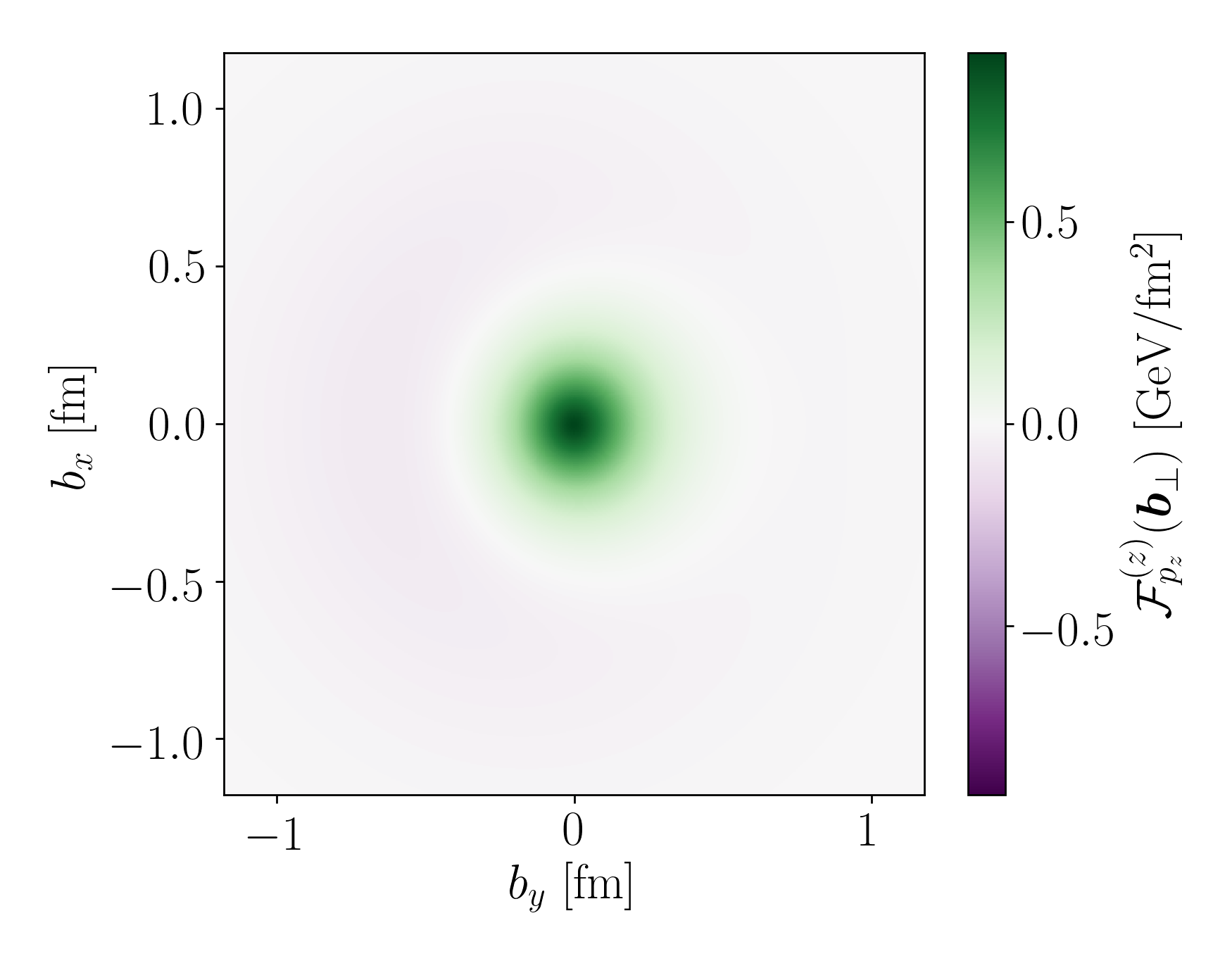}
  \includegraphics[width=0.33\textwidth]{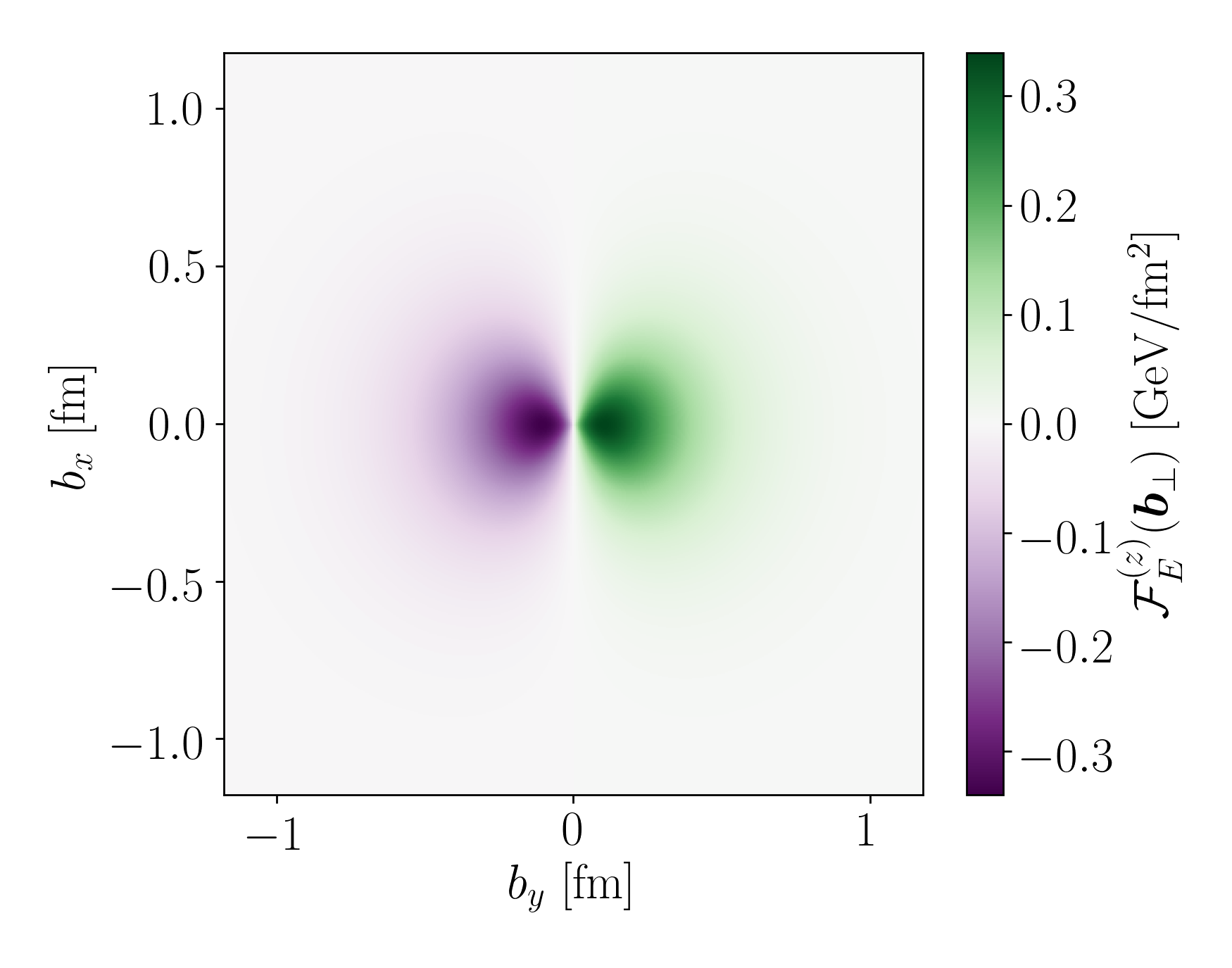}
  \caption{
    $P_z$ and energy flux densities in the $z$ direction.
    The (left panel) shows the
    $P_z$ flux density for a longitudinally polarized state,
    while the (middle panel) shows the same for a
    transversely-polarized state with spin up along the $x$ axis.
    The (right panel) shows the longitudinal energy flux for this
    same transversely-polarized state.
    The longitudinal energy flux is identically zero for a
    longitudinally-polarized state
    (since energy flux is carried by angular momentum),
    and is thus not plotted.
    In these plots, the $x$-axis is oriented vertically
    and the $y$-axis horizontally so that the $z$ axis points into the page,
    allowing plots to mimic what an observer would see
    at fixed light front time.
    Accordingly, positive flux is into the page
    (away from the observer)
    and negative flux is out of the page
    (towards the observer).
  }
  \label{fig:proton:flux}
\end{figure}

The longitudinal $P_z$ flux, which can also be interpreted
as pressure in the $z$ direction
(since it is a normal stress)
can be obtained by putting Eq.~(\ref{eqn:gff:holography})
into Eq.~(\ref{eqn:stress:half}) with $i=j=3$.
One can similarly obtain the longitudinal energy flux
by plugging Eq.~(\ref{eqn:gff:holography})
into Eq.~(\ref{eqn:flux:half}) with $i=3$.
Numerical results of taking these 2D Fourier transforms
are presented in Fig.~\ref{fig:proton:flux}.

\begin{figure}
  \includegraphics[width=0.49\textwidth]{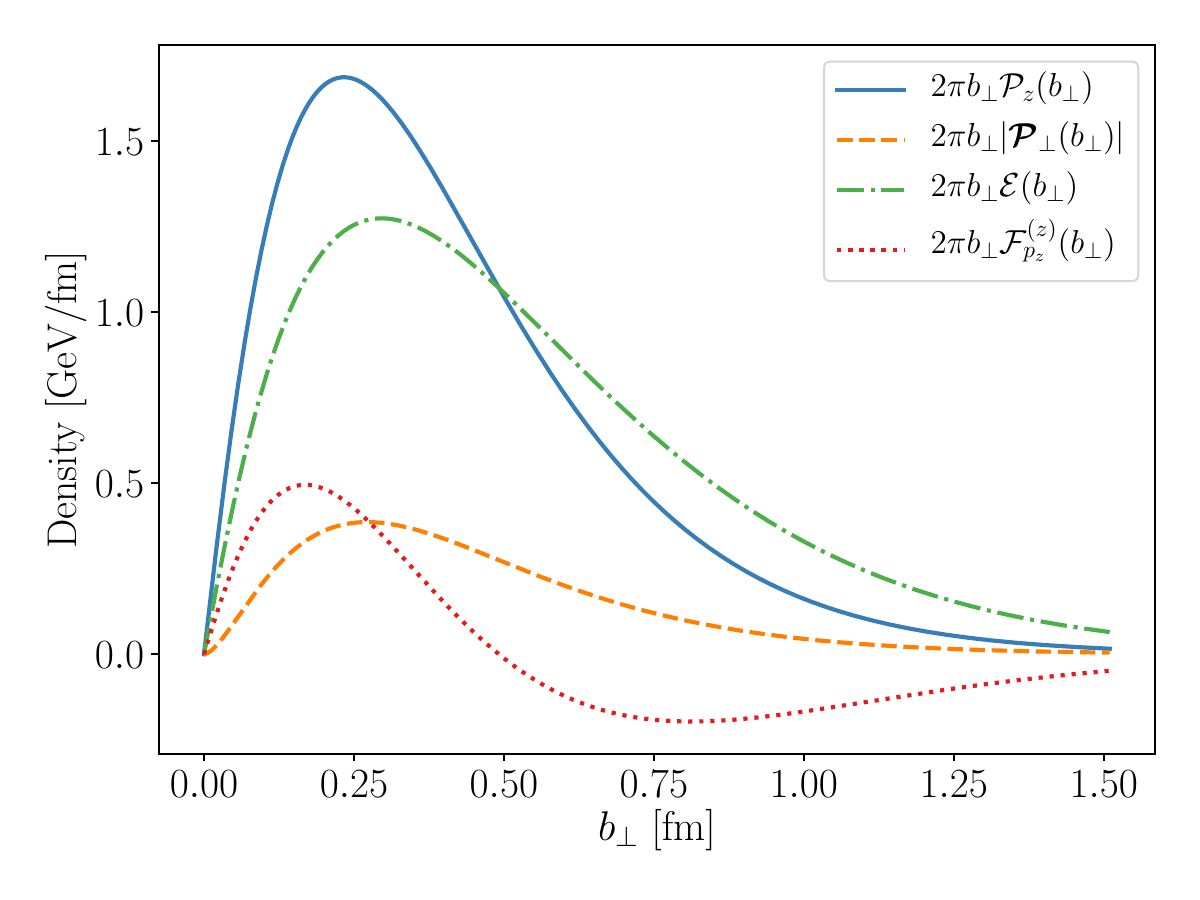}
  \caption{
    One-dimensional reductions of several transverse
    densities in the proton.
    The solid blue line is the $P_z$ density;
    the dashed orange line is the magnitude
    of the $\bm{P}_\perp$ density;
    the dash-dotted green line is the energy density;
    and the dotted red line is the longitudinal $P_z$ flux.
  }
  \label{fig:proton:1D}
\end{figure}

The longitudinal flux densities all integrate to zero,
which may be difficult to see by eye in the $P_z$ flux plots;
the core of negative $P_z$ flux
is surrounded by a diffuse cloud of positive flux.
A one-dimensional reductions of $P_z$ flux density
for longitudinally polarized protons is presented later
in Fig.~\ref{fig:proton:1D},
where the vanishing of the net flux is easier to see.

\begin{figure}
  \centering
  \includegraphics[width=0.49\textwidth]{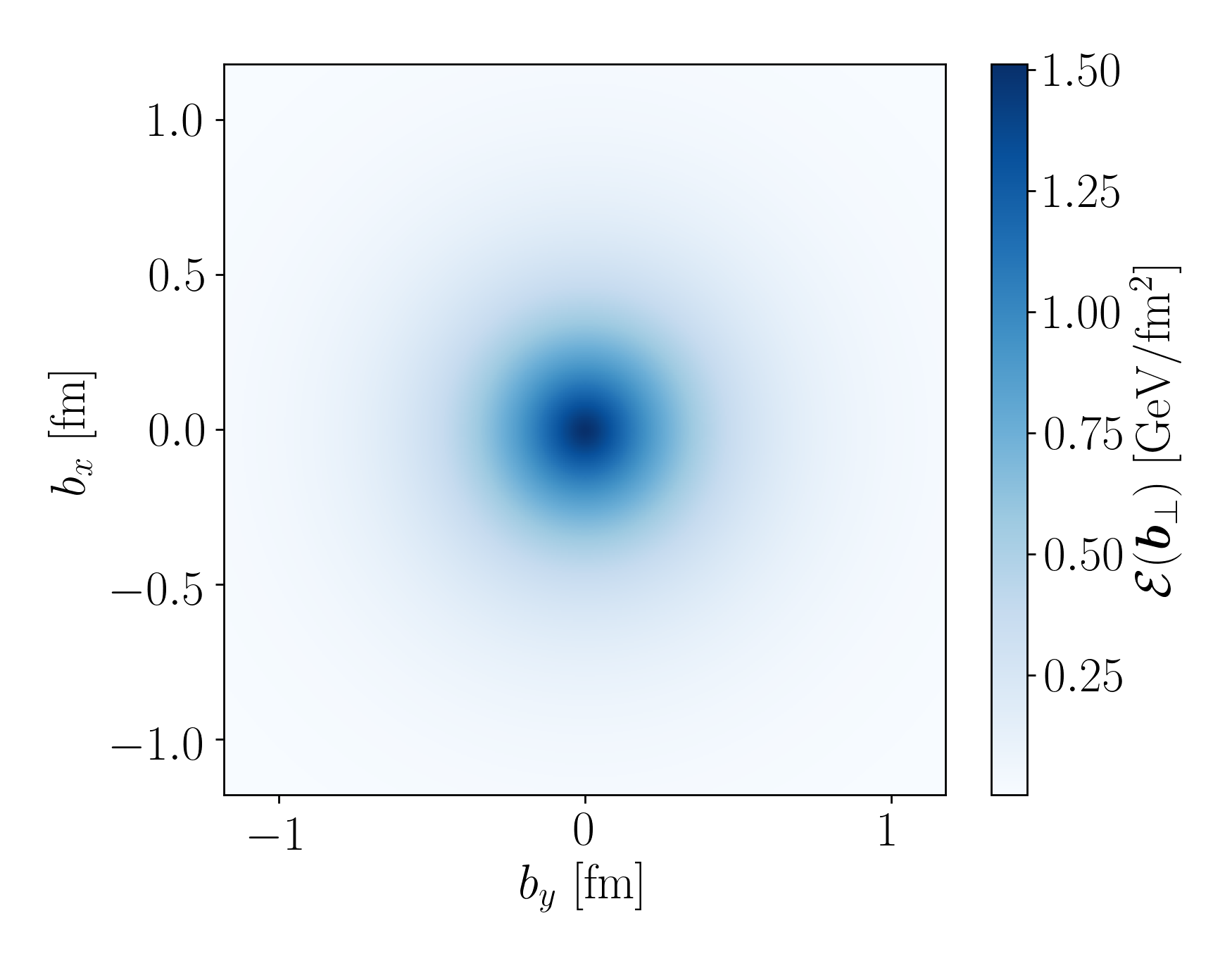}
  \includegraphics[width=0.49\textwidth]{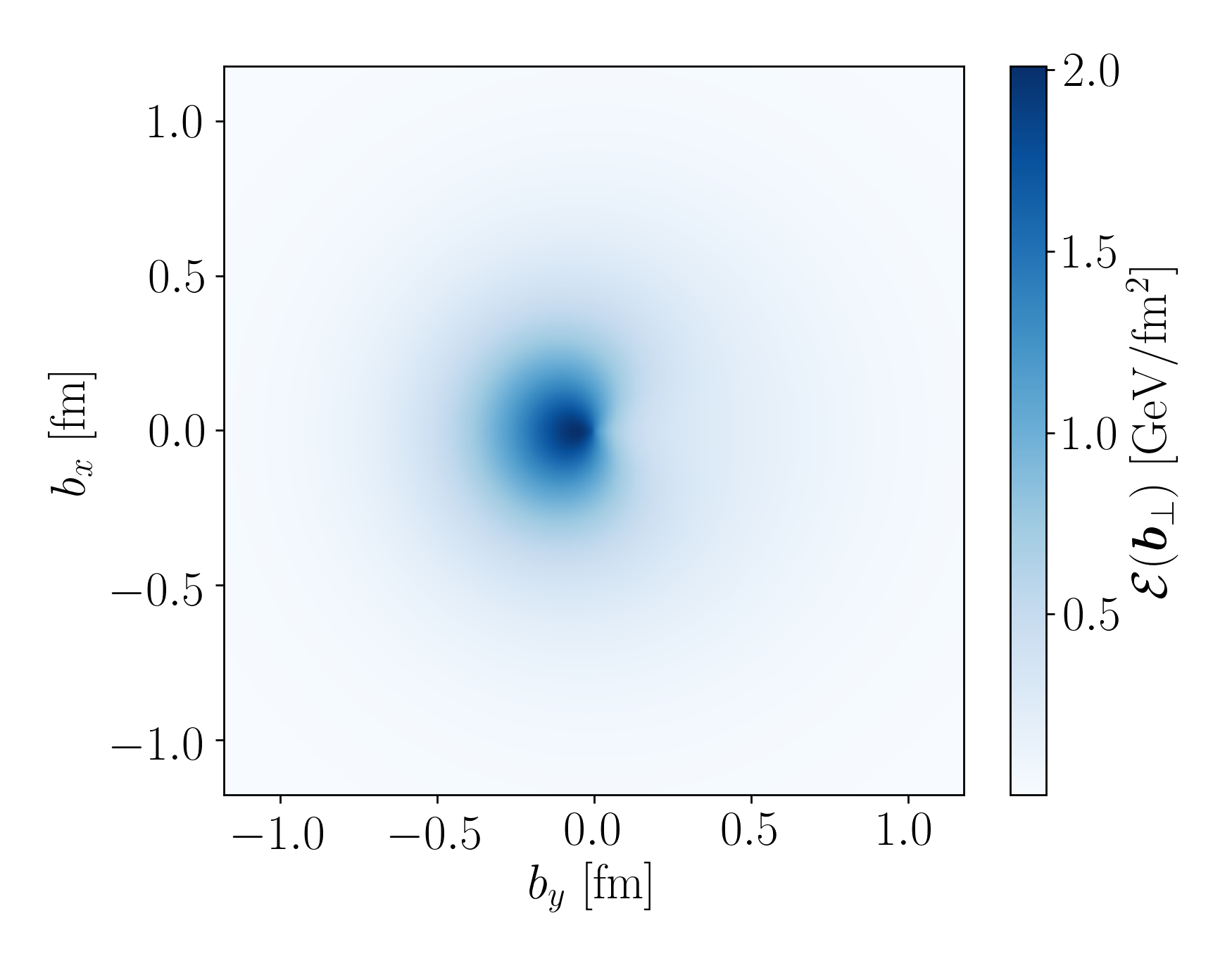}
  \caption{
    Energy density
    for protons polarized with spin up along the $z$ axis
    (left panel)
    and spin up along the $x$ axis
    (right panel).
    In these plots, the $x$-axis is oriented vertically
    and the $y$-axis horizontally so that the $z$ axis points into the page,
    allowing plots to mimic what an observer would see
    at fixed light front time.
  }
  \label{fig:proton:energy}
\end{figure}

Finally, from Eqs.~(\ref{eqn:energy:half})
and (\ref{eqn:gff:holography}) we obtain the energy density.
Numerical results for the energy density of both
longitudinally and transversely polarized states are presented in
Fig.~\ref{fig:proton:energy}.
The right panel in particular shows the energy density of
a transversely-polarized proton with its spin up along the
$x$-axis.
In these plots, the $x$-axis is vertical and the $y$-axis horizontal,
so that the $z$-axis points into the page by the right-hand rule.
This is done so that the plots are representative of what an observer
would actually see at fixed light front time.
The energy is lopsided on the side of the proton that is revolving
towards the observer,
in contrast to the modulations previously seen in the
proton's charge density~\cite{Freese:2023jcp}.
As explained in Sec.~\ref{sec:energy:half},
these modulations have a different cause than
the charge density modulations,
which were the result of clock rate modulations.
The energy density modulations are largely
an artifact of the proton's center in the light front formalism
being the center-of-$P_z$:
there are equal amounts of $P_z$ on both sides of the proton,
and given the tilted coordinate relation $p_z v_z = p_z - E$,
there must be more energy on the side with $v_z < 0$---i.e.,
the side moving towards the observer.

It is also instructive to consider one-dimensional reductions of the
densities and currents we have calculated.
Such reductions are presented in Fig.~\ref{fig:proton:1D},
specifically for the case of a longitudinally-polarized proton.
This figure illustrates several
interesting qualitative features of the energy and momentum densities.
First of all, the magnitude of the transverse momentum density is much smaller
than the $P_z$ or energy densities.
The $P_z$ density, it should be recalled,
is not a measure of motion in tilted coordinates,
but can better be interpreted as a measure of inertia,
and in fact the $P_z$ density here integrates to the mass.
It is thus not surprising that the $P_z$ density is much larger
than the $\bm{P}_\perp$ density.
On the contrary, the $\bm{P}_\perp$ density is quite large,
becoming as large as hundreds of MeV/fm,
which is indicative of ultrarelativistic motion within the proton.

Fig.~\ref{fig:proton:1D} also qualitatively illustrates
that the energy distribution in the proton is broader than the $P_z$ distribution.
Accordingly, the energy radius should be larger than the momentum radius.
Using Eqs.~(\ref{eqn:radius:momentum:half})
and (\ref{eqn:radius:energy:half}) for the radii, we find:
\begin{subequations}
  \begin{align}
    \langle \bm{b}_\perp^2 \rangle_{P_z}
    &=
    \big(
    0.50~\mathrm{fm}
    \big)^2
    \\
    \langle \bm{b}_\perp^2 \rangle_{E}
    &=
    \big(
    0.65 \pm 0.02~\mathrm{fm}
    \big)^2
    \,.
  \end{align}
\end{subequations}
We stress that these are just illustrative,
and not the result of a precision calculation
or extraction from precision empirical data.
Future experiments aimed at extracting
generalized parton distributions,
such as deeply virtual Compton scattering~\cite{Ji:1996nm,Radyushkin:1997ki,Belitsky:2005qn,Kriesten:2019jep}
and single-diffractive hard exclusive reactions~\cite{Qiu:2022pla},
must be carried out to provide both more realistic estimates
of the proton's gravitational form factors.

Lastly, it is more apparent by eye in Fig.~\ref{fig:proton:1D}
that the longitudinal $P_z$ flux integrates to zero than
in the left panel Fig.~\ref{fig:proton:flux}.

%%%%%%%%%%%%%%%%%%%%%%%%%%%%%%%%%%%%%%%%

\subsection{Changes when using the asymmetric EMT}
\label{sec:asym}

Before concluding, let us consider how the densities we have present
would be modified by including an antisymmetric piece in the EMT,
as defined in Eq.~(\ref{eqn:emt:anti}).
This would introduce an additional form factor
\begin{align}
  \langle \bm{p}',\lambda' |
  \hat{T}_{\text{A}}^{\mu\nu}(0)
  | \bm{p}, \lambda \rangle
  =
  \bar{u}(\bm{p}',\lambda')
  \gamma^{[\mu}P^{\nu]}
  u(\bm{p},\lambda)
  S(\varDelta^2)
\end{align}
which must be added to the breakdown in Eq.~(\ref{eqn:gff:half}).
Using formulas from Appendix A of Ref.~\cite{Freese:2023jcp},
we can explicitly evaluate matrix elements of this when $\varDelta_z = 0$ to be:
\begin{align}
  \langle \bm{p}',\lambda' |
  \hat{T}_{\text{A}}^{\mu\nu}(0)
  | \bm{p}, \lambda \rangle
  &=
  -
  \left\{
    -
    \frac{
      \i
      P^{[\mu}
      \epsilon^{\nu]\rho\sigma\tau}
      {n}_\rho
      {P}_\sigma {\varDelta}_\tau
    }{({P}\cdot{n})}
    (\sigma_3)_{\lambda'\lambda}
    +
    \frac{mP^{[\mu}{n}^{\nu]}}{({P}\cdot{n})}
    \i
    \epsilon^{\alpha\beta\gamma\delta}
    {n}_\alpha {\bar{n}}_\beta
    {\varDelta}_\gamma (\sigma_\delta)_{\lambda'\lambda}
    \right\}
  S(-\bm{\varDelta}_\perp^2)
  \,.
\end{align}
Using Eq.~(\ref{eqn:emt}), the additional contribution
of this antisymmetric piece to the intrinsic EMT is:
\begin{align}
  t^{\mu\nu}(\bm{b}_\perp;\lambda,\lambda')
  &=
  -
  m
  \int \frac{\d^2\bm{\varDelta}_\perp}{(2\pi)^2}
  \frac{
    \bar{n}^{[\mu}
    \big( \bm{\sigma}_{\lambda'\lambda} \times \i\bm{\varDelta}_\perp \big)^{\nu]}}{2m}
  S(-\bm{\varDelta}_\perp^2)
  \,.
\end{align}
Adding this to the intrinsic EMT would alter
several of the densities we explored above.
Firstly,
in the expression for the transverse momentum density
$\bm{\P}_\perp(\bm{b}_\perp,\hat{\bm{s}})$
in Eq.~(\ref{eqn:momentum:half}),
the form factor
$J(-\bm{\varDelta}_\perp^2)$
would be replaced by
${
  J(-\bm{\varDelta}_\perp^2)
  -
  S(-\bm{\varDelta}_\perp^2)
}$,
giving:
\begin{align}
  \label{eqn:momentum:asym}
  \bm{\P}_\perp^{(\text{asym})}(\bm{b}_\perp, \hat{\bm{b}}_\perp)
  =
  \bm{\P}_\perp(\bm{b}_\perp, \hat{\bm{b}}_\perp)
  -
  m
  (\hat{e}_z\cdot\hat{\bm{s}})
  \int \frac{\d^2\bm{\varDelta}_\perp}{(2\pi)^2}
  \frac{\hat{e}_z \times \i \bm{\varDelta}_\perp}{2m}
  S(-\bm{\varDelta}_\perp^2)
  \e^{-\i\bm{\varDelta}_\perp\cdot\bm{b}_\perp}
  \,.
\end{align}
Under the interpretation that
$S(-\bm{\varDelta}_\perp^2)$
encodes the spatial distribution of \emph{fermion} spin,
the $J-S$ difference encodes a combination of quark OAM
and gluon total angular momentum.
Accordingly, when using the asymmetric EMT,
the transverse momentum density tracks this particular mix
of contributions to the angular momentum.
If no gluons were present in the hadron,
${
  J(-\bm{\varDelta}_\perp^2)
  -
  S(-\bm{\varDelta}_\perp^2)
}$
would simply track OAM.
It is interesting to note that for a free fermion,
this form factor difference is zero,
and the transverse momentum density in a free fermion thus vanishes
for the asymmetric EMT.

Secondly,
in the energy flux density
$\bm{\mathcal{F}}_E(\bm{b}_\perp,\hat{\bm{s}})$
of Eq.~(\ref{eqn:flux:half}),
$J(-\bm{\varDelta}_\perp^2)$
would be replaced by
${
  J(-\bm{\varDelta}_\perp^2)
  +
  S(-\bm{\varDelta}_\perp^2)
}$, giving:
\begin{align}
  \label{eqn:flux:asym}
  \bm{\mathcal{F}}_E^{(\text{asym})}(\bm{b}_\perp, \hat{\bm{b}}_\perp)
  =
  \bm{\mathcal{F}}_E(\bm{b}_\perp, \hat{\bm{b}}_\perp)
  +
  m
  \int \frac{\d^2\bm{\varDelta}_\perp}{(2\pi)^2}
  \frac{\hat{\bm{s}} \times \i \bm{\varDelta}_\perp}{2m}
  S(-\bm{\varDelta}_\perp^2)
  \e^{-\i\bm{\varDelta}_\perp\cdot\bm{b}_\perp}
  \,.
\end{align}
Along the same vein, the e $P_z$ flux
is also modified by replacing
$J(-\bm{\varDelta}_\perp^2)$
with
${
  J(-\bm{\varDelta}_\perp^2)
  +
  S(-\bm{\varDelta}_\perp^2)
}$ in Eq.~(\ref{eqn:stress:half}) giving:
\begin{align}
  \label{eqn:stress:asym}
  \bm{\mathcal{F}}_{p_z}^{(\text{asym})}(\bm{b}_\perp, \hat{\bm{b}}_\perp)
  =
  \bm{\mathcal{F}}_{p_z}(\bm{b}_\perp, \hat{\bm{b}}_\perp)
  +
  m
  (\hat{e}_z\cdot\hat{\bm{s}})
  \int \frac{\d^2\bm{\varDelta}_\perp}{(2\pi)^2}
  \frac{\hat{e}_z \times \i \bm{\varDelta}_\perp}{2m}
  S(-\bm{\varDelta}_\perp^2)
  \e^{-\i\bm{\varDelta}_\perp\cdot\bm{b}_\perp}
  \,.
\end{align}
Notably, the transverse $P_z$ flux and transverse momentum densities are no
longer equal for the asymmetric EMT,
meaning small elements of matter inside the hadron no longer obey
the relation $\bm{p}_\perp = p_z \bm{v}_\perp$.
Of course, there is no formal constraint that formal elements of matter
(as opposed to on-shell particles)
must obey this relation, so the asymmetric EMT is not inconsistent for this.

Lastly, the energy density
$\E(\bm{b}_\perp,\hat{\bm{s}})$
is modified in Eq.~(\ref{eqn:energy:half})
by replacing
$J(-\bm{\varDelta}_\perp^2)$
with
${
  J(-\bm{\varDelta}_\perp^2)
  -
  S(-\bm{\varDelta}_\perp^2)
}$,
giving:
\begin{align}
  \label{eqn:energy:asym}
  \E^{(\text{asym})}(\bm{b}_\perp, \hat{\bm{b}}_\perp)
  =
  \E(\bm{b}_\perp, \hat{\bm{b}}_\perp)
  -
  m
  \int \frac{\d^2\bm{\varDelta}_\perp}{(2\pi)^2}
  \frac{\hat{e}_z\cdot(\hat{\bm{s}} \times \i \bm{\varDelta}_\perp)}{2m}
  S(-\bm{\varDelta}_\perp^2)
  \e^{-\i\bm{\varDelta}_\perp\cdot\bm{b}_\perp}
  \,.
\end{align}
It is interesting to note that for a free elementary fermion,
$\E^{(\text{asym})}(\bm{b}_\perp, \hat{\bm{b}}_\perp)
= m \delta^{(2)}(\bm{b}_\perp)$
regardless of polarization;
since
${
  J_{\text{free}}(-\bm{\varDelta}_\perp^2)
  =
  S_{\text{free}}(-\bm{\varDelta}_\perp^2)
}$,
the antisymmetric contribution to the energy density cancels the
angular modulations that occurred in the symmetric EMT.
Accordingly, for the asymmetric EMT,
a synchronization-induced energy dipole moment arises from internal dynamics
rather than being universally present in all fermions.
The induced energy dipole moment in this case is:
\begin{align}
  \bm{d}_E^{(\text{asym})}
  =
  \left(-\frac{1}{4} + \frac{1}{2}S(0)\right)
  \hat{e}_z \times \hat{\bm{s}}
  \approx
  -
  0.148
  \,
  \hat{e}_z \times \hat{\bm{s}}
  \,,
\end{align}
if we use $S(0) = \frac{1}{2}\Big(\Delta u^+ + \Delta d^+\Big)$
from JAM estimates~\cite{Cocuzza:2022jye}.

One benefit of the asymmetric EMT is that the energy and momentum
\emph{densities} are all trivial for a free elementary fermion,
which appeals to intuition about the behavior of pointlike particles.
The symmetric EMT, by contrast, suggests that pointlike fermions
have non-trivial distributions of energy and momentum.
To be sure, this picture seems more reasonable when recalling that
fields are the fundamental objects of quantum field theories
rather than particles;
it is not farfetched to imagine that the fermion field can carry
momentum around the center of an excitation in the field.
For that matter, the flux densities in the asymmetric EMT
of a free fermion are not zero,
meaning that even the asymmetric EMT describes flows of
energy and momentum---but flows that themselves contain zero momentum.
In this regard, the symmetric EMT paints a more straightforward picture.

\section{Conclusions}
\label{sec:end}

In this work, we constructed and explored a formalism for obtaining exact,
two-dimensional relativistic rest frame energy-momentum densities and currents
for spin-zero and spin-half targets.
We derived a general expression for these densities in terms of matrix elements
of the EMT operator in Eq.~(\ref{eqn:emt}),
and subsequently obtained more explicit formulas for specific EMT components
in terms of the gravitational form factors.
Additionally, we provided numerical estimates for what these densities
may look like for a proton,
using the holographic model of Mamo and Zahed~\cite{Mamo:2019mka,Mamo:2021krl}
in light of the limited empirical data that exist.

The densities were obtained under a non-standard time synchronization convention
via tilted light front coordinates, which results in angular modulations
in the densities of transversely-polarized systems.
These are similar to previously-known modulations in the
light front charge density~\cite{Carlson:2007xd,Freese:2023jcp},
which arise at fixed light front time due to clock rate modulations
for quarks moving towards or away from the observer.
The energy density of transversely-polarized targets has the peculiarity
that its $\sin\phi$ modulations have the opposite sign from the modulations
in any other density---in particular, energy bunches on the side of the target
moving towards the observer rather than away.
As discussed in Sec.~\ref{sec:energy:half},
this is an artifact of the hadron's barycenter being the center-of-$P_z$
rather than the center-of-energy.
Curiously, the synchronization-induced energy dipole moment of a spin-half
target is universally $-1/4$ if the symmetric Belinfante EMT is used to
define the energy density---even for pointlike fermions.
However, if the asymmetric EMT
(defined by Leader and Lorc\'{e}~\cite{Leader:2013jra})
is used instead, the energy dipole moment is non-universal
and vanishes for pointlike fermions (see Sec.~\ref{sec:asym}).

That the energy density should differ depending on whether the symmetric
or asymmetric EMT is used may be the most peculiar and interesting
of our results.
It is not clear whether the energy density can be directly measured
(as opposed to indirectly obtained by taking Fourier transforms of
empirical gravitational form factors)
so it is unclear that these cases
can be empirically distinguished.
Nonetheless, the energy density is one of the most preeminent desired
quantities for describing hadron structure---being closely tied in with
the mass decomposition and mass origin questions---and that the different
EMT operators should entail different energy densities
suggests that we ought to seriously consider which is more appropriate to use.

%%%%%%%%%%%%%%%%%%%%%%%%%%%%%%%%%%%%%%%%

\begin{acknowledgments}
  We would like to thank
  Ian Clo\"{e}t, Wim Cosyn, Yang Li,
  C\'{e}dric Lorc\'{e}, and Andreas Metz
  for thought-provoking conversations that helped contribute to this work.
  This work was supported by the U.S.\ Department of Energy
  Office of Science, Office of Nuclear Physics under Award Number
  DE-FG02-97ER-41014.
  AF was additionally supported by
  the U.S.\ Department of Energy contract No.\ DE-AC05-06OR23177,
  under which Jefferson Science Associates, LLC operates Jefferson Lab.
  This work has been stimulated by the Quark Gluon Topical Collaboration of the
  U.S.\ Department of Energy.
\end{acknowledgments}

%%%%%%%%%%%%%%%%%%%%%%%%%%%%%%%%%%%%%%%%%%%%%%%%%%%%%%%%%%%%%%%%%%%%%%%%%%%%%%%%

\appendix

\section{Basic identities in tilted light front coordinates}
\label{sec:basic}

For convenience, we reproduce here several identities involving
tilted light front coordinates from our previous work~\cite{Freese:2023jcp}.
In this work, we do not use tildes to signify tilted coordinates,
and expressions without explicit indication of the coordinate system
should be assumed to be in tilted coordinates.
By contrast, we explicitly signify instant form coordinates
with a subscript or superscript IF.

Tilted light front coordinates are defined in terms of Minkowski
(or instant form) coordinates as:
\begin{subequations}
  \label{eqn:appa:tilted}
  \begin{align}
    \tau = x^0 &\equiv t_{\text{IF}} + z_{\text{IF}} \\
    x    = x^1 &\equiv x_{\text{IF}} \\
    y    = x^2 &\equiv y_{\text{IF}} \\
    z    = x^3 &\equiv z_{\text{IF}}
    \,.
  \end{align}
\end{subequations}
In this way, tilted coordinates operationally correspond to
a change in the way that spatially distant clocks are
synchronized~\cite{reichenbach2012philosophy,Anderson:1998mu,Veritasium:2020oct,Freese:2023jcp}.
The metric tensor and its inverse are:
\begin{subequations}
  \label{eqn:appa:metric}
  \begin{align}
    g_{\mu\nu}
    &=
    \frac{\partial x_{\text{IF}}^\alpha}{\partial {x}^\mu}
    \frac{\partial x_{\text{IF}}^\beta }{\partial {x}^\nu}
    g_{\alpha\beta}
    =
    \left[
      \begin{array}{cccc}
        \phantom{-}1 & \phantom{-}0 & \phantom{-}0 & -1 \\
        \phantom{-}0 & -1 & \phantom{-}0 & \phantom{-}0 \\
        \phantom{-}0 & \phantom{-}0 & -1 & \phantom{-}0 \\
        -1 & \phantom{-}0 & \phantom{-}0 & \phantom{-}0
      \end{array}
      \right]
    \\
    g^{\mu\nu}
    &=
    \frac{\partial {x}^\mu}{\partial x_{\text{IF}}^\alpha}
    \frac{\partial {x}^\nu}{\partial x_{\text{IF}}^\beta }
    g^{\alpha\beta}
    =
    \left[
      \begin{array}{cccc}
        \phantom{-}0 & \phantom{-}0 & \phantom{-}0 & -1 \\
        \phantom{-}0 & -1 & \phantom{-}0 & \phantom{-}0 \\
        \phantom{-}0 & \phantom{-}0 & -1 & \phantom{-}0 \\
        -1 & \phantom{-}0 & \phantom{-}0 & -1
      \end{array}
      \right]
  \end{align}
\end{subequations}
Covariant (lower-index) and contravariant (upper-index)
four-vector components are related by:
\begin{subequations}
  \label{eqn:appa:covcon}
  \begin{align}
    &
    {A}_\mu = {g}_{\mu\nu} {A}^\nu
    &
    {A}^\mu = {g}^{\mu\nu} {A}_\nu
    \,,
  \end{align}
  which in terms of individual components gives:
  \begin{align}
    &
    {A}_0
    =
    {A}^0 - {A}^3
    &
    {A}^0
    =
    -
    {A}_3
    \,\phantom{.}
    \\
    &
    {A}_1
    =
    -{A}^1
    &
    {A}^1
    =
    -{A}_1
    \,\phantom{.}
    \\
    &
    {A}_2
    =
    -{A}^2
    &
    {A}^2
    =
    -{A}_2
    \,\phantom{.}
    \\
    &
    {A}_3
    =
    -{A}^0
    &
    {A}^3
    =
    -{A}_0 - {A}_3
    \,.
  \end{align}
\end{subequations}

The energy and momentum are defined to be time and space translation generators:
\begin{subequations}
  \begin{align}
    i[{E},\hat{O}(x)]
    &=
    {\partial}_0 \hat{O}(x)
    \\
    -i[{\bm{p}},\hat{O}(x)]
    &=
    {\bm{\nabla}} \hat{O}(x)
    \,,
  \end{align}
\end{subequations}
meaning they are related to covariant (lower-index)
components of the four-momentum $p_\mu$:
\begin{align}
  {p}_\mu
  \equiv
  ({E};-{p}_x,-{p}_y,-{p}_z)
  \,.
\end{align}
The tilted coordinate energy and momentum have the following relationships
to instant form energy and momentum:
\begin{subequations}
  \begin{align}
    {E}
    &=
    E^{\text{IF}}
    \\
    {p}_x
    &=
    p_x^{\text{IF}}
    \\
    {p}_y
    &=
    p_y^{\text{IF}}
    \\
    {p}_z
    &=
    E^{\text{IF}} + p_z^{\text{IF}}
    \,.
  \end{align}
\end{subequations}
The energy of a particle with mass $m$ is given by:
\begin{align}
  \label{eqn:appa:energy}
  {E}
  &=
  \frac{m^2 + {\bm{p}}^2}{2 {p}_z}
  \,,
\end{align}
The momentum and velocity are related in the following way:
\begin{subequations}
  \label{eqn:appa:velocity}
  \begin{align}
    {v}_x
    &=
    \frac{{p}_x}{{p}_z}
    \\
    {v}_y
    &=
    \frac{{p}_y}{{p}_z}
    \\
    {v}_z
    &=
    1
    -
    \frac{{E}}{{p}_z}
    \,.
  \end{align}
\end{subequations}
Notably, at rest, one has $\bm{p}_{\text{rest}} = (0,0,m)$.
This occurs because $p_z$ is defined to be the generator of translations
rather than to be proportional to velocity.

As usual, boosts transform contravariant four-vectors
according to the formula:
\begin{align}
  A'^\mu
  =
  \varLambda^{\mu}_{\phantom{\mu}\nu}
  A^\nu
  \,.
\end{align}
An active transverse boost can be written in matrix form as:
\begin{align}
  \label{eqn:appa:boost:transverse}
  (\varLambda_\perp)^{\mu}_{\phantom{\mu}\nu}
  =
  \left[
    \begin{array}{cccc}
      1 & 0 & 0 & 0 \\
      \beta_x & 1 & 0 & 0 \\
      \beta_y & 0 & 1 & 0 \\
      -\bm{\beta}_\perp^2/2 & -\beta_x & -\beta_y & 1
    \end{array}
    \right]
  \,,
\end{align}
where $\bm{\beta}_\perp = (\beta_x, \beta_y)$ is the velocity of the boost.
An active longitudinal boost can be written:
\begin{align}
  \label{eqn:appa:boost:longitudinal}
  (\varLambda_\parallel)^{\mu}_{\phantom{\mu}\nu}
  =
  \left[
    \begin{array}{cccc}
      \e^\eta & 0 & 0 & 0 \\
      0 & 1 & 0 & 0 \\
      0 & 0 & 1 & 0 \\
      \sinh\eta & 0 & 0 & \e^{-\eta}
    \end{array}
    \right]
  \,.
\end{align}
Here, $\eta$ is the rapidity of the longitudinal boost,
and is related to the velocity $v_z$ of the boost by:
\begin{align}
  \label{eqn:appa:rapidity}
  \beta_z^\parallel
  =
  \e^{-\eta} \sinh(\eta)
  =
  \frac{1}{2}\big( 1 - \e^{-2\eta} \big)
  \,.
\end{align}
As discussed in the main text, we consider states with arbitrary momentum $\bm{p}$
to be reached from the rest state through a longitudinal boost followed by a transverse boost.
This combined boost can be written:
\begin{align}
  (\varLambda)^{\mu}_{\phantom{\mu}\nu}
  =
  (\varLambda_\perp \varLambda_\parallel)^{\mu}_{\phantom{\mu}\nu}
  =
  \left[
    \begin{array}{cccc}
      \e^\eta & 0 & 0 & 0 \\
      \e^\eta \beta_x & 1 & 0 & 0 \\
      \e^\eta \beta_y & 0 & 1 & 0 \\
      \e^\eta \beta_z & -\beta_x & -\beta_y & \e^{-\eta}
    \end{array}
    \right]
  \,.
\end{align}
Here,
\begin{align}
  \beta_z
  =
  \e^{-\eta} \sinh(\eta)
  -
  \frac{\bm{\beta}_\perp^2}{2}
  =
  \beta_z^\parallel
  -
  \frac{\bm{\beta}_\perp^2}{2}
  \,,
\end{align}
as light front transverse boosts impart longitudinal velocity to the system,
so the total longitudinal velocity of the combined boost differs
from that of the longitudinal boost alone.
Notably, the transverse boosts are defined to leave $p_z$ invariant,
but the relationship between $p_z$ and $v_z$ (see Eq.~(\ref{eqn:appa:velocity}))
means that $v_z$ must change.
From Eqs.~(\ref{eqn:appa:energy}) and (\ref{eqn:appa:velocity}),
the boost that takes a system from rest to an arbitrary momentum $\bm{p}$
can be written in terms of its energy and momentum as:
\begin{align}
  \label{eqn:appa:boost}
  \varLambda^{\mu}_{\phantom{\mu}\nu}
  &=
  \left[
    \begin{array}{cccc}
      p_z/m & 0 & 0 & 0 \\
      p_x/m & 1 & 0 & 0 \\
      p_y/m & 0 & 1 & 0 \\
      (p_z-E)/m & -p_x/p_z & -p_y/p_z & m/p_z
    \end{array}
    \right]
  \,.
\end{align}

%%%%%%%%%%%%%%%%%%%%%%%%%%%%%%%%%%%%%%%%%%%%%%%%%%%%%%%%%%%%%%%%%%%%%%%%%%%%%%%%

\section{Proofs of intrinsic density formulas}
\label{sec:proof}

In this Appendix, we prove the intrinsic density formulas
Eqs.~(\ref{eqn:current}) and (\ref{eqn:emt}).
The premises underlying the proof are the universality (target independence)
of the smearing functions in Eqs.~(\ref{eqn:conv:j}) and (\ref{eqn:conv:emt}),
certain reasonable expectations for the intrinsic densities of point particles,
and the requirement that the smearing does not mix
different irreducible representations of the Lorentz group.

Our strategy is to utilize the universality of the smearing functions.
By finding the physical densities and the intrinsic densities in simple cases
where both are already known,
components of the smearing functions can be deduced.
Since the smearing functions are target-independent,
these same smearing functions are also applicable to hadrons.
We thus begin by proving Eqs.~(\ref{eqn:smear:current})
and (\ref{eqn:smear:emt}).

Once the smearing functions have been obtained,
Eqs.~(\ref{eqn:conv:j}) and (\ref{eqn:conv:emt}) can be inverted
to obtain general formulas for the intrinsic densities.
Thus, after obtaining the smearing functions,
we prove that Eqs.~(\ref{eqn:current}) and (\ref{eqn:emt})
are the results of this inversion.

Throughout this proof, we find it especially helpful to work
with Fourier transforms, utilizing the convolution theorem.
The Fourier tranforms of Eqs.~(\ref{eqn:conv:j})
and (\ref{eqn:conv:emt}) are:
\begin{subequations}
  \label{eqn:smear:fourier}
  \begin{align}
    \langle J^\mu(\bm{\varDelta}_\perp,\tau) \rangle_{\text{2D}}
    &=
    \sum_{\lambda,\lambda'}
    \mathscr{P}^\mu_{\phantom{\mu}\nu}(\bm{\varDelta}_\perp,\tau;\lambda,\lambda')
    j^\nu(\bm{\varDelta}_\perp;\lambda,\lambda')
    \Bigg|_{\varDelta_z=0}
    \\
    \langle T^{\mu\nu}(\bm{\varDelta}_\perp,\tau) \rangle_{\text{2D}}
    &=
    \sum_{\lambda,\lambda'}
    \mathscr{Q}^{\mu\nu}_{\phantom{\mu\nu}\alpha\beta}(\bm{\varDelta}_\perp,\tau;\lambda,\lambda')
    t^{\alpha\beta}(\bm{\varDelta}_\perp;\lambda,\lambda')
    \Bigg|_{\varDelta_z=0}
  \end{align}
\end{subequations}
where $\bm{\varDelta}_\perp$ is the Fourier conjugate of the position argument.
When these are written as functions of $\bm{\varDelta}_\perp$,
it should be implicitly assumed
that we mean the Fourier transform of this function.

Given Eqs.~(\ref{eqn:j:phys}) and (\ref{eqn:conv:emt}),
the Fourier transforms of the physical current and EMT densities can be written:
\begin{subequations}
  \begin{align}
    \langle
    J^\mu(\bm{\varDelta}_\perp)
    \rangle_{\text{2D}}
    &=
    \sum_{\lambda\lambda'}
    \int \frac{\d^3\bm{P}}{2P_z(2\pi)^3}
    \langle \bm{p},\lambda | \hat{\rho} | \bm{p}',\lambda' \rangle
    \frac{
      \langle \bm{p}',\lambda' | \hat{J}^\mu(0) | \bm{p},\lambda \rangle
    }{
      2P_z
    }
    \e^{\i \varDelta_0\tau}
    \Bigg|_{\varDelta_z=0}
    \\
    \langle
    T^{\mu\nu}(\bm{\varDelta}_\perp)
    \rangle_{\text{2D}}
    &=
    \sum_{\lambda\lambda'}
    \int \frac{\d^3\bm{P}}{2P_z(2\pi)^3}
    \langle \bm{p},\lambda | \hat{\rho} | \bm{p}',\lambda' \rangle
    \frac{
      \langle \bm{p}',\lambda' | \hat{T}^{\mu\nu}(0) | \bm{p},\lambda \rangle
    }{
      2P_z
    }
    \e^{\i \varDelta_0\tau}
    \Bigg|_{\varDelta_z=0}
    \,.
  \end{align}
\end{subequations}
where $\varDelta_0 = p'_0 - p_0 = (\bm{P}\cdot\bm{\varDelta}_\perp)/P_z$.
If we define non-script counterparts to the smearing funcitons via:
\begin{subequations}
  \begin{align}
    \mathscr{P}^\mu_{\phantom{\mu}\nu}(\bm{\varDelta}_\perp,\tau;\lambda,\lambda')
    &=
    \int \frac{\d^3\bm{P}}{2P_z(2\pi)^3}
    \langle \bm{p},\lambda | \hat{\rho} | \bm{p}',\lambda' \rangle
    \frac{m}{P_z}
    %\mathfrak{p}^\mu_{\phantom{\mu}\nu}(\bm{\varDelta}_\perp,\tau;\lambda,\lambda')
    \mathfrak{p}^\mu_{\phantom{\mu}\nu}(\bm{P},\bm{\varDelta}_\perp)
    \e^{\i \varDelta_0\tau}
    \Bigg|_{\varDelta_z=0}
    \\
    \mathscr{Q}^{\mu\nu}_{\phantom{\mu\nu}\alpha\beta}(\bm{\varDelta}_\perp,\tau;\lambda,\lambda')
    &=
    \int \frac{\d^3\bm{P}}{2P_z(2\pi)^3}
    \langle \bm{p},\lambda | \hat{\rho} | \bm{p}',\lambda' \rangle
    \frac{m}{P_z}
    %\mathfrak{q}^{\mu\nu}_{\phantom{\mu\nu}\alpha\beta}(\bm{\varDelta}_\perp,\tau;\lambda,\lambda')
    \mathfrak{q}^{\mu\nu}_{\phantom{\mu\nu}\alpha\beta}(\bm{P},\bm{\varDelta}_\perp)
    \e^{\i \varDelta_0\tau}
    \Bigg|_{\varDelta_z=0}
    \,,
  \end{align}
\end{subequations}
then Eq.~(\ref{eqn:smear:fourier}) can be reduced to:
\begin{subequations}
  \label{eqn:smear:reduced}
  \begin{align}
    \frac{
      \langle \bm{p}',\lambda' | \hat{J}^\mu(0) | \bm{p},\lambda \rangle
    }{
      2m
    }
    \Bigg|_{\varDelta_z=0}
    &=
    %\mathfrak{p}^\mu_{\phantom{\mu}\nu}(\bm{\varDelta}_\perp,\tau;\lambda,\lambda')
    \mathfrak{p}^\mu_{\phantom{\mu}\nu}(\bm{P},\bm{\varDelta}_\perp)
    j^\nu(\bm{\varDelta}_\perp;\lambda,\lambda')
    %%%\Bigg|_{\varDelta_z=0}
    \\
    \frac{
      \langle \bm{p}',\lambda' | \hat{T}^{\mu\nu}(0) | \bm{p},\lambda \rangle
    }{
      2m
    }
    \Bigg|_{\varDelta_z=0}
    &=
    %\mathfrak{q}^{\mu\nu}_{\phantom{\mu\nu}\alpha\beta}(\bm{\varDelta}_\perp,\tau;\lambda,\lambda')
    \mathfrak{q}^{\mu\nu}_{\phantom{\mu\nu}\alpha\beta}(\bm{P},\bm{\varDelta}_\perp)
    t^{\alpha\beta}(\bm{\varDelta}_\perp;\lambda,\lambda')
    %%%\Bigg|_{\varDelta_z=0}
    \,.
  \end{align}
\end{subequations}
Thus Eq.~(\ref{eqn:smear:current}) is true if
$
  \mathfrak{p}^\mu_{\phantom{\mu}\nu}
  =
  \bar{\varLambda}^\mu_{\phantom{\mu}\nu}
$
and Eq.~(\ref{eqn:smear:emt}) is true if
$
  \mathfrak{q}^{\mu\nu}_{\phantom{\mu\nu}\alpha\beta}
  =
  \left(
  \bar{\varLambda}^\mu_{\phantom{\mu}\alpha}
  \bar{\varLambda}^\nu_{\phantom{\nu}\beta}
  -
  \frac{\bm{\varDelta}_\perp^2}{4P_z^2}
  \delta^\mu_3
  \delta^\nu_3
  \delta^3_\alpha
  \delta^3_\beta
  \right)
$.
We shall proceed to prove these relations.

%%%%%%%%%%%%%%%%%%%%%%%%%%%%%%%%%%%%%%%%

\subsection{Constraints from Galilean densities}

The components of the intrinsic densities with indices
0, 1 and 2 when written in contravariant form
(i.e., with all indices raised)
are known as Galilean densities~\cite{Lorce:2018egm,Freese:2021czn}
and we refer to here them as the Galilean components
of $j^\mu$ and $t^{\mu\nu}$.
The remaining non-Galilean components
(those with a 3 as an index in contravariant form)
do not mix into the Galilean components
under transformations in the Galilean subgroup
of the Poincar\'{e} group
(see discussions in Refs.~\cite{Soper:1972xc,Burkardt:2002hr,Lorce:2018egm,Freese:2023jcp}).
For these components, the intrinsic transverse densities are
already known in the literature~\cite{Burkardt:2002hr,Miller:2010nz,Lorce:2018egm,Miller:2018ybm,Freese:2021czn,Freese:2023jcp}.
To obtain them,
one simply applies the rest condition $\bm{P} = (0,0,m)$
(see Eq.~(\ref{eqn:appa:velocity}))
to the kinematic components of the average target momentum
in the standard hadronic matrix elements,
and sets $\varDelta_z = 0$ since the $z$ coordinate is integrated out:
\begin{subequations}
  \label{eqn:galileo:int}
  \begin{align}
    j^C(\bm{\varDelta}_\perp;\lambda,\lambda')
    &=
    \frac{
      \langle \bm{p}',\lambda' | \hat{J}^C(0) | \bm{p},\lambda \rangle
    }{
      2m
    }
    \Bigg|_{\bm{P}=(0,0,m),\varDelta_z=0}
    \\
    t^{CD}(\bm{\varDelta}_\perp;\lambda,\lambda')
    &=
    \frac{
      \langle \bm{p}',\lambda' | \hat{T}^{CD}(0) | \bm{p},\lambda \rangle
    }{
      2m
    }
    \Bigg|_{\bm{P}=(0,0,m),\varDelta_z=0}
  \end{align}
\end{subequations}
Here, we use uppercase Latin letters $\{A,B,C,D\}$
for indices constrained to $\{0,1,2\}$.
On the other hand, the Galilean components of the left-hand side
of Eq.~(\ref{eqn:smear:reduced}) are given by the same matrix elements
without the rest condition applied.
%and are accordingly related to the rest frame results
%through Lorentz boost matrices:
%\begin{subequations}
%  \label{eqn:galileo:ext}
%  \begin{align}
%    \frac{
%      \langle \bm{p}',\lambda' | \hat{J}^A(0) | \bm{p},\lambda \rangle
%    }{
%      2P_z
%    }
%    \Bigg|_{\varDelta_z=0}
%    &=
%    \frac{m}{P_z}
%    \\
%    \frac{
%      \langle \bm{p}',\lambda' | \hat{T}^{AB}(0) | \bm{p},\lambda \rangle
%    }{
%      2P_z
%    }
%    \Bigg|_{\varDelta_z=0}
%    &=
%    \frac{m}{P_z}
%  \end{align}
%\end{subequations}
The matrix elements in the rest frame and moving frame
are connected by Lorentz boosts, which in general requires
the application of Wigner rotations,
as observed by Lorc\'{e} for instance~\cite{Lorce:2020onh}.
However, the spin dependence appearing in these matrix elements
is through the light front helicity $\lambda$,
which is invariant under light front
boosts~\cite{Soper:1972xc,Freese:2023jcp}.
Thus, the Wigner rotations are trivial in this framework,
and we simply have:
\begin{subequations}
  \begin{align}
    \langle \bm{p}',\lambda' | \hat{J}^A(0) | \bm{p},\lambda \rangle
    \Big|_{\varDelta_z=0}
    &=
    \varLambda^{A}_{\phantom{A}C}(\bm{P})
    \,
    \langle \bm{p}',\lambda' | \hat{J}^C(0) | \bm{p},\lambda \rangle
    %\e^{\i \varDelta_0\tau}
    \Big|_{\bm{P}=(0,0,m),\varDelta_z=0}
    \\
    \langle \bm{p}',\lambda' | \hat{T}^{AB}(0) | \bm{p},\lambda \rangle
    \Big|_{\varDelta_z=0}
    &=
    \varLambda^{A}_{\phantom{A}C}(\bm{P})
    \varLambda^{B}_{\phantom{B}D}(\bm{P})
    \,
    \langle \bm{p}',\lambda' | \hat{T}^{CD}(0) | \bm{p},\lambda \rangle
    %\e^{\i \varDelta_0\tau}
    \Big|_{\bm{P}=(0,0,m),\varDelta_z=0}
  \end{align}
\end{subequations}
Crucially, the validity of these formulas follows from the fact
that non-Galilean components do not mix into Galilean components
under boosts; see Eq.~(\ref{eqn:appa:boost}).
Since for the Galilean components specifically,
$\varLambda^{A}_{\phantom{A}C}(\bm{P}) = \bar{\varLambda}^{A}_{\phantom{A}C}$
(compare Eqs.~(\ref{eqn:boost}) and (\ref{eqn:appa:boost})),
it follows that
$
  \mathfrak{p}^A_{\phantom{A}C}
  =
  \bar{\varLambda}^A_{\phantom{A}C}
$
and
$
  \mathfrak{q}^{AB}_{\phantom{AB}CD}
  =
  \bar{\varLambda}^A_{\phantom{A}C}
  \bar{\varLambda}^B_{\phantom{B}D}
$.
Additionally, since we have seen that the Galilean components
of the physical densities are independent of the non-Galilean
components of the intrinsic densities,
it follows that $\mathfrak{p}^A_{\phantom{A}3} = 0$ and
$\mathfrak{q}^{AB}_{\phantom{AB}\alpha 3} = \mathfrak{q}^{AB}_{\phantom{AB}3\beta} = 0$.

To deduce the remaining components of $\mathfrak{p}$ and $\mathfrak{q}$,
we consider pointlike targets.
Since the smearing functions are universal (i.e., target-independent),
the smearing functions obtained for pointlike targets
are also applicable to hadrons.

%%%%%%%%%%%%%%%%%%%%%%%%%%%%%%%%%%%%%%%%

\subsection{Remaining components of electromagnetic smearing function}

The components
$\mathfrak{p}^3_{\phantom{3}A}$
for the electromagnetic smearing function can be deduced by considering
a point particle with charge $Q$ and magnetic dipole moment
$\frac{\mu}{2m}$ in a definite-helicity state.
The intrinsic Galilean charge and current densities
are known, with the following Fourier transforms:
%\begin{subequations}
%  \begin{align}
%    j^0(\bm{b}_\perp)
%    &=
%    Q \delta^{(2)}(\bm{b}_\perp)
%    =
%    Q
%    \int \frac{\d^2\bm{\varDelta}_\perp}{(2\pi)^2}
%    \e^{-\i\bm{\varDelta}_\perp\cdot\bm{b}_\perp}
%    \\
%    \bm{j}_\perp(\bm{b}_\perp)
%    &=
%    -
%    \frac{\mu\lambda}{2m}
%    \hat{e}_z \times \bm{\nabla}
%    \delta^{(2)}(\bm{b}_\perp)
%    =
%    \frac{\mu\lambda}{2m}
%    \int \frac{\d^2\bm{\varDelta}_\perp}{(2\pi)^2}
%    \hat{e}_z \times \i\bm{\varDelta}_\perp
%    \e^{-\i\bm{\varDelta}_\perp\cdot\bm{b}_\perp}
%    \,.
%  \end{align}
%\end{subequations}
\begin{subequations}
  \begin{align}
    j^0(\bm{\varDelta}_\perp)
    &=
    Q
    \\
    \bm{j}_\perp(\bm{\varDelta}_\perp)
    &=
    \frac{\mu\lambda}{2m}
    (\hat{e}_z \times \i\bm{\varDelta}_\perp)
    \,.
  \end{align}
\end{subequations}
The matrix element appearing in the
physical densities are also known,
and using identities in Appendix A of Ref.~\cite{Freese:2023jcp}
gives:
\begin{align}
  \frac{
    \langle \bm{p}',\lambda' | \hat{J}^\mu(0) | \bm{p},\lambda \rangle
  }{2m}
  &=
  %\frac{m}{P_z}
  %\left\{
    Q
    \frac{P^\mu}{m}
    +
    \frac{\mu\lambda}{2m}
    (\hat{e}_z \times \i\bm{\varDelta}_\perp)^\mu
    -
    \frac{\mu\lambda}{2m}
    (\hat{e}_z \times \i\bm{\varDelta}_\perp)
    \cdot
    \frac{\bm{P}}{P_z}
    \delta^\mu_3
    %\right\}
  \,.
\end{align}
From this and Eq.~(\ref{eqn:boost}),
a little algebra can be used to show:
\begin{align}
  \frac{
    \langle \bm{p}',\lambda' | \hat{J}^\mu(0) | \bm{p},\lambda \rangle
  }{2m}
  =
  \bar{\varLambda}^\mu_{\phantom{\mu}\nu}
  j^\mu(\bm{\varDelta}_\perp)
  \,.
\end{align}
%\begin{subequations}
%  \begin{align}
%    Q
%    \frac{P^\mu}{m}
%    &=
%    \bar{\varLambda}^\mu_{\phantom{\mu}0}
%    \\
%    (\hat{e}_z \times \i\bm{\varDelta}_\perp)^\mu
%    -
%    (\hat{e}_z \times \i\bm{\varDelta}_\perp)
%    \cdot
%    \frac{\bm{P}}{P_z}
%    \delta^\mu_3
%    &=
%    \bar{\varLambda}^\mu_{\phantom{\mu}\nu}
%    (\hat{e}_z \times \i\bm{\varDelta}_\perp)^\nu
%    \,,
%  \end{align}
%\end{subequations}
%Eq.~(\ref{eqn:smear:current}) is now effectively proved
%for all components except $\mathscr{P}^3_{\phantom{3}3}$.
Comparing to Eq.~(\ref{eqn:smear:reduced}),
this shows that
$
  \mathfrak{p}^\mu_{\phantom{\mu}\nu}
  =
  \bar{\varLambda}^\mu_{\phantom{\mu}\nu}
$
as long as $\mu\neq3$ or $\nu\neq3$;
$\mathfrak{p}^3_{\phantom{3}3}$ remains undetermined
because $j^3(\bm{\varDelta}_\perp)=0$
in the case we just considered.

This last component can be obtained by considering the
point charge to be polarized in the transverse plane.
In this case, the charge density may (and in fact does)
contain synchronization artifacts due to the presence
of a $z$-direction current.
However, the intrinsic current density of a point magnetic dipole
is stationary and thus unaltered by synchronization effects.
Accordingly, the intrinsic $z$-direction current
has the Fourier transform:
\begin{align}
  \bm{j}(\bm{\varDelta}_\perp)
  &=
  \frac{\mu}{4m}
  \hat{\bm{s}}_\perp \times \i\bm{\varDelta}_\perp
  \,.
\end{align}
Using identities from Appendix A of Ref.~\cite{Freese:2023jcp},
the matrix element appearing in the physical four-current for this state is:
\begin{align}
  \frac{
    \langle \bm{p}',\lambda' | \hat{J}^\mu(0) | \bm{p},\lambda \rangle
  }{2m}
  =
  %\frac{m}{P_z}
  %\left\{
    \frac{P^\mu}{m}
    \left(
    Q
    +
    \frac{\mu}{2m}
    (\hat{\bm{s}}_\perp \times \i \bm{\varDelta}_\perp)\cdot\hat{e}_z
    \right)
    +
    \frac{\mu}{4P_z}
    (\bm{\hat{s}}_\perp \times \i\bm{\varDelta}_\perp)^\mu
    %\right\}
  \,.
\end{align}
One can read off the terms multiplying $\frac{P^\mu}{m}$
as coming from the intrinsic charge density.
The remaining term is related to the intrinsic current density via:
\begin{align}
  \frac{\mu}{4P_z}
  (\bm{\hat{s}}_\perp \times \i\bm{\varDelta}_\perp)^3
  =
  \frac{m}{P_z}
  \frac{\mu}{4m}
  (\hat{\bm{s}}_\perp \times \i\bm{\varDelta}_\perp)^3
  =
  \bar{\varLambda}^3_{\phantom{3}3}
  j^3(\bm{\varDelta}_\perp)
  \,,
\end{align}
meaning
$
  \mathfrak{p}^3_{\phantom{3}3}
  =
  \bar{\varLambda}^3_{\phantom{3}3}
$.
Thus
$
  \mathfrak{p}^\mu_{\phantom{\mu}\nu}
  =
  \bar{\varLambda}^\mu_{\phantom{\mu}\nu}
$
holds for all components,
from which Eq.~(\ref{eqn:smear:current}) follows.

%%%%%%%%%%%%%%%%%%%%%%%%%%%%%%%%%%%%%%%%

\subsection{Remaining components of EMT smearing function}

The remaining components of the EMT smearing function
can be derived in a similar manner to the electromagnetic smearing functions,
namely by considering pointlike particles.
The universality of the smearing functions again means
that the results are also applicable to hadrons.

As a first example, we consider a point mass without spin.
The intrinsic EMT density of this system has a Fourier transform:
\begin{align}
  t^{00}(\bm{\varDelta}_\perp)
  =
  m
  \,.
\end{align}
The matrix element appearing in
the physical EMT density of the point mass
(with zero D-term and no spin) is:
\begin{align}
  \frac{
    \langle \bm{p}'| \hat{T}^{\mu\nu}(0) |\bm{p}\rangle
  }{2m}
  &=
  m
  \frac{P^\mu P^\nu}{m^2}
  =
  \bar{\varLambda}^\mu_{\phantom{\mu}0}
  \bar{\varLambda}^\nu_{\phantom{\nu}0}
  t^{00}(\bm{\varDelta}_\perp)
  \,.
\end{align}
Thus,
$
  \mathfrak{q}^{\mu\nu}_{\phantom{\mu\nu}00}
  =
  \bar{\varLambda}^\mu_{\phantom{\mu}0}
  \bar{\varLambda}^\nu_{\phantom{\nu}0}
$.

Next, we consider a point particle with spin oriented along the $z$ axis.
In addition to the intrinsic $t^{00}$ density noted above,
there are additional momentum densities, with Fourier transforms:
\begin{align}
  t^{0i}(\bm{\varDelta}_\perp)
  =
  t^{i0}(\bm{\varDelta}_\perp)
  =
  \frac{\lambda}{2}
  (\hat{e}_z \times \i\bm{\varDelta}_\perp)^i
\end{align}
for $i=1,2$,
which reproduces a point distribution of total angular momentum
$\lambda = \pm\frac{1}{2}$ at the origin.
The matrix element appearing in
the physical EMT density of this point particle is given by:
\begin{align}
  \frac{
    \langle \bm{p}'| \hat{T}^{\mu\nu}(0) |\bm{p}\rangle
  }{2m}
  =
  m
  \left\{
    \frac{P^\mu P^\nu}{m^2}
    +
    \frac{\lambda}{2}
    \frac{P^{\{\mu} (\hat{e}_z \times \i\bm{\varDelta}_\perp)^{\nu\}} }{m}
    -
    \frac{\lambda}{2}
    (\hat{e}_z \times \i\bm{\varDelta}_\perp)
    \cdot
    \frac{\bm{P}}{P_z}
    \frac{P_{\phantom{3}}^{\{\mu} \delta^{\nu\}}_3 }{m}
    \right\}
  \,.
\end{align}
We already know $\frac{P^\mu P^\nu}{m^2}$ structure is the contribution
from the intrinsic $t^{00}$ density.
The new structure can be confirmed component-by-component to be equal to:
\begin{align}
  \frac{P^{\{\mu} (\hat{e}_z \times \i\bm{\varDelta}_\perp)^{\nu\}} }{m}
  -
  (\hat{e}_z \times \i\bm{\varDelta}_\perp)
  \cdot
  \frac{\bm{P}}{P_z}
  \frac{P_{\phantom{3}}^{\{\mu} \delta^{\nu\}}_3 }{m}
  =
  \bar{\varLambda}^\mu_{\phantom{\mu}\alpha}
  \bar{\varLambda}^\nu_{\phantom{\nu}\beta}
  \bar{n}^{\{\alpha}
  (\hat{e}_z \times \i\bm{\varDelta}_\perp)^{\beta\}}
  \,,
\end{align}
meaning that
$
  \mathfrak{q}^{\mu\nu}_{\phantom{\mu\nu}CD}
  =
  \bar{\varLambda}^\mu_{\phantom{\mu}C}
  \bar{\varLambda}^\nu_{\phantom{\nu}D}
$,
where $C,D\in\{0,1,2\}$.
Components of $\mathfrak{q}$ with 3 in the latter two indices
remain to be determined.

Just as with the electromagnetic case, we next consider a point mass with
spin oriented in the transverse plane.
The $t^{00}$ density can (and will) obtain angular modulations,
but the $t^{03}$ and $t^{30}$ densities---like the $j^3$ density in the
electromagnetic case---should be related to the longitudinally-polarized
$t^{0i}$ densities by rotation, since the angular momentum is stationary.
These Fourier transforms of densities are thus:
\begin{align}
  t^{03}(\bm{\varDelta}_\perp)
  =
  t^{30}(\bm{\varDelta}_\perp)
  =
  \frac{\lambda}{2}
  (\hat{\bm{s}} \times \i\bm{\varDelta}_\perp)\cdot\hat{e}_z
  \,.
\end{align}
The matrix element appearing in
the corresponding physical EMT is:
\begin{align}
  \frac{
    \langle \bm{p}'| \hat{T}^{\mu\nu}(0) |\bm{p}\rangle
  }{2m}
  =
  m
  \left\{
    \frac{P^\mu P^\nu}{m^2}
    \left(
    1
    +
    \frac{(\hat{\bm{s}}_\perp\times\i\bm{\varDelta}_\perp)\cdot\hat{e}_z}{2m}
    \right)
    +
    \frac{\lambda}{2}
    \frac{P^{\{\mu} (\hat{\bm{s}}_\perp \times \i\bm{\varDelta}_\perp)^{\nu\}} }{P_z}
    \right\}
  \,.
\end{align}
The $\frac{P^\mu P^\nu}{m^2}$ piece (which indeed has angular modulations) can
be read off as giving the intrinsic $t^{00}$ density.
For the remaining piece, one can confirm component-by-component that:
\begin{align}
  \frac{P^{\{\mu} (\hat{\bm{s}}_\perp \times \i\bm{\varDelta}_\perp)^{\nu\}} }{P_z}
  =
  \bar{\varLambda}^\mu_{\phantom{\mu}\alpha}
  \bar{\varLambda}^\nu_{\phantom{\nu}\beta}
  \bar{n}^{\{\alpha}
  (\hat{\bm{s}} \times \i\bm{\varDelta}_\perp)^{\beta\}}
  \,,
\end{align}
from which it follows that
$
  \mathfrak{q}^{\mu\nu}_{\phantom{\mu\nu}C3}
  =
  \bar{\varLambda}^\mu_{\phantom{\mu}C}
  \bar{\varLambda}^\nu_{\phantom{\nu}3}
$
and
$
  \mathfrak{q}^{\mu\nu}_{\phantom{\mu\nu}3D}
  =
  \bar{\varLambda}^\mu_{\phantom{\mu}3}
  \bar{\varLambda}^\nu_{\phantom{\nu}D}
$.
Only
$\mathfrak{q}^{33}_{\phantom{33}33}$
remains undetermined.

The final component of the EMT smearing function is obtained by
the requirement that different irreducible representations
of the Lorentz group do not mix under smearing.
The EMT operator can be decomposed into a pure trace part
in the $(0,0)$ representation and a traceless part in the
$(1,1)$ representation~\cite{Ji:1995sv},
and in principle the formalism considered in this work
should be applicable to these parts separately.

A necessary condition for this requirement to be observed
is that smearing maps the metric into itself:
\begin{align}
  \mathfrak{q}^{\mu\nu}_{\phantom{\mu\nu}\alpha\beta}
  g^{\alpha\beta}
  =
  g^{\mu\nu}
  \,.
\end{align}
Using Eq.~(\ref{eqn:boost}),
and the equation for the tilted coordinate metric
Eq.~(\ref{eqn:appa:metric}), we find:
\begin{align}
  \label{eqn:notboost}
  \bar{\varLambda}^{\mu}_{\phantom{\mu}\alpha}
  \bar{\varLambda}^{\nu}_{\phantom{\nu}\beta}
  g^{\alpha\beta}
  =
  g^{\mu\nu}
  +
  \delta^{\mu}_3
  \delta^{\nu}_3
  \frac{\bm{\varDelta}_\perp^2}{4P_z^2}
  \,.
\end{align}
(This demonstrates that $\bar{\varLambda}$ is not a Lorentz boost,
since Lorentz boosts leave the metric invariant.)
Thus we observe that
$
  \mathfrak{q}^{\mu\nu}_{\phantom{\mu\nu}\alpha\beta}
  \neq
  \bar{\varLambda}^\mu_{\phantom{\mu}\alpha}
  \bar{\varLambda}^\nu_{\phantom{\nu}\beta}
$
as has apparently been observed for all the other components,
but that there must be a departure from this
in order to preserve the metric.
The metric is preserved if
$
  \mathfrak{q}^{33}_{\phantom{33}33}
  =
  \left(
  \bar{\varLambda}^3_{\phantom{3}3}
  \bar{\varLambda}^3_{\phantom{3}3}
  -
  \frac{\bm{\varDelta}_\perp^2}{4P_z^2}
  \right)
$,
which means that
${
  \mathfrak{q}^{\mu\nu}_{\phantom{\mu\nu}\alpha\beta}
  =
  \left(
  \bar{\varLambda}^\mu_{\phantom{\mu}\alpha}
  \bar{\varLambda}^\nu_{\phantom{\nu}\beta}
  -
  \frac{\bm{\varDelta}_\perp^2}{4P_z^2}
  \delta^\mu_3
  \delta^\nu_3
  \delta^3_\alpha
  \delta^3_\beta
  \right)
}$.
With this rule, the metric is preserved,
and Eq.~(\ref{eqn:smear:emt}) follows.

%%%%%%%%%%%%%%%%%%%%%%%%%%%%%%%%%%%%%%%%

\subsection{Inversions of convolution formulas}

Since we know that the smearing functions are given by
Eqs.~(\ref{eqn:smear:current}) and Eq.~(\ref{eqn:smear:emt}),
the reduced convolution relations of Eq.~(\ref{eqn:smear:reduced})
can be written:
\begin{subequations}
  \begin{align}
    \frac{
      \langle \bm{p}',\lambda' | \hat{J}^\mu(0) | \bm{p},\lambda \rangle
    }{2m}
    &=
    \bar{\varLambda}^\mu_{\phantom{\mu}\nu}
    j^\nu(\bm{\varDelta}_\perp;\lambda,\lambda')
    \\
    \frac{
      \langle \bm{p}',\lambda' | \hat{T}^{\mu\nu}(0) | \bm{p},\lambda \rangle
    }{2m}
    &=
    \Big(
    \bar{\varLambda}^\mu_{\phantom{\mu}\alpha}
    \bar{\varLambda}^\nu_{\phantom{\nu}\beta}
    -
    \frac{\bm{\varDelta}_\perp^2}{4P_z^2}
    \delta^\mu_3
    \delta^\nu_3
    \delta^3_\alpha
    \delta^3_\beta
    \Big)
    t^{\alpha\beta}(\bm{\varDelta}_\perp;\lambda,\lambda')
    \,.
  \end{align}
\end{subequations}
Since $\bar{\varLambda}$ is an invertible matrix (its determinant being 1),
these equations can be inverted.
For the current, this immediately gives Eq.~(\ref{eqn:current}).
For the EMT, we have an additional step; matrix inversion gives:
\begin{align}
  t^{\mu\nu}(\bm{\varDelta}_\perp;\lambda,\lambda')
  -
  \frac{\bm{\varDelta}_\perp^2}{4m^2}
  \delta^\mu_3
  \delta^\nu_3
  t^{33}(\bm{\varDelta}_\perp;\lambda,\lambda')
  =
  \bar{\varLambda}^\mu_{\phantom{\mu}\alpha}
  \bar{\varLambda}^\nu_{\phantom{\nu}\beta}
  \frac{
    \langle \bm{p}',\lambda' | \hat{T}^{\alpha\beta}(0) | \bm{p},\lambda \rangle
  }{2m}
  \,.
\end{align}
For $\mu\neq3$ or $\nu\neq3$ the left-hand is already the desired quantity
$t^{\mu\nu}(\bm{\varDelta}_\perp;\lambda,\lambda')$,
while for $\mu=\nu=3$ the equation needs to be divided by
$\left(1 - \frac{\bm{\varDelta}_\perp^2}{4m^2}\right)$
to obtain the desired quantity.
Form this, Eq.~(\ref{eqn:emt}) follows.
This completes the proof.

\bibliography{references.bib}

\begin{thebibliography}{83}%
\makeatletter
\providecommand \@ifxundefined [1]{%
 \@ifx{#1\undefined}
}%
\providecommand \@ifnum [1]{%
 \ifnum #1\expandafter \@firstoftwo
 \else \expandafter \@secondoftwo
 \fi
}%
\providecommand \@ifx [1]{%
 \ifx #1\expandafter \@firstoftwo
 \else \expandafter \@secondoftwo
 \fi
}%
\providecommand \natexlab [1]{#1}%
\providecommand \enquote  [1]{``#1''}%
\providecommand \bibnamefont  [1]{#1}%
\providecommand \bibfnamefont [1]{#1}%
\providecommand \citenamefont [1]{#1}%
\providecommand \href@noop [0]{\@secondoftwo}%
\providecommand \href [0]{\begingroup \@sanitize@url \@href}%
\providecommand \@href[1]{\@@startlink{#1}\@@href}%
\providecommand \@@href[1]{\endgroup#1\@@endlink}%
\providecommand \@sanitize@url [0]{\catcode `\\12\catcode `\$12\catcode
  `\&12\catcode `\#12\catcode `\^12\catcode `\_12\catcode `\%12\relax}%
\providecommand \@@startlink[1]{}%
\providecommand \@@endlink[0]{}%
\providecommand \url  [0]{\begingroup\@sanitize@url \@url }%
\providecommand \@url [1]{\endgroup\@href {#1}{\urlprefix }}%
\providecommand \urlprefix  [0]{URL }%
\providecommand \Eprint [0]{\href }%
\providecommand \doibase [0]{http://dx.doi.org/}%
\providecommand \selectlanguage [0]{\@gobble}%
\providecommand \bibinfo  [0]{\@secondoftwo}%
\providecommand \bibfield  [0]{\@secondoftwo}%
\providecommand \translation [1]{[#1]}%
\providecommand \BibitemOpen [0]{}%
\providecommand \bibitemStop [0]{}%
\providecommand \bibitemNoStop [0]{.\EOS\space}%
\providecommand \EOS [0]{\spacefactor3000\relax}%
\providecommand \BibitemShut  [1]{\csname bibitem#1\endcsname}%
\let\auto@bib@innerbib\@empty
%</preamble>
\bibitem [{\citenamefont {Kobzarev}\ and\ \citenamefont
  {Okun}(1962)}]{Kobzarev:1962wt}%
  \BibitemOpen
  \bibfield  {author} {\bibinfo {author} {\bibfnamefont {I.~Yu.}\ \bibnamefont
  {Kobzarev}}\ and\ \bibinfo {author} {\bibfnamefont {L.~B.}\ \bibnamefont
  {Okun}},\ }\bibfield  {title} {\enquote {\bibinfo {title} {{Gravitational
  interaction of fermions}},}\ }\href@noop {} {\bibfield  {journal} {\bibinfo
  {journal} {Zh. Eksp. Teor. Fiz.}\ }\textbf {\bibinfo {volume} {43}},\
  \bibinfo {pages} {1904--1909} (\bibinfo {year} {1962})}\BibitemShut {NoStop}%
\bibitem [{\citenamefont {Ji}(1995{\natexlab{a}})}]{Ji:1994av}%
  \BibitemOpen
  \bibfield  {author} {\bibinfo {author} {\bibfnamefont {Xiang-Dong}\
  \bibnamefont {Ji}},\ }\bibfield  {title} {\enquote {\bibinfo {title} {{A QCD
  analysis of the mass structure of the nucleon}},}\ }\href {\doibase
  10.1103/PhysRevLett.74.1071} {\bibfield  {journal} {\bibinfo  {journal}
  {Phys. Rev. Lett.}\ }\textbf {\bibinfo {volume} {74}},\ \bibinfo {pages}
  {1071--1074} (\bibinfo {year} {1995}{\natexlab{a}})},\ \Eprint
  {http://arxiv.org/abs/hep-ph/9410274} {arXiv:hep-ph/9410274} \BibitemShut
  {NoStop}%
\bibitem [{\citenamefont {Ji}(1995{\natexlab{b}})}]{Ji:1995sv}%
  \BibitemOpen
  \bibfield  {author} {\bibinfo {author} {\bibfnamefont {Xiang-Dong}\
  \bibnamefont {Ji}},\ }\bibfield  {title} {\enquote {\bibinfo {title}
  {{Breakup of hadron masses and energy - momentum tensor of QCD}},}\ }\href
  {\doibase 10.1103/PhysRevD.52.271} {\bibfield  {journal} {\bibinfo  {journal}
  {Phys. Rev. D}\ }\textbf {\bibinfo {volume} {52}},\ \bibinfo {pages}
  {271--281} (\bibinfo {year} {1995}{\natexlab{b}})},\ \Eprint
  {http://arxiv.org/abs/hep-ph/9502213} {arXiv:hep-ph/9502213} \BibitemShut
  {NoStop}%
\bibitem [{\citenamefont {Lorc\'e}(2018{\natexlab{a}})}]{Lorce:2017xzd}%
  \BibitemOpen
  \bibfield  {author} {\bibinfo {author} {\bibfnamefont {C\'edric}\
  \bibnamefont {Lorc\'e}},\ }\bibfield  {title} {\enquote {\bibinfo {title}
  {{On the hadron mass decomposition}},}\ }\href {\doibase
  10.1140/epjc/s10052-018-5561-2} {\bibfield  {journal} {\bibinfo  {journal}
  {Eur. Phys. J. C}\ }\textbf {\bibinfo {volume} {78}},\ \bibinfo {pages} {120}
  (\bibinfo {year} {2018}{\natexlab{a}})},\ \Eprint
  {http://arxiv.org/abs/1706.05853} {arXiv:1706.05853 [hep-ph]} \BibitemShut
  {NoStop}%
\bibitem [{\citenamefont {Hatta}\ \emph {et~al.}(2018)\citenamefont {Hatta},
  \citenamefont {Rajan},\ and\ \citenamefont {Tanaka}}]{Hatta:2018sqd}%
  \BibitemOpen
  \bibfield  {author} {\bibinfo {author} {\bibfnamefont {Yoshitaka}\
  \bibnamefont {Hatta}}, \bibinfo {author} {\bibfnamefont {Abha}\ \bibnamefont
  {Rajan}}, \ and\ \bibinfo {author} {\bibfnamefont {Kazuhiro}\ \bibnamefont
  {Tanaka}},\ }\bibfield  {title} {\enquote {\bibinfo {title} {{Quark and gluon
  contributions to the QCD trace anomaly}},}\ }\href {\doibase
  10.1007/JHEP12(2018)008} {\bibfield  {journal} {\bibinfo  {journal} {JHEP}\
  }\textbf {\bibinfo {volume} {12}},\ \bibinfo {pages} {008} (\bibinfo {year}
  {2018})},\ \Eprint {http://arxiv.org/abs/1810.05116} {arXiv:1810.05116
  [hep-ph]} \BibitemShut {NoStop}%
\bibitem [{\citenamefont {Metz}\ \emph {et~al.}(2021)\citenamefont {Metz},
  \citenamefont {Pasquini},\ and\ \citenamefont {Rodini}}]{Metz:2020vxd}%
  \BibitemOpen
  \bibfield  {author} {\bibinfo {author} {\bibfnamefont {Andreas}\ \bibnamefont
  {Metz}}, \bibinfo {author} {\bibfnamefont {Barbara}\ \bibnamefont
  {Pasquini}}, \ and\ \bibinfo {author} {\bibfnamefont {Simone}\ \bibnamefont
  {Rodini}},\ }\bibfield  {title} {\enquote {\bibinfo {title} {{Revisiting the
  proton mass decomposition}},}\ }\href {\doibase 10.1103/PhysRevD.102.114042}
  {\bibfield  {journal} {\bibinfo  {journal} {Phys. Rev. D}\ }\textbf {\bibinfo
  {volume} {102}},\ \bibinfo {pages} {114042} (\bibinfo {year} {2021})},\
  \Eprint {http://arxiv.org/abs/2006.11171} {arXiv:2006.11171 [hep-ph]}
  \BibitemShut {NoStop}%
\bibitem [{\citenamefont {Ji}(2021)}]{Ji:2021mtz}%
  \BibitemOpen
  \bibfield  {author} {\bibinfo {author} {\bibfnamefont {Xiangdong}\
  \bibnamefont {Ji}},\ }\bibfield  {title} {\enquote {\bibinfo {title} {{Proton
  mass decomposition: naturalness and interpretations}},}\ }\href {\doibase
  10.1007/s11467-021-1065-x} {\bibfield  {journal} {\bibinfo  {journal} {Front.
  Phys. (Beijing)}\ }\textbf {\bibinfo {volume} {16}},\ \bibinfo {pages}
  {64601} (\bibinfo {year} {2021})},\ \Eprint {http://arxiv.org/abs/2102.07830}
  {arXiv:2102.07830 [hep-ph]} \BibitemShut {NoStop}%
\bibitem [{\citenamefont {Lorc\'e}\ \emph {et~al.}(2021)\citenamefont
  {Lorc\'e}, \citenamefont {Metz}, \citenamefont {Pasquini},\ and\
  \citenamefont {Rodini}}]{Lorce:2021xku}%
  \BibitemOpen
  \bibfield  {author} {\bibinfo {author} {\bibfnamefont {C\'edric}\
  \bibnamefont {Lorc\'e}}, \bibinfo {author} {\bibfnamefont {Andreas}\
  \bibnamefont {Metz}}, \bibinfo {author} {\bibfnamefont {Barbara}\
  \bibnamefont {Pasquini}}, \ and\ \bibinfo {author} {\bibfnamefont {Simone}\
  \bibnamefont {Rodini}},\ }\bibfield  {title} {\enquote {\bibinfo {title}
  {{Energy-momentum tensor in QCD: nucleon mass decomposition and mechanical
  equilibrium}},}\ }\href {\doibase 10.1007/JHEP11(2021)121} {\bibfield
  {journal} {\bibinfo  {journal} {JHEP}\ }\textbf {\bibinfo {volume} {11}},\
  \bibinfo {pages} {121} (\bibinfo {year} {2021})},\ \Eprint
  {http://arxiv.org/abs/2109.11785} {arXiv:2109.11785 [hep-ph]} \BibitemShut
  {NoStop}%
\bibitem [{\citenamefont {Ashman}\ \emph {et~al.}(1988)\citenamefont {Ashman}
  \emph {et~al.}}]{Ashman:1987hv}%
  \BibitemOpen
  \bibfield  {author} {\bibinfo {author} {\bibfnamefont {J.}~\bibnamefont
  {Ashman}} \emph {et~al.} (\bibinfo {collaboration} {European Muon}),\
  }\bibfield  {title} {\enquote {\bibinfo {title} {{A Measurement of the Spin
  Asymmetry and Determination of the Structure Function g(1) in Deep Inelastic
  Muon-Proton Scattering}},}\ }\href {\doibase 10.1016/0370-2693(88)91523-7}
  {\bibfield  {journal} {\bibinfo  {journal} {Phys. Lett. B}\ }\textbf
  {\bibinfo {volume} {206}},\ \bibinfo {pages} {364} (\bibinfo {year}
  {1988})}\BibitemShut {NoStop}%
\bibitem [{\citenamefont {Ji}(1997{\natexlab{a}})}]{Ji:1996ek}%
  \BibitemOpen
  \bibfield  {author} {\bibinfo {author} {\bibfnamefont {Xiang-Dong}\
  \bibnamefont {Ji}},\ }\bibfield  {title} {\enquote {\bibinfo {title}
  {{Gauge-Invariant Decomposition of Nucleon Spin}},}\ }\href {\doibase
  10.1103/PhysRevLett.78.610} {\bibfield  {journal} {\bibinfo  {journal} {Phys.
  Rev. Lett.}\ }\textbf {\bibinfo {volume} {78}},\ \bibinfo {pages} {610--613}
  (\bibinfo {year} {1997}{\natexlab{a}})},\ \Eprint
  {http://arxiv.org/abs/hep-ph/9603249} {arXiv:hep-ph/9603249} \BibitemShut
  {NoStop}%
\bibitem [{\citenamefont {Leader}\ and\ \citenamefont
  {Lorc\'e}(2014)}]{Leader:2013jra}%
  \BibitemOpen
  \bibfield  {author} {\bibinfo {author} {\bibfnamefont {E.}~\bibnamefont
  {Leader}}\ and\ \bibinfo {author} {\bibfnamefont {C.}~\bibnamefont
  {Lorc\'e}},\ }\bibfield  {title} {\enquote {\bibinfo {title} {{The angular
  momentum controversy: What\textquoteright{}s it all about and does it
  matter?}}}\ }\href {\doibase 10.1016/j.physrep.2014.02.010} {\bibfield
  {journal} {\bibinfo  {journal} {Phys. Rept.}\ }\textbf {\bibinfo {volume}
  {541}},\ \bibinfo {pages} {163--248} (\bibinfo {year} {2014})},\ \Eprint
  {http://arxiv.org/abs/1309.4235} {arXiv:1309.4235 [hep-ph]} \BibitemShut
  {NoStop}%
\bibitem [{\citenamefont {Wakamatsu}(2014)}]{Wakamatsu:2014zza}%
  \BibitemOpen
  \bibfield  {author} {\bibinfo {author} {\bibfnamefont {Masashi}\ \bibnamefont
  {Wakamatsu}},\ }\bibfield  {title} {\enquote {\bibinfo {title} {{Is
  gauge-invariant complete decomposition of the nucleon spin possible?}}}\
  }\href {\doibase 10.1142/S0217751X14300129} {\bibfield  {journal} {\bibinfo
  {journal} {Int. J. Mod. Phys. A}\ }\textbf {\bibinfo {volume} {29}},\
  \bibinfo {pages} {1430012} (\bibinfo {year} {2014})},\ \Eprint
  {http://arxiv.org/abs/1402.4193} {arXiv:1402.4193 [hep-ph]} \BibitemShut
  {NoStop}%
\bibitem [{\citenamefont {Ji}\ \emph {et~al.}(2021)\citenamefont {Ji},
  \citenamefont {Yuan},\ and\ \citenamefont {Zhao}}]{Ji:2020ena}%
  \BibitemOpen
  \bibfield  {author} {\bibinfo {author} {\bibfnamefont {Xiangdong}\
  \bibnamefont {Ji}}, \bibinfo {author} {\bibfnamefont {Feng}\ \bibnamefont
  {Yuan}}, \ and\ \bibinfo {author} {\bibfnamefont {Yong}\ \bibnamefont
  {Zhao}},\ }\bibfield  {title} {\enquote {\bibinfo {title} {{What we know and
  what we don\textquoteright{}t know about the proton spin after 30 years}},}\
  }\href {\doibase 10.1038/s42254-020-00248-4} {\bibfield  {journal} {\bibinfo
  {journal} {Nature Rev. Phys.}\ }\textbf {\bibinfo {volume} {3}},\ \bibinfo
  {pages} {27--38} (\bibinfo {year} {2021})},\ \Eprint
  {http://arxiv.org/abs/2009.01291} {arXiv:2009.01291 [hep-ph]} \BibitemShut
  {NoStop}%
\bibitem [{\citenamefont {Polyakov}(2003)}]{Polyakov:2002yz}%
  \BibitemOpen
  \bibfield  {author} {\bibinfo {author} {\bibfnamefont {M.~V.}\ \bibnamefont
  {Polyakov}},\ }\bibfield  {title} {\enquote {\bibinfo {title} {{Generalized
  parton distributions and strong forces inside nucleons and nuclei}},}\ }\href
  {\doibase 10.1016/S0370-2693(03)00036-4} {\bibfield  {journal} {\bibinfo
  {journal} {Phys. Lett. B}\ }\textbf {\bibinfo {volume} {555}},\ \bibinfo
  {pages} {57--62} (\bibinfo {year} {2003})},\ \Eprint
  {http://arxiv.org/abs/hep-ph/0210165} {arXiv:hep-ph/0210165} \BibitemShut
  {NoStop}%
\bibitem [{\citenamefont {Polyakov}\ and\ \citenamefont
  {Schweitzer}(2018)}]{Polyakov:2018zvc}%
  \BibitemOpen
  \bibfield  {author} {\bibinfo {author} {\bibfnamefont {Maxim~V.}\
  \bibnamefont {Polyakov}}\ and\ \bibinfo {author} {\bibfnamefont {Peter}\
  \bibnamefont {Schweitzer}},\ }\bibfield  {title} {\enquote {\bibinfo {title}
  {{Forces inside hadrons: pressure, surface tension, mechanical radius, and
  all that}},}\ }\href {\doibase 10.1142/S0217751X18300259} {\bibfield
  {journal} {\bibinfo  {journal} {Int. J. Mod. Phys. A}\ }\textbf {\bibinfo
  {volume} {33}},\ \bibinfo {pages} {1830025} (\bibinfo {year} {2018})},\
  \Eprint {http://arxiv.org/abs/1805.06596} {arXiv:1805.06596 [hep-ph]}
  \BibitemShut {NoStop}%
\bibitem [{\citenamefont {Lorc\'e}\ \emph {et~al.}(2019)\citenamefont
  {Lorc\'e}, \citenamefont {Moutarde},\ and\ \citenamefont
  {Trawi\'nski}}]{Lorce:2018egm}%
  \BibitemOpen
  \bibfield  {author} {\bibinfo {author} {\bibfnamefont {C\'edric}\
  \bibnamefont {Lorc\'e}}, \bibinfo {author} {\bibfnamefont {Herv\'e}\
  \bibnamefont {Moutarde}}, \ and\ \bibinfo {author} {\bibfnamefont
  {Arkadiusz~P.}\ \bibnamefont {Trawi\'nski}},\ }\bibfield  {title} {\enquote
  {\bibinfo {title} {{Revisiting the mechanical properties of the nucleon}},}\
  }\href {\doibase 10.1140/epjc/s10052-019-6572-3} {\bibfield  {journal}
  {\bibinfo  {journal} {Eur. Phys. J. C}\ }\textbf {\bibinfo {volume} {79}},\
  \bibinfo {pages} {89} (\bibinfo {year} {2019})},\ \Eprint
  {http://arxiv.org/abs/1810.09837} {arXiv:1810.09837 [hep-ph]} \BibitemShut
  {NoStop}%
\bibitem [{\citenamefont {Freese}\ and\ \citenamefont
  {Miller}(2021{\natexlab{a}})}]{Freese:2021czn}%
  \BibitemOpen
  \bibfield  {author} {\bibinfo {author} {\bibfnamefont {Adam}\ \bibnamefont
  {Freese}}\ and\ \bibinfo {author} {\bibfnamefont {Gerald~A.}\ \bibnamefont
  {Miller}},\ }\bibfield  {title} {\enquote {\bibinfo {title} {{Forces within
  hadrons on the light front}},}\ }\href {\doibase 10.1103/PhysRevD.103.094023}
  {\bibfield  {journal} {\bibinfo  {journal} {Phys. Rev. D}\ }\textbf {\bibinfo
  {volume} {103}},\ \bibinfo {pages} {094023} (\bibinfo {year}
  {2021}{\natexlab{a}})},\ \Eprint {http://arxiv.org/abs/2102.01683}
  {arXiv:2102.01683 [hep-ph]} \BibitemShut {NoStop}%
\bibitem [{\citenamefont {Ji}\ and\ \citenamefont {Liu}(2022)}]{Ji:2021mfb}%
  \BibitemOpen
  \bibfield  {author} {\bibinfo {author} {\bibfnamefont {Xiangdong}\
  \bibnamefont {Ji}}\ and\ \bibinfo {author} {\bibfnamefont {Yizhuang}\
  \bibnamefont {Liu}},\ }\bibfield  {title} {\enquote {\bibinfo {title}
  {{Momentum-Current Gravitational Multipoles of Hadrons}},}\ }\href {\doibase
  10.1103/PhysRevD.106.034028} {\bibfield  {journal} {\bibinfo  {journal}
  {Phys. Rev. D}\ }\textbf {\bibinfo {volume} {106}},\ \bibinfo {pages}
  {034028} (\bibinfo {year} {2022})},\ \Eprint
  {http://arxiv.org/abs/2110.14781} {arXiv:2110.14781 [hep-ph]} \BibitemShut
  {NoStop}%
\bibitem [{\citenamefont {Boer}\ \emph {et~al.}(2011)\citenamefont {Boer} \emph
  {et~al.}}]{Boer:2011fh}%
  \BibitemOpen
  \bibfield  {author} {\bibinfo {author} {\bibfnamefont {Daniel}\ \bibnamefont
  {Boer}} \emph {et~al.},\ }\bibfield  {title} {\enquote {\bibinfo {title}
  {{Gluons and the quark sea at high energies: Distributions, polarization,
  tomography}},}\ }\href@noop {} {\  (\bibinfo {year} {2011})},\ \Eprint
  {http://arxiv.org/abs/1108.1713} {arXiv:1108.1713 [nucl-th]} \BibitemShut
  {NoStop}%
%%CITATION = ARXIV:1108.1713;%%
\bibitem [{\citenamefont {Accardi}\ \emph {et~al.}(2016)\citenamefont {Accardi}
  \emph {et~al.}}]{Accardi:2012qut}%
  \BibitemOpen
  \bibfield  {author} {\bibinfo {author} {\bibfnamefont {A.}~\bibnamefont
  {Accardi}} \emph {et~al.},\ }\bibfield  {title} {\enquote {\bibinfo {title}
  {{Electron Ion Collider: The Next QCD Frontier}: {Understanding the glue that
  binds us all}},}\ }\href {\doibase 10.1140/epja/i2016-16268-9} {\bibfield
  {journal} {\bibinfo  {journal} {Eur. Phys. J. A}\ }\textbf {\bibinfo {volume}
  {52}},\ \bibinfo {pages} {268} (\bibinfo {year} {2016})},\ \Eprint
  {http://arxiv.org/abs/1212.1701} {arXiv:1212.1701 [nucl-ex]} \BibitemShut
  {NoStop}%
\bibitem [{\citenamefont {Abdul~Khalek}\ \emph {et~al.}(2021)\citenamefont
  {Abdul~Khalek} \emph {et~al.}}]{AbdulKhalek:2021gbh}%
  \BibitemOpen
  \bibfield  {author} {\bibinfo {author} {\bibfnamefont {R.}~\bibnamefont
  {Abdul~Khalek}} \emph {et~al.},\ }\bibfield  {title} {\enquote {\bibinfo
  {title} {{Science Requirements and Detector Concepts for the Electron-Ion
  Collider: EIC Yellow Report}},}\ }\href@noop {} {\  (\bibinfo {year}
  {2021})},\ \Eprint {http://arxiv.org/abs/2103.05419} {arXiv:2103.05419
  [physics.ins-det]} \BibitemShut {NoStop}%
\bibitem [{\citenamefont {Ji}(1997{\natexlab{b}})}]{Ji:1996nm}%
  \BibitemOpen
  \bibfield  {author} {\bibinfo {author} {\bibfnamefont {Xiang-Dong}\
  \bibnamefont {Ji}},\ }\bibfield  {title} {\enquote {\bibinfo {title} {{Deeply
  virtual Compton scattering}},}\ }\href {\doibase 10.1103/PhysRevD.55.7114}
  {\bibfield  {journal} {\bibinfo  {journal} {Phys. Rev. D}\ }\textbf {\bibinfo
  {volume} {55}},\ \bibinfo {pages} {7114--7125} (\bibinfo {year}
  {1997}{\natexlab{b}})},\ \Eprint {http://arxiv.org/abs/hep-ph/9609381}
  {arXiv:hep-ph/9609381} \BibitemShut {NoStop}%
\bibitem [{\citenamefont {Radyushkin}(1997)}]{Radyushkin:1997ki}%
  \BibitemOpen
  \bibfield  {author} {\bibinfo {author} {\bibfnamefont {A.~V.}\ \bibnamefont
  {Radyushkin}},\ }\bibfield  {title} {\enquote {\bibinfo {title} {{Nonforward
  parton distributions}},}\ }\href {\doibase 10.1103/PhysRevD.56.5524}
  {\bibfield  {journal} {\bibinfo  {journal} {Phys. Rev. D}\ }\textbf {\bibinfo
  {volume} {56}},\ \bibinfo {pages} {5524--5557} (\bibinfo {year} {1997})},\
  \Eprint {http://arxiv.org/abs/hep-ph/9704207} {arXiv:hep-ph/9704207}
  \BibitemShut {NoStop}%
\bibitem [{\citenamefont {Belitsky}\ and\ \citenamefont
  {Radyushkin}(2005)}]{Belitsky:2005qn}%
  \BibitemOpen
  \bibfield  {author} {\bibinfo {author} {\bibfnamefont {A.~V.}\ \bibnamefont
  {Belitsky}}\ and\ \bibinfo {author} {\bibfnamefont {A.~V.}\ \bibnamefont
  {Radyushkin}},\ }\bibfield  {title} {\enquote {\bibinfo {title} {{Unraveling
  hadron structure with generalized parton distributions}},}\ }\href {\doibase
  10.1016/j.physrep.2005.06.002} {\bibfield  {journal} {\bibinfo  {journal}
  {Phys. Rept.}\ }\textbf {\bibinfo {volume} {418}},\ \bibinfo {pages} {1--387}
  (\bibinfo {year} {2005})},\ \Eprint {http://arxiv.org/abs/hep-ph/0504030}
  {arXiv:hep-ph/0504030} \BibitemShut {NoStop}%
\bibitem [{\citenamefont {Fleming}(1974)}]{Fleming:1974af}%
  \BibitemOpen
  \bibfield  {author} {\bibinfo {author} {\bibfnamefont {Gordon~N.}\
  \bibnamefont {Fleming}},\ }\enquote {\bibinfo {title} {{Charge Distributions
  from Relativistic Form-Factors}},}\ in\ \href {\doibase
  10.1007/978-94-010-2274-3_22} {\emph {\bibinfo {booktitle} {{Physical reality
  and mathematical description}: {Festschrift Jauch (Josef Maria) on his 60th
  birthday}}}},\ \bibinfo {editor} {edited by\ \bibinfo {editor} {\bibfnamefont
  {Charles~P.}\ \bibnamefont {Enz}}\ and\ \bibinfo {editor} {\bibfnamefont
  {Jagdish}\ \bibnamefont {Mehra}}}\ (\bibinfo {year} {1974})\ pp.\ \bibinfo
  {pages} {357--374}\BibitemShut {NoStop}%
\bibitem [{\citenamefont {Burkardt}(2000)}]{Burkardt:2000za}%
  \BibitemOpen
  \bibfield  {author} {\bibinfo {author} {\bibfnamefont {Matthias}\
  \bibnamefont {Burkardt}},\ }\bibfield  {title} {\enquote {\bibinfo {title}
  {{Impact parameter dependent parton distributions and off forward parton
  distributions for zeta ---\ensuremath{>} 0}},}\ }\href {\doibase
  10.1103/PhysRevD.62.071503} {\bibfield  {journal} {\bibinfo  {journal} {Phys.
  Rev. D}\ }\textbf {\bibinfo {volume} {62}},\ \bibinfo {pages} {071503}
  (\bibinfo {year} {2000})},\ \bibinfo {note} {[Erratum: Phys.Rev.D 66, 119903
  (2002)]},\ \Eprint {http://arxiv.org/abs/hep-ph/0005108}
  {arXiv:hep-ph/0005108} \BibitemShut {NoStop}%
\bibitem [{\citenamefont {Miller}(2019)}]{Miller:2018ybm}%
  \BibitemOpen
  \bibfield  {author} {\bibinfo {author} {\bibfnamefont {Gerald~A.}\
  \bibnamefont {Miller}},\ }\bibfield  {title} {\enquote {\bibinfo {title}
  {{Defining the proton radius: A unified treatment}},}\ }\href {\doibase
  10.1103/PhysRevC.99.035202} {\bibfield  {journal} {\bibinfo  {journal} {Phys.
  Rev. C}\ }\textbf {\bibinfo {volume} {99}},\ \bibinfo {pages} {035202}
  (\bibinfo {year} {2019})},\ \Eprint {http://arxiv.org/abs/1812.02714}
  {arXiv:1812.02714 [nucl-th]} \BibitemShut {NoStop}%
\bibitem [{\citenamefont {Epelbaum}\ \emph {et~al.}(2022)\citenamefont
  {Epelbaum}, \citenamefont {Gegelia}, \citenamefont {Lange}, \citenamefont
  {Mei\ss{}ner},\ and\ \citenamefont {Polyakov}}]{Epelbaum:2022fjc}%
  \BibitemOpen
  \bibfield  {author} {\bibinfo {author} {\bibfnamefont {E.}~\bibnamefont
  {Epelbaum}}, \bibinfo {author} {\bibfnamefont {J.}~\bibnamefont {Gegelia}},
  \bibinfo {author} {\bibfnamefont {N.}~\bibnamefont {Lange}}, \bibinfo
  {author} {\bibfnamefont {U.~G.}\ \bibnamefont {Mei\ss{}ner}}, \ and\ \bibinfo
  {author} {\bibfnamefont {M.~V.}\ \bibnamefont {Polyakov}},\ }\bibfield
  {title} {\enquote {\bibinfo {title} {{Definition of Local Spatial Densities
  in Hadrons}},}\ }\href {\doibase 10.1103/PhysRevLett.129.012001} {\bibfield
  {journal} {\bibinfo  {journal} {Phys. Rev. Lett.}\ }\textbf {\bibinfo
  {volume} {129}},\ \bibinfo {pages} {012001} (\bibinfo {year} {2022})},\
  \Eprint {http://arxiv.org/abs/2201.02565} {arXiv:2201.02565 [hep-ph]}
  \BibitemShut {NoStop}%
\bibitem [{\citenamefont {Li}\ \emph {et~al.}(2023)\citenamefont {Li},
  \citenamefont {Dong}, \citenamefont {Yin}, \citenamefont {Wang},\ and\
  \citenamefont {Vary}}]{Li:2022ldb}%
  \BibitemOpen
  \bibfield  {author} {\bibinfo {author} {\bibfnamefont {Yang}\ \bibnamefont
  {Li}}, \bibinfo {author} {\bibfnamefont {Wen-bo}\ \bibnamefont {Dong}},
  \bibinfo {author} {\bibfnamefont {Yi-liang}\ \bibnamefont {Yin}}, \bibinfo
  {author} {\bibfnamefont {Qun}\ \bibnamefont {Wang}}, \ and\ \bibinfo {author}
  {\bibfnamefont {James~P.}\ \bibnamefont {Vary}},\ }\bibfield  {title}
  {\enquote {\bibinfo {title} {{Minkowski's lost legacy and hadron
  electromagnetism}},}\ }\href {\doibase 10.1016/j.physletb.2023.137676}
  {\bibfield  {journal} {\bibinfo  {journal} {Phys. Lett. B}\ }\textbf
  {\bibinfo {volume} {838}},\ \bibinfo {pages} {137676} (\bibinfo {year}
  {2023})},\ \Eprint {http://arxiv.org/abs/2206.12903} {arXiv:2206.12903
  [hep-ph]} \BibitemShut {NoStop}%
\bibitem [{\citenamefont {Chen}\ and\ \citenamefont
  {Lorc\'e}(2022)}]{Chen:2022smg}%
  \BibitemOpen
  \bibfield  {author} {\bibinfo {author} {\bibfnamefont {Yi}~\bibnamefont
  {Chen}}\ and\ \bibinfo {author} {\bibfnamefont {C\'edric}\ \bibnamefont
  {Lorc\'e}},\ }\bibfield  {title} {\enquote {\bibinfo {title} {{Pion and
  nucleon relativistic electromagnetic four-current distributions}},}\ }\href
  {\doibase 10.1103/PhysRevD.106.116024} {\bibfield  {journal} {\bibinfo
  {journal} {Phys. Rev. D}\ }\textbf {\bibinfo {volume} {106}},\ \bibinfo
  {pages} {116024} (\bibinfo {year} {2022})},\ \Eprint
  {http://arxiv.org/abs/2210.02908} {arXiv:2210.02908 [hep-ph]} \BibitemShut
  {NoStop}%
\bibitem [{\citenamefont {Freese}\ and\ \citenamefont
  {Miller}(2023{\natexlab{a}})}]{Freese:2022fat}%
  \BibitemOpen
  \bibfield  {author} {\bibinfo {author} {\bibfnamefont {Adam}\ \bibnamefont
  {Freese}}\ and\ \bibinfo {author} {\bibfnamefont {Gerald~A.}\ \bibnamefont
  {Miller}},\ }\bibfield  {title} {\enquote {\bibinfo {title} {{Convolution
  formalism for defining densities of hadrons}},}\ }\href {\doibase
  10.1103/PhysRevD.108.034008} {\bibfield  {journal} {\bibinfo  {journal}
  {Phys. Rev. D}\ }\textbf {\bibinfo {volume} {108}},\ \bibinfo {pages}
  {034008} (\bibinfo {year} {2023}{\natexlab{a}})},\ \Eprint
  {http://arxiv.org/abs/2210.03807} {arXiv:2210.03807 [hep-ph]} \BibitemShut
  {NoStop}%
\bibitem [{\citenamefont {Panteleeva}\ \emph
  {et~al.}(2022{\natexlab{a}})\citenamefont {Panteleeva}, \citenamefont
  {Epelbaum}, \citenamefont {Gegelia},\ and\ \citenamefont
  {Mei\ss{}ner}}]{Panteleeva:2022uii}%
  \BibitemOpen
  \bibfield  {author} {\bibinfo {author} {\bibfnamefont {J.~Yu.}\ \bibnamefont
  {Panteleeva}}, \bibinfo {author} {\bibfnamefont {E.}~\bibnamefont
  {Epelbaum}}, \bibinfo {author} {\bibfnamefont {J.}~\bibnamefont {Gegelia}}, \
  and\ \bibinfo {author} {\bibfnamefont {U.~G.}\ \bibnamefont {Mei\ss{}ner}},\
  }\bibfield  {title} {\enquote {\bibinfo {title} {{Definition of gravitational
  local spatial densities for spin-0 and spin-1/2 systems}},}\ }\href@noop {}
  {\  (\bibinfo {year} {2022}{\natexlab{a}})},\ \Eprint
  {http://arxiv.org/abs/2211.09596} {arXiv:2211.09596 [hep-ph]} \BibitemShut
  {NoStop}%
\bibitem [{\citenamefont {Freese}\ and\ \citenamefont
  {Miller}(2023{\natexlab{b}})}]{Freese:2023jcp}%
  \BibitemOpen
  \bibfield  {author} {\bibinfo {author} {\bibfnamefont {Adam}\ \bibnamefont
  {Freese}}\ and\ \bibinfo {author} {\bibfnamefont {Gerald~A.}\ \bibnamefont
  {Miller}},\ }\bibfield  {title} {\enquote {\bibinfo {title} {{Light front
  synchronization and rest frame densities of the proton: Electromagnetic
  densities}},}\ }\href {\doibase 10.1103/PhysRevD.107.074036} {\bibfield
  {journal} {\bibinfo  {journal} {Phys. Rev. D}\ }\textbf {\bibinfo {volume}
  {107}},\ \bibinfo {pages} {074036} (\bibinfo {year} {2023}{\natexlab{b}})},\
  \Eprint {http://arxiv.org/abs/2302.09171} {arXiv:2302.09171 [hep-ph]}
  \BibitemShut {NoStop}%
\bibitem [{\citenamefont {Chen}\ and\ \citenamefont
  {Lorc\'e}(2023)}]{Chen:2023dxp}%
  \BibitemOpen
  \bibfield  {author} {\bibinfo {author} {\bibfnamefont {Yi}~\bibnamefont
  {Chen}}\ and\ \bibinfo {author} {\bibfnamefont {C\'edric}\ \bibnamefont
  {Lorc\'e}},\ }\bibfield  {title} {\enquote {\bibinfo {title} {{Nucleon
  relativistic polarization and magnetization distributions}},}\ }\href
  {\doibase 10.1103/PhysRevD.107.096003} {\bibfield  {journal} {\bibinfo
  {journal} {Phys. Rev. D}\ }\textbf {\bibinfo {volume} {107}},\ \bibinfo
  {pages} {096003} (\bibinfo {year} {2023})},\ \Eprint
  {http://arxiv.org/abs/2302.04672} {arXiv:2302.04672 [hep-ph]} \BibitemShut
  {NoStop}%
\bibitem [{\citenamefont {Panteleeva}\ \emph {et~al.}(2023)\citenamefont
  {Panteleeva}, \citenamefont {Epelbaum}, \citenamefont {Gegelia},\ and\
  \citenamefont {Mei\ss{}ner}}]{Panteleeva:2023evj}%
  \BibitemOpen
  \bibfield  {author} {\bibinfo {author} {\bibfnamefont {J.~Yu.}\ \bibnamefont
  {Panteleeva}}, \bibinfo {author} {\bibfnamefont {E.}~\bibnamefont
  {Epelbaum}}, \bibinfo {author} {\bibfnamefont {J.}~\bibnamefont {Gegelia}}, \
  and\ \bibinfo {author} {\bibfnamefont {U.~G.}\ \bibnamefont {Mei\ss{}ner}},\
  }\bibfield  {title} {\enquote {\bibinfo {title} {{Electromagnetic and
  gravitational local spatial densities for spin-1 systems}},}\ }\href@noop {}
  {\  (\bibinfo {year} {2023})},\ \Eprint {http://arxiv.org/abs/2305.01491}
  {arXiv:2305.01491 [hep-ph]} \BibitemShut {NoStop}%
\bibitem [{\citenamefont {Burkardt}(2003)}]{Burkardt:2002hr}%
  \BibitemOpen
  \bibfield  {author} {\bibinfo {author} {\bibfnamefont {Matthias}\
  \bibnamefont {Burkardt}},\ }\bibfield  {title} {\enquote {\bibinfo {title}
  {{Impact parameter space interpretation for generalized parton
  distributions}},}\ }\href {\doibase 10.1142/S0217751X03012370} {\bibfield
  {journal} {\bibinfo  {journal} {Int. J. Mod. Phys. A}\ }\textbf {\bibinfo
  {volume} {18}},\ \bibinfo {pages} {173--208} (\bibinfo {year} {2003})},\
  \Eprint {http://arxiv.org/abs/hep-ph/0207047} {arXiv:hep-ph/0207047}
  \BibitemShut {NoStop}%
\bibitem [{\citenamefont {Miller}(2010)}]{Miller:2010nz}%
  \BibitemOpen
  \bibfield  {author} {\bibinfo {author} {\bibfnamefont {Gerald~A.}\
  \bibnamefont {Miller}},\ }\bibfield  {title} {\enquote {\bibinfo {title}
  {{Transverse Charge Densities}},}\ }\href {\doibase
  10.1146/annurev.nucl.012809.104508} {\bibfield  {journal} {\bibinfo
  {journal} {Ann. Rev. Nucl. Part. Sci.}\ }\textbf {\bibinfo {volume} {60}},\
  \bibinfo {pages} {1--25} (\bibinfo {year} {2010})},\ \Eprint
  {http://arxiv.org/abs/1002.0355} {arXiv:1002.0355 [nucl-th]} \BibitemShut
  {NoStop}%
\bibitem [{\citenamefont {Lorc\'e}(2020)}]{Lorce:2020onh}%
  \BibitemOpen
  \bibfield  {author} {\bibinfo {author} {\bibfnamefont {C\'edric}\
  \bibnamefont {Lorc\'e}},\ }\bibfield  {title} {\enquote {\bibinfo {title}
  {{Charge Distributions of Moving Nucleons}},}\ }\href {\doibase
  10.1103/PhysRevLett.125.232002} {\bibfield  {journal} {\bibinfo  {journal}
  {Phys. Rev. Lett.}\ }\textbf {\bibinfo {volume} {125}},\ \bibinfo {pages}
  {232002} (\bibinfo {year} {2020})},\ \Eprint
  {http://arxiv.org/abs/2007.05318} {arXiv:2007.05318 [hep-ph]} \BibitemShut
  {NoStop}%
\bibitem [{\citenamefont {Blunden}\ \emph {et~al.}(2000)\citenamefont
  {Blunden}, \citenamefont {Burkardt},\ and\ \citenamefont
  {Miller}}]{Blunden:1999wb}%
  \BibitemOpen
  \bibfield  {author} {\bibinfo {author} {\bibfnamefont {P.~G.}\ \bibnamefont
  {Blunden}}, \bibinfo {author} {\bibfnamefont {M.}~\bibnamefont {Burkardt}}, \
  and\ \bibinfo {author} {\bibfnamefont {G.~A.}\ \bibnamefont {Miller}},\
  }\bibfield  {title} {\enquote {\bibinfo {title} {{Light front nuclear
  physics: Toy models, static sources and tilted light front coordinates}},}\
  }\href {\doibase 10.1103/PhysRevC.61.025206} {\bibfield  {journal} {\bibinfo
  {journal} {Phys. Rev. C}\ }\textbf {\bibinfo {volume} {61}},\ \bibinfo
  {pages} {025206} (\bibinfo {year} {2000})},\ \Eprint
  {http://arxiv.org/abs/nucl-th/9908067} {arXiv:nucl-th/9908067} \BibitemShut
  {NoStop}%
\bibitem [{\citenamefont {Einstein}(1905)}]{Einstein:1905ve}%
  \BibitemOpen
  \bibfield  {author} {\bibinfo {author} {\bibfnamefont {Albert}\ \bibnamefont
  {Einstein}},\ }\bibfield  {title} {\enquote {\bibinfo {title} {{On the
  electrodynamics of moving bodies}},}\ }\href {\doibase
  10.1002/andp.200590006} {\bibfield  {journal} {\bibinfo  {journal} {Annalen
  Phys.}\ }\textbf {\bibinfo {volume} {17}},\ \bibinfo {pages} {891--921}
  (\bibinfo {year} {1905})}\BibitemShut {NoStop}%
\bibitem [{\citenamefont {Reichenbach}(2012)}]{reichenbach2012philosophy}%
  \BibitemOpen
  \bibfield  {author} {\bibinfo {author} {\bibfnamefont {H.}~\bibnamefont
  {Reichenbach}},\ }\href {https://books.google.com/books?id=E\_DDAgAAQBAJ}
  {\emph {\bibinfo {title} {The Philosophy of Space and Time}}},\ Dover Books
  on Physics\ (\bibinfo  {publisher} {Dover Publications},\ \bibinfo {year}
  {2012})\BibitemShut {NoStop}%
\bibitem [{\citenamefont {Gr{\"u}nbaum}(2012)}]{gruenbaum2012philosophical}%
  \BibitemOpen
  \bibfield  {author} {\bibinfo {author} {\bibfnamefont {A.}~\bibnamefont
  {Gr{\"u}nbaum}},\ }\href {https://books.google.com/books?id=m3ugBwAAQBAJ}
  {\emph {\bibinfo {title} {Philosophical Problems of Space and Time: Second,
  enlarged edition}}},\ Boston Studies in the Philosophy and History of
  Science\ (\bibinfo  {publisher} {Springer Netherlands},\ \bibinfo {year}
  {2012})\BibitemShut {NoStop}%
\bibitem [{\citenamefont {Zhang}(1995)}]{Zhang:1995test}%
  \BibitemOpen
  \bibfield  {author} {\bibinfo {author} {\bibfnamefont {Yuan~Zhong}\
  \bibnamefont {Zhang}},\ }\bibfield  {title} {\enquote {\bibinfo {title} {Test
  theories of special relativity},}\ }\href@noop {} {\bibfield  {journal}
  {\bibinfo  {journal} {General Relativity and Gravitation}\ }\textbf {\bibinfo
  {volume} {27}},\ \bibinfo {pages} {475--493} (\bibinfo {year}
  {1995})}\BibitemShut {NoStop}%
\bibitem [{\citenamefont {Anderson}\ \emph {et~al.}(1998)\citenamefont
  {Anderson}, \citenamefont {Stedman},\ and\ \citenamefont
  {Vetharaniam}}]{Anderson:1998mu}%
  \BibitemOpen
  \bibfield  {author} {\bibinfo {author} {\bibfnamefont {Ronald}\ \bibnamefont
  {Anderson}}, \bibinfo {author} {\bibfnamefont {Geoffrey~E.}\ \bibnamefont
  {Stedman}}, \ and\ \bibinfo {author} {\bibfnamefont {I.}~\bibnamefont
  {Vetharaniam}},\ }\bibfield  {title} {\enquote {\bibinfo {title}
  {{Conventionality of synchronisation, gauge dependence and test theories of
  relativity}},}\ }\href {\doibase 10.1016/S0370-1573(97)00051-3} {\bibfield
  {journal} {\bibinfo  {journal} {Phys. Rept.}\ }\textbf {\bibinfo {volume}
  {295}},\ \bibinfo {pages} {93--180} (\bibinfo {year} {1998})}\BibitemShut
  {NoStop}%
\bibitem [{\citenamefont {Muller}\ and\ \citenamefont
  {Lebedev}(2020)}]{Veritasium:2020oct}%
  \BibitemOpen
  \bibfield  {author} {\bibinfo {author} {\bibfnamefont {Derek}\ \bibnamefont
  {Muller}}\ and\ \bibinfo {author} {\bibfnamefont {Petr}\ \bibnamefont
  {Lebedev}},\ }\href {https://www.youtube.com/watch?v=pTn6Ewhb27k} {\enquote
  {\bibinfo {title} {{Why No One Has Measured The Speed Of Light}},}\ }
  (\bibinfo {year} {2020})\BibitemShut {NoStop}%
\bibitem [{\citenamefont {Panteleeva}\ \emph
  {et~al.}(2022{\natexlab{b}})\citenamefont {Panteleeva}, \citenamefont
  {Epelbaum}, \citenamefont {Gegelia},\ and\ \citenamefont
  {Mei\ss{}ner}}]{Panteleeva:2022khw}%
  \BibitemOpen
  \bibfield  {author} {\bibinfo {author} {\bibfnamefont {J.~Yu.}\ \bibnamefont
  {Panteleeva}}, \bibinfo {author} {\bibfnamefont {E.}~\bibnamefont
  {Epelbaum}}, \bibinfo {author} {\bibfnamefont {J.}~\bibnamefont {Gegelia}}, \
  and\ \bibinfo {author} {\bibfnamefont {U.~G.}\ \bibnamefont {Mei\ss{}ner}},\
  }\bibfield  {title} {\enquote {\bibinfo {title} {{On the definition of
  electromagnetic local spatial densities for composite spin-$1/2$ systems}},}\
  }\href@noop {} {\  (\bibinfo {year} {2022}{\natexlab{b}})},\ \Eprint
  {http://arxiv.org/abs/2205.15061} {arXiv:2205.15061 [hep-ph]} \BibitemShut
  {NoStop}%
\bibitem [{\citenamefont {Freese}\ and\ \citenamefont
  {Miller}(2022)}]{Freese:2021mzg}%
  \BibitemOpen
  \bibfield  {author} {\bibinfo {author} {\bibfnamefont {Adam}\ \bibnamefont
  {Freese}}\ and\ \bibinfo {author} {\bibfnamefont {Gerald~A.}\ \bibnamefont
  {Miller}},\ }\bibfield  {title} {\enquote {\bibinfo {title} {{Unified
  formalism for electromagnetic and gravitational probes: Densities}},}\ }\href
  {\doibase 10.1103/PhysRevD.105.014003} {\bibfield  {journal} {\bibinfo
  {journal} {Phys. Rev. D}\ }\textbf {\bibinfo {volume} {105}},\ \bibinfo
  {pages} {014003} (\bibinfo {year} {2022})},\ \Eprint
  {http://arxiv.org/abs/2108.03301} {arXiv:2108.03301 [hep-ph]} \BibitemShut
  {NoStop}%
\bibitem [{\citenamefont {Kugo}\ and\ \citenamefont
  {Ojima}(1979)}]{Kugo:1979gm}%
  \BibitemOpen
  \bibfield  {author} {\bibinfo {author} {\bibfnamefont {Taichiro}\
  \bibnamefont {Kugo}}\ and\ \bibinfo {author} {\bibfnamefont {Izumi}\
  \bibnamefont {Ojima}},\ }\bibfield  {title} {\enquote {\bibinfo {title}
  {{Local Covariant Operator Formalism of Nonabelian Gauge Theories and Quark
  Confinement Problem}},}\ }\href {\doibase 10.1143/PTPS.66.1} {\bibfield
  {journal} {\bibinfo  {journal} {Prog. Theor. Phys. Suppl.}\ }\textbf
  {\bibinfo {volume} {66}},\ \bibinfo {pages} {1--130} (\bibinfo {year}
  {1979})}\BibitemShut {NoStop}%
\bibitem [{\citenamefont {{Belinfante}}(1939)}]{Belinfante:1939emt}%
  \BibitemOpen
  \bibfield  {author} {\bibinfo {author} {\bibfnamefont {F.~J.}\ \bibnamefont
  {{Belinfante}}},\ }\bibfield  {title} {\enquote {\bibinfo {title} {{On the
  spin angular momentum of mesons}},}\ }\href {\doibase
  10.1016/S0031-8914(39)90090-X} {\bibfield  {journal} {\bibinfo  {journal}
  {Physica}\ }\textbf {\bibinfo {volume} {6}},\ \bibinfo {pages} {887--898}
  (\bibinfo {year} {1939})}\BibitemShut {NoStop}%
\bibitem [{\citenamefont {Freese}(2022)}]{Freese:2021jqs}%
  \BibitemOpen
  \bibfield  {author} {\bibinfo {author} {\bibfnamefont {Adam}\ \bibnamefont
  {Freese}},\ }\bibfield  {title} {\enquote {\bibinfo {title} {{Noether's
  theorems and the energy-momentum tensor in quantum gauge theories}},}\ }\href
  {\doibase 10.1103/PhysRevD.106.125012} {\bibfield  {journal} {\bibinfo
  {journal} {Phys. Rev. D}\ }\textbf {\bibinfo {volume} {106}},\ \bibinfo
  {pages} {125012} (\bibinfo {year} {2022})},\ \Eprint
  {http://arxiv.org/abs/2112.00047} {arXiv:2112.00047 [hep-th]} \BibitemShut
  {NoStop}%
\bibitem [{\citenamefont {Fetter}\ and\ \citenamefont
  {Walecka}(1980)}]{Fetter:1980str}%
  \BibitemOpen
  \bibfield  {author} {\bibinfo {author} {\bibfnamefont {Alexander~L}\
  \bibnamefont {Fetter}}\ and\ \bibinfo {author} {\bibfnamefont {John~Dirk}\
  \bibnamefont {Walecka}},\ }\href {https://cds.cern.ch/record/1596004} {\emph
  {\bibinfo {title} {{Theoretical mechanics of particles and continua}}}},\
  International series in pure and applied physics\ (\bibinfo  {publisher}
  {McGraw-Hill},\ \bibinfo {address} {New York, NY},\ \bibinfo {year}
  {1980})\BibitemShut {NoStop}%
\bibitem [{\citenamefont {Batchelor}\ and\ \citenamefont
  {Batchelor}(2000)}]{Batchelor:2000int}%
  \BibitemOpen
  \bibfield  {author} {\bibinfo {author} {\bibfnamefont {C.K.}\ \bibnamefont
  {Batchelor}}\ and\ \bibinfo {author} {\bibfnamefont {G.K.}\ \bibnamefont
  {Batchelor}},\ }\href {https://books.google.com/books?id=Rla7OihRvUgC} {\emph
  {\bibinfo {title} {An Introduction to Fluid Dynamics}}},\ Cambridge
  Mathematical Library\ (\bibinfo  {publisher} {Cambridge University Press},\
  \bibinfo {year} {2000})\BibitemShut {NoStop}%
\bibitem [{\citenamefont {Irgens}(2008)}]{Irgens:2008con}%
  \BibitemOpen
  \bibfield  {author} {\bibinfo {author} {\bibfnamefont {F.}~\bibnamefont
  {Irgens}},\ }\href {https://books.google.com/books?id=q5dB7Gf4bIoC} {\emph
  {\bibinfo {title} {Continuum Mechanics}}}\ (\bibinfo  {publisher} {Springer
  Berlin Heidelberg},\ \bibinfo {year} {2008})\BibitemShut {NoStop}%
\bibitem [{\citenamefont {Friedman}\ and\ \citenamefont
  {Stergioulas}(2013)}]{Friedman:2013xza}%
  \BibitemOpen
  \bibfield  {author} {\bibinfo {author} {\bibfnamefont {John~L.}\ \bibnamefont
  {Friedman}}\ and\ \bibinfo {author} {\bibfnamefont {Nikolaos}\ \bibnamefont
  {Stergioulas}},\ }\href {\doibase 10.1017/CBO9780511977596} {\emph {\bibinfo
  {title} {{Rotating Relativistic Stars}}}},\ Cambridge Monographs on
  Mathematical Physics\ (\bibinfo  {publisher} {Cambridge University Press},\
  \bibinfo {year} {2013})\BibitemShut {NoStop}%
\bibitem [{\citenamefont {Fl{\"u}gge}(2013)}]{flugge2013tensor}%
  \BibitemOpen
  \bibfield  {author} {\bibinfo {author} {\bibfnamefont {W.}~\bibnamefont
  {Fl{\"u}gge}},\ }\href {https://books.google.com/books?id=00P1CAAAQBAJ}
  {\emph {\bibinfo {title} {Tensor Analysis and Continuum Mechanics}}}\
  (\bibinfo  {publisher} {Springer Berlin Heidelberg},\ \bibinfo {year}
  {2013})\BibitemShut {NoStop}%
\bibitem [{\citenamefont {Liu}(2013)}]{liu2013continuum}%
  \BibitemOpen
  \bibfield  {author} {\bibinfo {author} {\bibfnamefont {I.S.}\ \bibnamefont
  {Liu}},\ }\href {https://books.google.com/books?id=fzrvCAAAQBAJ} {\emph
  {\bibinfo {title} {Continuum Mechanics}}},\ Advanced Texts in Physics\
  (\bibinfo  {publisher} {Springer Berlin Heidelberg},\ \bibinfo {year}
  {2013})\BibitemShut {NoStop}%
\bibitem [{\citenamefont {Bower}(2009)}]{bower2009applied}%
  \BibitemOpen
  \bibfield  {author} {\bibinfo {author} {\bibfnamefont {A.F.}\ \bibnamefont
  {Bower}},\ }\href {https://books.google.com/books?id=1\_sCzMNZMEsC} {\emph
  {\bibinfo {title} {Applied Mechanics of Solids}}}\ (\bibinfo  {publisher}
  {CRC Press},\ \bibinfo {year} {2009})\BibitemShut {NoStop}%
\bibitem [{\citenamefont {Ericksen}(1961)}]{10.1122/1.548883}%
  \BibitemOpen
  \bibfield  {author} {\bibinfo {author} {\bibfnamefont {J.~L.}\ \bibnamefont
  {Ericksen}},\ }\bibfield  {title} {\enquote {\bibinfo {title} {{Conservation
  Laws for Liquid Crystals}},}\ }\href {\doibase 10.1122/1.548883} {\bibfield
  {journal} {\bibinfo  {journal} {Transactions of The Society of Rheology}\
  }\textbf {\bibinfo {volume} {5}},\ \bibinfo {pages} {23--34} (\bibinfo {year}
  {1961})},\ \Eprint
  {http://arxiv.org/abs/https://pubs.aip.org/sor/jor/article-pdf/5/1/23/12261492/1\_548883.pdf}
  {https://pubs.aip.org/sor/jor/article-pdf/5/1/23/12261492/1\_548883.pdf}
  \BibitemShut {NoStop}%
\bibitem [{\citenamefont {Forster}\ \emph {et~al.}(1971)\citenamefont
  {Forster}, \citenamefont {Lubensky}, \citenamefont {Martin}, \citenamefont
  {Swift},\ and\ \citenamefont {Pershan}}]{PhysRevLett.26.1016}%
  \BibitemOpen
  \bibfield  {author} {\bibinfo {author} {\bibfnamefont {Dieter}\ \bibnamefont
  {Forster}}, \bibinfo {author} {\bibfnamefont {Tom~C.}\ \bibnamefont
  {Lubensky}}, \bibinfo {author} {\bibfnamefont {Paul~C.}\ \bibnamefont
  {Martin}}, \bibinfo {author} {\bibfnamefont {Jack}\ \bibnamefont {Swift}}, \
  and\ \bibinfo {author} {\bibfnamefont {P.~S.}\ \bibnamefont {Pershan}},\
  }\bibfield  {title} {\enquote {\bibinfo {title} {Hydrodynamics of liquid
  crystals},}\ }\href {\doibase 10.1103/PhysRevLett.26.1016} {\bibfield
  {journal} {\bibinfo  {journal} {Phys. Rev. Lett.}\ }\textbf {\bibinfo
  {volume} {26}},\ \bibinfo {pages} {1016--1019} (\bibinfo {year}
  {1971})}\BibitemShut {NoStop}%
\bibitem [{\citenamefont {Brand}\ and\ \citenamefont
  {Pleiner}(2021)}]{PhysRevE.103.012705}%
  \BibitemOpen
  \bibfield  {author} {\bibinfo {author} {\bibfnamefont {Helmut~R.}\
  \bibnamefont {Brand}}\ and\ \bibinfo {author} {\bibfnamefont {Harald}\
  \bibnamefont {Pleiner}},\ }\bibfield  {title} {\enquote {\bibinfo {title}
  {Two-fluid model for the breakdown of flow alignment in nematic liquid
  crystals},}\ }\href {\doibase 10.1103/PhysRevE.103.012705} {\bibfield
  {journal} {\bibinfo  {journal} {Phys. Rev. E}\ }\textbf {\bibinfo {volume}
  {103}},\ \bibinfo {pages} {012705} (\bibinfo {year} {2021})}\BibitemShut
  {NoStop}%
\bibitem [{\citenamefont {Friedman}\ \emph {et~al.}(1992)\citenamefont
  {Friedman}, \citenamefont {Ipser},\ and\ \citenamefont
  {Chandrasekhar}}]{doi:10.1098/rsta.1992.0074}%
  \BibitemOpen
  \bibfield  {author} {\bibinfo {author} {\bibfnamefont {John~L.}\ \bibnamefont
  {Friedman}}, \bibinfo {author} {\bibfnamefont {James~R.}\ \bibnamefont
  {Ipser}}, \ and\ \bibinfo {author} {\bibfnamefont {Subrahmanyan}\
  \bibnamefont {Chandrasekhar}},\ }\bibfield  {title} {\enquote {\bibinfo
  {title} {Rapidly rotating relativistic stars},}\ }\href {\doibase
  10.1098/rsta.1992.0074} {\bibfield  {journal} {\bibinfo  {journal}
  {Philosophical Transactions of the Royal Society of London. Series A:
  Physical and Engineering Sciences}\ }\textbf {\bibinfo {volume} {340}},\
  \bibinfo {pages} {391--422} (\bibinfo {year} {1992})},\ \Eprint
  {http://arxiv.org/abs/https://royalsocietypublishing.org/doi/pdf/10.1098/rsta.1992.0074}
  {https://royalsocietypublishing.org/doi/pdf/10.1098/rsta.1992.0074}
  \BibitemShut {NoStop}%
\bibitem [{\citenamefont {Martin}\ \emph {et~al.}(1972)\citenamefont {Martin},
  \citenamefont {Parodi},\ and\ \citenamefont {Pershan}}]{PhysRevA.6.2401}%
  \BibitemOpen
  \bibfield  {author} {\bibinfo {author} {\bibfnamefont {P.~C.}\ \bibnamefont
  {Martin}}, \bibinfo {author} {\bibfnamefont {O.}~\bibnamefont {Parodi}}, \
  and\ \bibinfo {author} {\bibfnamefont {P.~S.}\ \bibnamefont {Pershan}},\
  }\bibfield  {title} {\enquote {\bibinfo {title} {Unified hydrodynamic theory
  for crystals, liquid crystals, and normal fluids},}\ }\href {\doibase
  10.1103/PhysRevA.6.2401} {\bibfield  {journal} {\bibinfo  {journal} {Phys.
  Rev. A}\ }\textbf {\bibinfo {volume} {6}},\ \bibinfo {pages} {2401--2420}
  (\bibinfo {year} {1972})}\BibitemShut {NoStop}%
\bibitem [{\citenamefont {Freese}\ and\ \citenamefont
  {Miller}(2021{\natexlab{b}})}]{Freese:2021qtb}%
  \BibitemOpen
  \bibfield  {author} {\bibinfo {author} {\bibfnamefont {Adam}\ \bibnamefont
  {Freese}}\ and\ \bibinfo {author} {\bibfnamefont {Gerald~A.}\ \bibnamefont
  {Miller}},\ }\bibfield  {title} {\enquote {\bibinfo {title} {{Genuine
  empirical pressure within the proton}},}\ }\href {\doibase
  10.1103/PhysRevD.104.014024} {\bibfield  {journal} {\bibinfo  {journal}
  {Phys. Rev. D}\ }\textbf {\bibinfo {volume} {104}},\ \bibinfo {pages}
  {014024} (\bibinfo {year} {2021}{\natexlab{b}})},\ \Eprint
  {http://arxiv.org/abs/2104.03213} {arXiv:2104.03213 [hep-ph]} \BibitemShut
  {NoStop}%
\bibitem [{\citenamefont {Lorc\'e}(2018{\natexlab{b}})}]{Lorce:2018zpf}%
  \BibitemOpen
  \bibfield  {author} {\bibinfo {author} {\bibfnamefont {C\'edric}\
  \bibnamefont {Lorc\'e}},\ }\bibfield  {title} {\enquote {\bibinfo {title}
  {{The relativistic center of mass in field theory with spin}},}\ }\href
  {\doibase 10.1140/epjc/s10052-018-6249-3} {\bibfield  {journal} {\bibinfo
  {journal} {Eur. Phys. J. C}\ }\textbf {\bibinfo {volume} {78}},\ \bibinfo
  {pages} {785} (\bibinfo {year} {2018}{\natexlab{b}})},\ \Eprint
  {http://arxiv.org/abs/1805.05284} {arXiv:1805.05284 [hep-ph]} \BibitemShut
  {NoStop}%
\bibitem [{\citenamefont {Teryaev}(1999)}]{Teryaev:1999su}%
  \BibitemOpen
  \bibfield  {author} {\bibinfo {author} {\bibfnamefont {O.~V.}\ \bibnamefont
  {Teryaev}},\ }\bibfield  {title} {\enquote {\bibinfo {title} {{Spin structure
  of nucleon and equivalence principle}},}\ }\href@noop {} {\  (\bibinfo {year}
  {1999})},\ \Eprint {http://arxiv.org/abs/hep-ph/9904376}
  {arXiv:hep-ph/9904376} \BibitemShut {NoStop}%
\bibitem [{\citenamefont {Hudson}\ and\ \citenamefont
  {Schweitzer}(2018)}]{Hudson:2017oul}%
  \BibitemOpen
  \bibfield  {author} {\bibinfo {author} {\bibfnamefont {Jonathan}\
  \bibnamefont {Hudson}}\ and\ \bibinfo {author} {\bibfnamefont {Peter}\
  \bibnamefont {Schweitzer}},\ }\bibfield  {title} {\enquote {\bibinfo {title}
  {{Dynamic origins of fermionic D-terms}},}\ }\href {\doibase
  10.1103/PhysRevD.97.056003} {\bibfield  {journal} {\bibinfo  {journal} {Phys.
  Rev. D}\ }\textbf {\bibinfo {volume} {97}},\ \bibinfo {pages} {056003}
  (\bibinfo {year} {2018})},\ \Eprint {http://arxiv.org/abs/1712.05317}
  {arXiv:1712.05317 [hep-ph]} \BibitemShut {NoStop}%
\bibitem [{\citenamefont {Mamo}\ and\ \citenamefont
  {Zahed}(2020)}]{Mamo:2019mka}%
  \BibitemOpen
  \bibfield  {author} {\bibinfo {author} {\bibfnamefont {Kiminad~A.}\
  \bibnamefont {Mamo}}\ and\ \bibinfo {author} {\bibfnamefont {Ismail}\
  \bibnamefont {Zahed}},\ }\bibfield  {title} {\enquote {\bibinfo {title}
  {{Diffractive photoproduction of $J/\psi$ and $\Upsilon$ using holographic
  QCD: gravitational form factors and GPD of gluons in the proton}},}\ }\href
  {\doibase 10.1103/PhysRevD.101.086003} {\bibfield  {journal} {\bibinfo
  {journal} {Phys. Rev. D}\ }\textbf {\bibinfo {volume} {101}},\ \bibinfo
  {pages} {086003} (\bibinfo {year} {2020})},\ \Eprint
  {http://arxiv.org/abs/1910.04707} {arXiv:1910.04707 [hep-ph]} \BibitemShut
  {NoStop}%
\bibitem [{\citenamefont {Mamo}\ and\ \citenamefont
  {Zahed}(2021)}]{Mamo:2021krl}%
  \BibitemOpen
  \bibfield  {author} {\bibinfo {author} {\bibfnamefont {Kiminad~A.}\
  \bibnamefont {Mamo}}\ and\ \bibinfo {author} {\bibfnamefont {Ismail}\
  \bibnamefont {Zahed}},\ }\bibfield  {title} {\enquote {\bibinfo {title}
  {{Nucleon mass radii and distribution: Holographic QCD, Lattice QCD and GlueX
  data}},}\ }\href {\doibase 10.1103/PhysRevD.103.094010} {\bibfield  {journal}
  {\bibinfo  {journal} {Phys. Rev. D}\ }\textbf {\bibinfo {volume} {103}},\
  \bibinfo {pages} {094010} (\bibinfo {year} {2021})},\ \Eprint
  {http://arxiv.org/abs/2103.03186} {arXiv:2103.03186 [hep-ph]} \BibitemShut
  {NoStop}%
\bibitem [{\citenamefont {Fujita}\ \emph {et~al.}(2022)\citenamefont {Fujita},
  \citenamefont {Hatta}, \citenamefont {Sugimoto},\ and\ \citenamefont
  {Ueda}}]{Fujita:2022jus}%
  \BibitemOpen
  \bibfield  {author} {\bibinfo {author} {\bibfnamefont {Mitsutoshi}\
  \bibnamefont {Fujita}}, \bibinfo {author} {\bibfnamefont {Yoshitaka}\
  \bibnamefont {Hatta}}, \bibinfo {author} {\bibfnamefont {Shigeki}\
  \bibnamefont {Sugimoto}}, \ and\ \bibinfo {author} {\bibfnamefont {Takahiro}\
  \bibnamefont {Ueda}},\ }\bibfield  {title} {\enquote {\bibinfo {title}
  {{Nucleon D-term in holographic quantum chromodynamics}},}\ }\href {\doibase
  10.1093/ptep/ptac110} {\bibfield  {journal} {\bibinfo  {journal} {PTEP}\
  }\textbf {\bibinfo {volume} {2022}},\ \bibinfo {pages} {093B06} (\bibinfo
  {year} {2022})},\ \Eprint {http://arxiv.org/abs/2206.06578} {arXiv:2206.06578
  [hep-th]} \BibitemShut {NoStop}%
\bibitem [{\citenamefont {Sakurai}(1960)}]{Sakurai:1960ju}%
  \BibitemOpen
  \bibfield  {author} {\bibinfo {author} {\bibfnamefont {J.~J.}\ \bibnamefont
  {Sakurai}},\ }\bibfield  {title} {\enquote {\bibinfo {title} {{Theory of
  strong interactions}},}\ }\href {\doibase 10.1016/0003-4916(60)90126-3}
  {\bibfield  {journal} {\bibinfo  {journal} {Annals Phys.}\ }\textbf {\bibinfo
  {volume} {11}},\ \bibinfo {pages} {1--48} (\bibinfo {year}
  {1960})}\BibitemShut {NoStop}%
\bibitem [{\citenamefont {Bauer}\ \emph {et~al.}(1978)\citenamefont {Bauer},
  \citenamefont {Spital}, \citenamefont {Yennie},\ and\ \citenamefont
  {Pipkin}}]{Bauer:1977iq}%
  \BibitemOpen
  \bibfield  {author} {\bibinfo {author} {\bibfnamefont {T.~H.}\ \bibnamefont
  {Bauer}}, \bibinfo {author} {\bibfnamefont {R.~D.}\ \bibnamefont {Spital}},
  \bibinfo {author} {\bibfnamefont {D.~R.}\ \bibnamefont {Yennie}}, \ and\
  \bibinfo {author} {\bibfnamefont {F.~M.}\ \bibnamefont {Pipkin}},\ }\bibfield
   {title} {\enquote {\bibinfo {title} {{The Hadronic Properties of the Photon
  in High-Energy Interactions}},}\ }\href {\doibase 10.1103/RevModPhys.50.261}
  {\bibfield  {journal} {\bibinfo  {journal} {Rev. Mod. Phys.}\ }\textbf
  {\bibinfo {volume} {50}},\ \bibinfo {pages} {261} (\bibinfo {year} {1978})},\
  \bibinfo {note} {[Erratum: Rev.Mod.Phys. 51, 407 (1979)]}\BibitemShut
  {NoStop}%
\bibitem [{\citenamefont {Hashimoto}\ \emph {et~al.}(2008)\citenamefont
  {Hashimoto}, \citenamefont {Sakai},\ and\ \citenamefont
  {Sugimoto}}]{Hashimoto:2008zw}%
  \BibitemOpen
  \bibfield  {author} {\bibinfo {author} {\bibfnamefont {Koji}\ \bibnamefont
  {Hashimoto}}, \bibinfo {author} {\bibfnamefont {Tadakatsu}\ \bibnamefont
  {Sakai}}, \ and\ \bibinfo {author} {\bibfnamefont {Shigeki}\ \bibnamefont
  {Sugimoto}},\ }\bibfield  {title} {\enquote {\bibinfo {title} {{Holographic
  Baryons: Static Properties and Form Factors from Gauge/String Duality}},}\
  }\href {\doibase 10.1143/PTP.120.1093} {\bibfield  {journal} {\bibinfo
  {journal} {Prog. Theor. Phys.}\ }\textbf {\bibinfo {volume} {120}},\ \bibinfo
  {pages} {1093--1137} (\bibinfo {year} {2008})},\ \Eprint
  {http://arxiv.org/abs/0806.3122} {arXiv:0806.3122 [hep-th]} \BibitemShut
  {NoStop}%
\bibitem [{\citenamefont {Shanahan}\ and\ \citenamefont
  {Detmold}(2019)}]{Shanahan:2018pib}%
  \BibitemOpen
  \bibfield  {author} {\bibinfo {author} {\bibfnamefont {P.~E.}\ \bibnamefont
  {Shanahan}}\ and\ \bibinfo {author} {\bibfnamefont {W.}~\bibnamefont
  {Detmold}},\ }\bibfield  {title} {\enquote {\bibinfo {title} {{Gluon
  gravitational form factors of the nucleon and the pion from lattice QCD}},}\
  }\href {\doibase 10.1103/PhysRevD.99.014511} {\bibfield  {journal} {\bibinfo
  {journal} {Phys. Rev. D}\ }\textbf {\bibinfo {volume} {99}},\ \bibinfo
  {pages} {014511} (\bibinfo {year} {2019})},\ \Eprint
  {http://arxiv.org/abs/1810.04626} {arXiv:1810.04626 [hep-lat]} \BibitemShut
  {NoStop}%
\bibitem [{\citenamefont {Duran}\ \emph {et~al.}(2023)\citenamefont {Duran}
  \emph {et~al.}}]{Duran:2022xag}%
  \BibitemOpen
  \bibfield  {author} {\bibinfo {author} {\bibfnamefont {B.}~\bibnamefont
  {Duran}} \emph {et~al.},\ }\bibfield  {title} {\enquote {\bibinfo {title}
  {{Determining the gluonic gravitational form factors of the proton}},}\
  }\href {\doibase 10.1038/s41586-023-05730-4} {\bibfield  {journal} {\bibinfo
  {journal} {Nature}\ }\textbf {\bibinfo {volume} {615}},\ \bibinfo {pages}
  {813--816} (\bibinfo {year} {2023})},\ \Eprint
  {http://arxiv.org/abs/2207.05212} {arXiv:2207.05212 [nucl-ex]} \BibitemShut
  {NoStop}%
\bibitem [{\citenamefont {Masjuan}\ \emph {et~al.}(2013)\citenamefont
  {Masjuan}, \citenamefont {Ruiz~Arriola},\ and\ \citenamefont
  {Broniowski}}]{Masjuan:2012sk}%
  \BibitemOpen
  \bibfield  {author} {\bibinfo {author} {\bibfnamefont {Pere}\ \bibnamefont
  {Masjuan}}, \bibinfo {author} {\bibfnamefont {Enrique}\ \bibnamefont
  {Ruiz~Arriola}}, \ and\ \bibinfo {author} {\bibfnamefont {Wojciech}\
  \bibnamefont {Broniowski}},\ }\bibfield  {title} {\enquote {\bibinfo {title}
  {{Meson dominance of hadron form factors and large-$N_c$ phenomenology}},}\
  }\href {\doibase 10.1103/PhysRevD.87.014005} {\bibfield  {journal} {\bibinfo
  {journal} {Phys. Rev. D}\ }\textbf {\bibinfo {volume} {87}},\ \bibinfo
  {pages} {014005} (\bibinfo {year} {2013})},\ \Eprint
  {http://arxiv.org/abs/1210.0760} {arXiv:1210.0760 [hep-ph]} \BibitemShut
  {NoStop}%
\bibitem [{{\relax DLMF}()}]{NIST:DLMF}%
  \BibitemOpen
  {\relax DLMF},\ \href {http://dlmf.nist.gov/} {\enquote {\bibinfo {title}
  {{\it NIST Digital Library of Mathematical Functions}},}\ }\bibinfo
  {howpublished} {http://dlmf.nist.gov/, Release 1.1.0 of 2020-12-15},\
  \bibinfo {note} {f.~W.~J. Olver, A.~B. {Olde Daalhuis}, D.~W. Lozier, B.~I.
  Schneider, R.~F. Boisvert, C.~W. Clark, B.~R. Miller, B.~V. Saunders, H.~S.
  Cohl, and M.~A. McClain, eds.}\BibitemShut {Stop}%
\bibitem [{\citenamefont {Miller}(2009)}]{Miller:2009qu}%
  \BibitemOpen
  \bibfield  {author} {\bibinfo {author} {\bibfnamefont {Gerald~A.}\
  \bibnamefont {Miller}},\ }\bibfield  {title} {\enquote {\bibinfo {title}
  {{Singular Charge Density at the Center of the Pion?}}}\ }\href {\doibase
  10.1103/PhysRevC.79.055204} {\bibfield  {journal} {\bibinfo  {journal} {Phys.
  Rev. C}\ }\textbf {\bibinfo {volume} {79}},\ \bibinfo {pages} {055204}
  (\bibinfo {year} {2009})},\ \Eprint {http://arxiv.org/abs/0901.1117}
  {arXiv:0901.1117 [nucl-th]} \BibitemShut {NoStop}%
\bibitem [{\citenamefont {Panteleeva}\ and\ \citenamefont
  {Polyakov}(2021)}]{Panteleeva:2021iip}%
  \BibitemOpen
  \bibfield  {author} {\bibinfo {author} {\bibfnamefont {Julia~Yu.}\
  \bibnamefont {Panteleeva}}\ and\ \bibinfo {author} {\bibfnamefont {Maxim~V.}\
  \bibnamefont {Polyakov}},\ }\bibfield  {title} {\enquote {\bibinfo {title}
  {{Forces inside the nucleon on the light front from 3D Breit frame force
  distributions: Abel tomography case}},}\ }\href {\doibase
  10.1103/PhysRevD.104.014008} {\bibfield  {journal} {\bibinfo  {journal}
  {Phys. Rev. D}\ }\textbf {\bibinfo {volume} {104}},\ \bibinfo {pages}
  {014008} (\bibinfo {year} {2021})},\ \Eprint
  {http://arxiv.org/abs/2102.10902} {arXiv:2102.10902 [hep-ph]} \BibitemShut
  {NoStop}%
\bibitem [{\citenamefont {Kriesten}\ \emph {et~al.}(2020)\citenamefont
  {Kriesten}, \citenamefont {Liuti}, \citenamefont {Calero-Diaz}, \citenamefont
  {Keller}, \citenamefont {Meyer}, \citenamefont {Goldstein},\ and\
  \citenamefont {Osvaldo Gonzalez-Hernandez}}]{Kriesten:2019jep}%
  \BibitemOpen
  \bibfield  {author} {\bibinfo {author} {\bibfnamefont {Brandon}\ \bibnamefont
  {Kriesten}}, \bibinfo {author} {\bibfnamefont {Simonetta}\ \bibnamefont
  {Liuti}}, \bibinfo {author} {\bibfnamefont {Liliet}\ \bibnamefont
  {Calero-Diaz}}, \bibinfo {author} {\bibfnamefont {Dustin}\ \bibnamefont
  {Keller}}, \bibinfo {author} {\bibfnamefont {Andrew}\ \bibnamefont {Meyer}},
  \bibinfo {author} {\bibfnamefont {Gary~R.}\ \bibnamefont {Goldstein}}, \ and\
  \bibinfo {author} {\bibfnamefont {J.}~\bibnamefont {Osvaldo
  Gonzalez-Hernandez}},\ }\bibfield  {title} {\enquote {\bibinfo {title}
  {{Extraction of Generalized Parton Distribution Observables from Deeply
  Virtual Electron Proton Scattering Experiments}},}\ }\href {\doibase
  10.1103/PhysRevD.101.054021} {\bibfield  {journal} {\bibinfo  {journal}
  {Phys. Rev. D}\ }\textbf {\bibinfo {volume} {101}},\ \bibinfo {pages}
  {054021} (\bibinfo {year} {2020})},\ \Eprint
  {http://arxiv.org/abs/1903.05742} {arXiv:1903.05742 [hep-ph]} \BibitemShut
  {NoStop}%
\bibitem [{\citenamefont {Qiu}\ and\ \citenamefont {Yu}(2023)}]{Qiu:2022pla}%
  \BibitemOpen
  \bibfield  {author} {\bibinfo {author} {\bibfnamefont {Jian-Wei}\
  \bibnamefont {Qiu}}\ and\ \bibinfo {author} {\bibfnamefont {Zhite}\
  \bibnamefont {Yu}},\ }\bibfield  {title} {\enquote {\bibinfo {title} {{Single
  diffractive hard exclusive processes for the study of generalized parton
  distributions}},}\ }\href {\doibase 10.1103/PhysRevD.107.014007} {\bibfield
  {journal} {\bibinfo  {journal} {Phys. Rev. D}\ }\textbf {\bibinfo {volume}
  {107}},\ \bibinfo {pages} {014007} (\bibinfo {year} {2023})},\ \Eprint
  {http://arxiv.org/abs/2210.07995} {arXiv:2210.07995 [hep-ph]} \BibitemShut
  {NoStop}%
\bibitem [{\citenamefont {Cocuzza}\ \emph {et~al.}(2022)\citenamefont
  {Cocuzza}, \citenamefont {Melnitchouk}, \citenamefont {Metz},\ and\
  \citenamefont {Sato}}]{Cocuzza:2022jye}%
  \BibitemOpen
  \bibfield  {author} {\bibinfo {author} {\bibfnamefont {C.}~\bibnamefont
  {Cocuzza}}, \bibinfo {author} {\bibfnamefont {W.}~\bibnamefont
  {Melnitchouk}}, \bibinfo {author} {\bibfnamefont {A.}~\bibnamefont {Metz}}, \
  and\ \bibinfo {author} {\bibfnamefont {N.}~\bibnamefont {Sato}} (\bibinfo
  {collaboration} {Jefferson Lab Angular Momentum (JAM)}),\ }\bibfield  {title}
  {\enquote {\bibinfo {title} {{Polarized antimatter in the proton from a
  global QCD analysis}},}\ }\href {\doibase 10.1103/PhysRevD.106.L031502}
  {\bibfield  {journal} {\bibinfo  {journal} {Phys. Rev. D}\ }\textbf {\bibinfo
  {volume} {106}},\ \bibinfo {pages} {L031502} (\bibinfo {year} {2022})},\
  \Eprint {http://arxiv.org/abs/2202.03372} {arXiv:2202.03372 [hep-ph]}
  \BibitemShut {NoStop}%
\bibitem [{\citenamefont {Carlson}\ and\ \citenamefont
  {Vanderhaeghen}(2008)}]{Carlson:2007xd}%
  \BibitemOpen
  \bibfield  {author} {\bibinfo {author} {\bibfnamefont {Carl~E.}\ \bibnamefont
  {Carlson}}\ and\ \bibinfo {author} {\bibfnamefont {Marc}\ \bibnamefont
  {Vanderhaeghen}},\ }\bibfield  {title} {\enquote {\bibinfo {title}
  {{Empirical transverse charge densities in the nucleon and the
  nucleon-to-Delta transition}},}\ }\href {\doibase
  10.1103/PhysRevLett.100.032004} {\bibfield  {journal} {\bibinfo  {journal}
  {Phys. Rev. Lett.}\ }\textbf {\bibinfo {volume} {100}},\ \bibinfo {pages}
  {032004} (\bibinfo {year} {2008})},\ \Eprint {http://arxiv.org/abs/0710.0835}
  {arXiv:0710.0835 [hep-ph]} \BibitemShut {NoStop}%
\bibitem [{\citenamefont {Soper}(1972)}]{Soper:1972xc}%
  \BibitemOpen
  \bibfield  {author} {\bibinfo {author} {\bibfnamefont {D.~E.}\ \bibnamefont
  {Soper}},\ }\bibfield  {title} {\enquote {\bibinfo {title}
  {{Infinite-momentum helicity states}},}\ }\href {\doibase
  10.1103/PhysRevD.5.1956} {\bibfield  {journal} {\bibinfo  {journal} {Phys.
  Rev. D}\ }\textbf {\bibinfo {volume} {5}},\ \bibinfo {pages} {1956--1962}
  (\bibinfo {year} {1972})}\BibitemShut {NoStop}%
\end{thebibliography}%

\end{document}